\documentclass[letterpaper,usenatbib]{mn2e}
\usepackage{amsmath}
\usepackage{epsfig}
\usepackage{times}
\usepackage{verbatim}
\def\gtrsim{~\rlap{$>$}{\lower 1.0ex\hbox{$\sim$}}}
\def\ltsim{~\rlap{$<$}{\lower 1.0ex\hbox{$\sim$}}}

\title[Dust heating mechanisms in nearby galaxies]{The identification
  of dust heating mechanisms in nearby galaxies using {\it Herschel} 160/250 and
  250/350~$\mu$m surface brightness ratios}

\author[G. J. Bendo et al.]
    {G. J. Bendo$^1$, M. Baes$^2$, S. Bianchi$^3$, M. Boquien$^4$, 
     A. Boselli$^5$, A. Cooray$^6$, L. Cortese$^7$, \newauthor
     I. De Looze$^2$, S.  di Serego Alighieri$^3$, J. Fritz$^2$, 
     G. Gentile$^{2,8}$, T. M. Hughes$^2$, N. Lu$^{9}$,\newauthor 
     C. Pappalardo$^{10}$, M. W. L. Smith$^{11}$, 
     L. Spinoglio$^{12}$, S. Viaene$^2$, C. Vlahakis$^{13}$ \\
    $^1$   UK ALMA Regional Centre Node, Jodrell Bank Centre for Astrophysics,
           School of Physics and Astronomy, University of Manchester, 
           Oxford Road,\\ Manchester M13 9PL, United Kingdom\\
    $^2$   Sterrenkundig Observatorium, Universiteit Gent, Krijgslaan 281 S9,  
           B-9000 Gent, Belgium\\
    $^3$   INAF-Osservatorio Astrofisico di Arcetri, Largo Enrico Fermi 5, 
           I-50125 Firenze, Italy\\
    $^4$   Institute of Astronomy, University of Cambridge, Madingley Road, 
           Cambridge CB3 0HA, United Kingdom \\
    $^5 $  Laboratoire d’Astrophysique de Marseille - LAM, Universit\'e 
           d'Aix-Marseille \& CNRS, UMR7326, 38 rue F. Joliot-Curie, 
           F-13388 Marseille Cedex 13,\\ France\\
    $^6$   Department of Physics and Astronomy, University of California, 
           Irvine, CA, 92697, USA\\
    $^7$   Centre for Astrophysics \& Supercomputing, Swinburne University 
           of Technology, Mail H30, PO Box 218, Hawthorn, VIC 3122, Australia\\
    $^8$   Department of Physics and Astrophysics, Vrije Universiteit Brussels, 
           Pleinlaan 2, B-1050 Brussels, Belgium\\
    $^9$   NASA Herschel Science Center, MS 100-22, California Institute of 
           Technology, Pasadena, CA 91125, USA\\
    $^{10}$ CAAUL, Observat\'orio Astron\'omico de Lisboa, Universidade de 
           Lisboa, Tapada da Ajuda, P-1349-018 Lisboa, Portugal\\
    $^{11}$ School of Physics and Astronomy, Cardiff University, 
           Queens Buildings, The Parade, Cardiff CF24 3AA, United Kingdom\\
    $^{12}$ Istituto di Fisica dello Spazio Interplanetario, INAF, Via del 
           Fosso del Cavaliere 100, I-00133 Roma, Italy\\
    $^{13}$ Joint ALMA Observatory/European Southern Observatory,
           Alonso de Cordova 3107, Vitacura, Santiago, Chile
}
\date{}
\pagerange{\pageref{firstpage}--\pageref{lastpage}}
\pubyear{}

\begin{document}
\label{firstpage}
\maketitle

\begin{abstract}
We examined variations in the 160/250 and 250/350~$\mu$m surface
brightness ratios within 24 nearby ($<30$~Mpc) face-on spiral galaxies
observed with the {\it Herschel} Space Observatory to identify the
heating mechanisms for dust emitting at these wavelengths.  The
analysis consisted of both qualitative and quantitative comparisons of
the 160/250 and 250/350~$\mu$m ratios to H$\alpha$ and 24~$\mu$m
surface brightnesses, which trace the light from star forming regions,
and 3.6~$\mu$m emission, which traces the light from the older stellar
populations of the galaxies.  We find broad variations in the heating
mechanisms for the dust.  In one subset of galaxies, we found evidence
that emission at $\leq160$~$\mu$m (and in rare cases potentially at
$\leq 350$~$\mu$m) originates from dust heated by star forming
regions.  In another subset, we found that the emission at $\geq
250$~$\mu$m (and sometimes at $\geq 160$~$\mu$m) originates from dust
heated by the older stellar population.  In the rest of the sample,
either the results are indeterminate or both of these stellar
populations may contribute equally to the global dust heating.  The
observed variations in dust heating mechanisms does not necessarily
match what has been predicted by dust emission and radiative transfer
models, which could lead to overestimated dust temperatures,
underestimated dust masses, false detections of variability in dust
emissivity, and inaccurate star formation rate measurements.
\end{abstract}


\section{Introduction}
\label{s_intro}
\addtocounter{footnote}{13}

The {\it Herschel} Space Observatory \citep{pilbratt10} was the first
telescope that had both the sensitivity and spatial resolution to map
250-500~$\mu$m dust emission from both compact and diffuse regions
within nearby spiral galaxies (galaxies within a distance of 30~Mpc)
on spatial scales on the order of 1~kpc.  Consequently, the telescope
has detected a heretofore unseen dust component in nearby spiral
galaxies that is heated primarily by the evolved stellar
  population and that emits primarily at $>$250~$\mu$m.  This
component was first identified in M81 by comparing 70/160, 160/250,
250/350, and 350/500~$\mu$m surface brightness ratios measured within
0.5~kpc subregions to tracers of star formation and the evolved
stellar population\footnote{\citet{bendo12a} refer to this as the
  ``total stellar population'', as they did not remove the
  contribution of younger stars to the near-infrared band used to
  trace this population.  We explicitly demonstrate that photoionising
  stars have a minimal effect on the near-infrared data we use, but
  non-photoionising intermediate-aged stars with ages $<$100~Myr may
  still strongly influence the band.  For simplicity, we refer to this
  as the ``evolved stellar population'' in this
  paper}. \citep{bendo10, bendo12a} This analysis was more capable
than various spectral energy distribution (SED) fitting techniques in
detecting this cold component for multiple reasons.  First, the
analysis utilised the spatial information about both the dust emission
and its heating sources.  Second, it eliminated many assumptions about
the SED of the illuminating radiation fields or the physical
properties of the dust grains.  Third, it made no assumptions about
the expected SED shape from the dust emission.  Fourth, it focused on
the analysis of subsegments of individual sections of the dust SED
rather than attempting to simultaneously describe emission at all
wavelengths, which could lead to false assumptions or inferences on
the relation between dust seen at shorter and longer wavelengths.

The same methods were later used to detect dust heated by the evolved
stellar population in M31 \citep{planck14}, M33 \citep{boquien11}, M83
\citep{bendo12a}, and NGC~2403 \citep{bendo12a}.  In contrast,
\citet{galametz10} and \citet{hughes14} identified nearby galaxies
where the infrared surface brightness ratios clearly indicated that
dust emission seen at $\leq 350$~$\mu$m was heated by young stars
(generally photoionising stars $\leq3$ Myr in age) in star forming
regions.  Unfortunately, when \citet{boselli12} used global infrared
surface brightness ratios for spiral galaxies in the {\it Herschel}
Reference Survey \citep[HRS; ][]{boselli10}, the results were more
ambiguous; the ratios were correlated as well with emission from star
formation tracers as from the evolved stellar population, probably
because stellar mass and star formation rate are correlated when
integrated over global scales.

Spectral energy distribution (SED) fitting techniques have also been
used to study dust heating on local and global scales, typically by
blindly fitting a model with two or more thermal components (either
modified blackbodies or more complex templates that includes dust
heated locally by star formation and dust heated by the diffuse
ISRF) while ignoring any spatial information on dust heating sources
\citep[e.g. ][]{draine07b, dacunha08, rowanrobinson10, aniano12,
  dale12, mentuchcooper12, ciesla14}.  These approaches may sometimes
but not always reproduce the results seen from the analysis of
infrared surface brightness ratios.  For example, the
\citet{draine07b} dust emission models and the \citet{bendo12a}
infrared ratio analysis both find for M81 that the 70-500~$\mu$m
emission is dominated by a component heated by the diffuse ISRF.  For
NGC~2403, however, the \citet{draine07b} models predict that dust
heated by the diffuse ISRF is seen at $>$70~$\mu$m, whereas the
\citet{bendo12a} analysis found that dust heated locally by star
forming regions dominates the SED at $<$160~$\mu$m.  This example
demonstrated that, while these dust models may accurately replicate
the magnitudes of the observed infrared surface brightnesses of
extragalactic sources, it is unclear that they can replicate the
infrared colours of these sources, which may indicate deeper problems
with how the dust absorption and emission is modelled as well as how
the dust mass is estimated.

Radiative transfer models that reproduce the dust attenuation in
ultraviolet and optical wavelengths as well as the emission at
infrared wavelengths \citep[e.g.][]{silva98, misisriotis01, bianchi08,
  baes11, popescu11, domingueztenreiro14} can also be used to examine
dust heating.  These models can quantify dust heating by different
stellar populations and can also account for the propagation of
radiation through the ISM, which allows them to characterise non-local
dust heating.  However, such models are generally more easily applied
to edge-on galaxies where structures are integrated along the line of
sight and information about the spatial distribution of stars and dust
is lost, although De Looze et al. (2014, in preparation) has performed
radiative transfer modelling of the face-on spiral galaxy M51.
Moreover, they are heavily dependent upon properly characterising
both the stellar populations and the dust properties, whereas more
empirically-based analyses relating colour variations to dust heating
sources are not strongly affected these issues.

It is also common to estimate dust temperatures and masses by fitting
all far-infrared data at $\geq 100$ or $\geq 160$~$\mu$m using a
single blackbody modified by an emissivity law that scales as
$\lambda^{-\beta}$ \citep[e.g. ][]{smith10, foyle12, smith12b,
  davies12, galametz12, groves12, auld13, tabatabaei14, cortese14}.
This often relies upon the a priori assumption that all emission seen
within a wavelength range in all locations in all galaxies can be
attributed to a single thermal component with a single heating source.
However, as can be seen from the results from \citet{bendo12a}, the
wavelength at which the SED transitions between dust heated by the
ISRF and dust heated by star forming regions may be different for
different galaxies.  In some cases, the transition occurs between 160
and 250~$\mu$m (as seen in M33, M83 and NGC 2403), while in other
cases, the transition may occur at $<$160~$\mu$m (as seen in the outer
disc of M81), and in some locations, it is possible to see dust
emission over the 70-500~$\mu$m range heated by a single thermal
component (as seen in the centre of M81).  

Hence, while single thermal component fits to 100-500~$\mu$m data may
accurately describe the shape of the galaxy SEDs, the dust
temperatures and masses as well as dust emissivity variations implied
by the fits may not represent the true physical properties of the
coldest dust within the galaxies.  Furthermore, measured dust
emissivity variations among or within galaxies may actually be
indicative of issues with properly separating emission from warmer and
colder dust components \citep[e.g. ][]{kirkpatrick14}.  Moreover,
\citet{galliano11} and \citet{xilouris12} have demonstrated that
colder diffuse dust could effectively be masked by warmer components
in SED fits unless special care was taken to spatially separate the
components.

It is critically important to properly identify the heating sources
for dust in galaxies and to properly separate dust into individual
thermal components with different heating sources so that the
temperature of the colder dust can be determined accurately.  Since
the colder thermal component typically contains most of the dust mass,
the accuracy of the colder dust temperature has an effect on the
accuracy of dust mass measurements.  Accurate dust temperature and
mass measurements may be critically important if dust mass is used as
a proxy for gas mass in scenarios where emission from gas cannot be
directly measured or in studies where estimates of the gas mass may be
uncertain, such as when the conversion from CO line emission to
molecular gas mass is suspect.

It is also important to identify the heating sources of dust so as to
accurately model radiative transfer within galaxies.  Moreover, it is
possible that far-infrared emission is related to star formation not
because the dust is heated by young stars but instead because the dust
traces the gas fueling star formation, as suggested by
\citet{boquien11} and \citet{bendo12a}.  If so, then this could affect
how star formation is calculated using far-infrared flux measurements
and could also affect the analyses of things such as gas-depletion
times in normal and starburst galaxies, the far-infrared to radio
correlation, and the ``main sequence'' relation between star formation
rate and stellar mass.

To better understand dust heating mechanisms within nearby galaxies,
we present here an expanded comparison of infrared surface brightness
ratios to dust heating sources within a set of 24 face-on spiral
galaxies with well-resolved far-infrared emission that consists of 11
galaxies from the HRS, 10 galaxies from the Key Insights on Nearby
Galaxies: A Far-Infrared Survey with {\it Herschel} \citep[KINGFISH;
][]{kennicutt11}, and the three galaxies from \citet{bendo12a}.  This
larger sample of galaxies can be used to determine whether the cold
dust component heated by the evolved stellar population is
consistently seen among all spiral galaxies or if, in some spiral
galaxies, star formation may be the dominant heating source for the
dust seen at all wavelengths.  Additionally, with the larger sample of
galaxies, it will be possible to look at variations in the relative
contributions of star forming regions and the evolved stellar
populations among a large sample of galaxies (as already seen within
the three galaxies studied by \citet{bendo12a}) and to possibly link
these variations with other physical properties of the galaxies.

The analysis primarily focuses on the 160/250 and 250/350~$\mu$m
surface brightness ratios based on 160~$\mu$m data taken with the
Photodetector Array Camera and Spectrometer \citep[PACS;
][]{poglitsch10} and 250 and 350~$\mu$m data taken with the Spectral
and Photometric Imaging REceiver \citep[SPIRE; ][]{griffin10}.  These
data cover the transition between warmer dust heated by star forming
regions and colder dust heated by the evolved stellar population
found previously by \citet{bendo10}, \citet{boquien11}, and
\citet{bendo12a}, so this is a good part of the far-infrared waveband
to examine.  We had PACS 70 and 100~$\mu$m data for most galaxies in
our sample but did not use it in most of our analysis because we had
difficulty detecting emission above a $3\sigma$ threshold from
  diffuse low surface brightness regions, particularly in the outer
  parts of the optical discs.  SPIRE 500~$\mu$m data were also
available, but we did not use the data mainly because the 500~$\mu$m
PSF has a full-width at half-maximum (FWHM) of 36~arcsec, which would
cause problems with attempting to resolve structures within many of
the sample galaxies.  Moreover, although \citet{bendo12a} was still
able to relate the 350/500~$\mu$m variations in their galaxies to
heating sources, it was clear that the 350/500~$\mu$m ratios were
relatively insensitive to temperature variations compared to shorter
wavelength ratios.  For these reasons, we limit our analyses to the
160/250 and 250/350~$\mu$m ratios.

Following descriptions of the sample in Section~\ref{s_sample} and the
data in Section~\ref{s_data}, we present three different parts of the
analysis.  Section~\ref{s_qualanalysis} presents a qualitative
analysis in which we compare 160/250 and 250/350~$\mu$m surface
brightness ratio maps to tracers of star forming regions and the evolved
stellar population.  Section~\ref{s_singlesourceanalysis} presents a
quantitative comparison of the 160/250 and 250/350~$\mu$m ratios to
the tracers of these two different heating sources.
Section~\ref{s_decompositionanalysis} includes an analytical approach
first proposed by \citep{bendo12a} in which the 160/250 and
250/350~$\mu$m ratios are fitted as a function of both the evolved
stellar population and star formation; the results of the fits can be
used to determine the fraction of the variation in the surface
brightness ratios that can be attributed to a dust heating source.
After this analysis, we discuss the implications of the results for
modelling dust emission and for measuring star formation rates in
Section~\ref{s_discussion}.  A summary of our results for each galaxy
as well as for the sample as a whole is presented in
Section~\ref{s_conclusions}.

\section{Sample}
\label{s_sample}

As stated in the introduction, the sample used for this analysis is
drawn from three other samples: HRS, KINGFISH, and the
\citet{bendo12a} sample, which is a subsample of galaxies from the
Very Nearby Galaxies Survey (VNGS; PI: C. Wilson).  The three surveys
have different purposes.  The HRS is a complete sample of 323 galaxies
that fall within distances between 15 and 25 Mpc that are also brighter
than K-band total magnitude of 12 (applied to Sa-Im and blue compact
dwarf galaxies) or 8.7 (applied to E, S0, and S0/a galaxies).  This
survey is intended to provide statistical information about the
properties of galaxies (e.g. far-infrared colours, far-infrared
luminosities, dust temperatures, dust masses, dust emissivities, etc.).
KINGFISH is a continuation of the {\it Spitzer} Infrared Nearby
Galaxies Survey \citep[SINGS; ][]{kennicutt03} and consists of a
sample of galaxies selected to span a range of morphologies, infrared
luminosities, and infrared/optical luminosity ratios.  The sample does
not have the same applications as a volume- or flux-limited sample,
but since SINGS and KINGFISH have also gathered additional infrared
spectral data and other ancillary data for the galaxies, the sample is
very useful for multiwavelength analyses.  The VNGS is a project that
observed 13 galaxies with {\it Herschel}, and while the sample is not
a statistically complete or representative sample, it includes many
very well-studied nearby galaxies, many of which are archetypal
representatives of a class of objects (e.g. Arp 220, Cen A, M82).

Using data from all three of these surveys is not ideal, as the
resulting sample is somewhat heterogeneous.  However, given the
stringent selection criteria that we needed to apply, it was necessary
to select galaxies from multiple surveys to build up our sample.  The
HRS and KINGFISH samples were used mainly because they are two of
the largest photometric surveys performed with {\it Herschel} that
included spiral galaxies.  The VNGS galaxies were included mainly
because the data from the previous analysis are already in hand.

To select galaxies for our analysis, we applied the following criteria:
\medbreak
\noindent 1. The galaxies must have morphological types between Sa and
Sd.  Irregular galaxies have low infrared surface brightnesses, making
it difficult for us to apply our analysis methods.  Emission from
E-S0/a galaxies is generally compact \citep[e.g.][]{bendo07,
  munozmateos09, smith12a}, so the infrared emission observed by {\it
  Herschel} is usually unresolved or marginally resolved, and our
analysis will not work.  In cases where emission is detected in
extended structures in E-S0/a galaxies (such as NGC 1291 or NGC 4406),
it is impractical to work with the data because of the relatively low
surface brightness of the extended emission.
\medbreak
\noindent 2. The galaxies must be face on (with minor/major axis
ratios of $>$0.5).  This eliminates edge-on and steeply inclined
galaxies where integration along the line of sight may complicate the
analysis.  
\medbreak
\noindent 3. The galaxies must have optical major axes that are
$>$5~arcmin.  This ensures that sufficient data are available to
examine substructures within the galaxies at the resolution of the
350~$\mu$m data.  
\medbreak
\noindent 4. After the application of the data preparation steps in
Section~\ref{s_data_preparation}, the 250/350~$\mu$m surface
brightness ratio must be measured at the $3\sigma$ level within a
$\geq7$~arcmin$^2$ area (which is equivalent to the area of a circle
with a diameter of 3~arcmin).  This ensures that enough data from each
galaxy are available for the analysis.  (The HRS galaxies NGC 3227,
3338, and 4450 and the KINGFISH galaxies NGC 1512 and 4826 satisfy
criteria 1-3 but not criterion 4.)
\medbreak
\noindent 5. The galaxies should not contain infrared-bright
point-like nuclei that contain significant fractions ($\gtrsim$30\%) of
the total infrared flux.  This is partly because the PSF matching
steps we apply in Section~\ref{s_data_preparation} produces severe
artefacts in the ratio maps around such bright sources; excluding the
artefacts also results in excluding large fractions of the
infrared emission from the galaxies.  High contrast between the
central region and the rest of the disc also produce ambiguous results
when comparing infrared surface brightness ratios to emission from
multiple dust heating sources.  This criteria excludes NGC~1097 and
3351 from the sample.

\begin{table*}
\centering
\begin{minipage}{165mm}
\caption{Sample galaxies.}
\label{t_sample}
\begin{tabular}{@{}lcccccc@{}}
\hline
Galaxy &      
    Survey$^a$ &
    Hubble &
    Size of optical &
    Distance &
    Distance &
    Physical size of \\
& 
    &
    type$^b$ &
    disc (arcmin$^b$) &
    (Mpc) &
    Reference &
    24 arcsec (kpc) \\
\hline
NGC 628 (M74) &
    KINGFISH &
    SA(s)c &
    $10.5 \times 9.5$ &
    $9.9 \pm 1.2$ &
    \citet{olivares10} &
    1.2\\
NGC 925 &
    KINGFISH &
    SAB(s)d &
    $10.4 \times 5.8$ &
    $9.2 \pm 0.2$ &
    \citet{freedman01} &
    1.1\\
NGC 2403 &
    VNGS &
    SAB(s)cd &
    $21.9 \times 12.3$ &
    $3.2 \pm 0.4$ &
    \citet{freedman01} &
    0.4\\
NGC 3031 (M81) &
    VNGS &
    SA(s)ab &
    $26.9 \times 14.1$ &
    $3.6 \pm 0.1$ &
    \citet{freedman01} &
    0.4\\
NGC 3184 &
    KINGFISH &
    SAB(rs)cd &
    $7.4 \times 6.9$ &
    $12.6 \pm 1.6$ &
    \citet{olivares10} &
    1.5\\
NGC 3621 &
    KINGFISH &
    SA(s)d &
    $12.3 \times 7.1$ &
    $6.6 \pm 0.2$ &
    \citet{freedman01} &
    0.8\\
NGC 3631 &
    HRS &
    SA(s)c &
    $5.0 \times 3.2$ &
    $17.5 \pm 3.2$ &
    \citet{thereau07} &
    2.0\\
NGC 3938 &
    KINGFISH &
    SA(s)c &
    $5.3 \times 4.9$ &
    $17.9 \pm 1.1$ &
    \citet{poznanski09} &
    2.1\\
NGC 3953 &
    HRS &
    SB(r)bc &
    $6.9 \times 3.5$ &
    $17.5 \pm 0.8$ &
    \citet{dessart08} &
    2.0\\
NGC 4254 (M99) &
    HeViCS/HRS/KINGFISH &
    SA(s)c &
    $5.3 \times 4.6$ &
    $14.6 \pm 1.9$ &
    \citet{poznanski09} &
    1.7\\
NGC 4303 (M61) &
    HeViCS/HRS &
    SAB(rs)bc &
    $6.4 \times 5.7$ &
    $12.2 \pm 1.9$ &
    \citet{roy11} &
    1.4\\
NGC 4321 (M100) &
    HeViCS/HRS/KINGFISH &
    SAB(s)bc &
    $7.4 \times 6.3$ &
    $15.2 \pm 0.5$ &
    \citet{freedman01} &
    1.8\\
NGC 4501 (M88) &
    HeViCS/HRS &
    SA(rs)b &
    $6.9 \times 3.7$ &
    $11.7 \pm 1.4$ &
    \citet{amanullah10} &
    1.4\\
NGC 4535 &
    HeViCS/HRS &
    SAB(s)c &
    $7.1 \times 5.0$ &
    $15.8 \pm 3.6$ &
    \citet{freedman01} &
    1.8\\
NGC 4548 (M91) &
    HeViCS/HRS &
    SB(rs)b &
    $5.4 \times 4.3$ &
    $16.2 \pm 0.4$ &
    \citet{freedman01} &
    1.9\\
NGC 4579 (M58) &
    HeViCS/HRS/KINGFISH &
    SAB(rs)b &
    $5.9 \times 4.9$ &
    $18.8 \pm 3.0$ &
    \citet{tully09} &
    2.2\\
NGC 4725 &
    HRS/KINGFISH &
    SAB(r)ab pec &
    $10.7 \times 7.6$ &
    $12.4 \pm 0.3$ &
    \citet{freedman01} &
    1.4\\
NGC 4736 (M94) &
    KINGFISH &
    RSA(r)ab &
    $11.2 \times 9.1$ &
    $5.2 \pm 0.4$ &
    \citet{tonry01} &
    0.6\\
NGC 5055 (M63) &
    KINGFISH &
    SA(rs)bc &
    $12.6 \times 7.2$ &
    $7.9 \pm 1.3$ &
    \citet{tully09}  &
    0.9\\
NGC 5236 (M83) &
    VNGS &
    SAB(s)c &
    $12.9 \times 11.5$ &
    $4.6 \pm 0.3$ &
    \citet{saha06} &
    0.5\\
NGC 5364 &
    HRS &
    SA(rs)bc pec &
    $6.8 \times 4.4$ &
    $21.6 \pm 4.0$ &
    \citet{thereau07} &
    2.5\\
NGC 5457 (M101) &
    KINGFISH &
    SAB(rs)cd &
    $28.8 \times 26.9$ &
    $6.7 \pm 0.3$ &
    \citet{freedman01} &
    0.8\\
NGC 6946 &
    KINGFISH &
    SAB(rs)cd &
    $11.5 \times 9.8$ &
    $4.7 \pm 0.7$ &
    \citet{olivares10} &
    0.5\\
NGC 7793 &
    KINGFISH &
    SA(s)d &
    $9.3 \times 6.3$ &
    $3.4 \pm 0.1$ &
    \citet{pietrzynski10} &
    0.4\\
\hline
\end{tabular}
$^a$ The surveys that include the galaxy in this list.  In some cases, 
     galaxies appear in multiple surveys.\\
$^b$ Data are taken from \citet{devaucouleurs91}.\\
\end{minipage}
\end{table*}

\medskip
A total of 24 galaxies meet all five of the above criteria.  The
sample galaxies along with the survey from which the data were taken
and morphological, optical disc, and distance information are listed
(with the galaxies sorted by NGC number) in Table~\ref{t_sample}.
Seven of the HRS galaxies fall within the $\sim$64~deg$^2$ fields
mapped by the {\it Herschel} Virgo Cluster Survey \citep[HeViCS;
][]{davies10} at 100-500~$\mu$m; we used the HeViCS data for some of
these targets.  We also note that 4 galaxies fall within both the HRS
and KINGFISH surveys.

\section{Data}
\label{s_data}

\subsection{Far-infrared data}
\label{s_data_fir}

As stated in Section~\ref{s_intro}, the analysis mainly relies upon
160~$\mu$m data acquired with PACS and 250 and 350~$\mu$m data acquire
with SPIRE.  However, we did reprocess the PACS 100~$\mu$m images of
NGC~628 and 5457 for performing global photometry measurements on
those two galaxies (see Section~\ref{s_discussion_sed}).  The PACS and
SPIRE data were obtained by the surveys specified in
Table~\ref{t_sample}.  We used the HeViCS parallel-mode scan map data
(which simultaneously produced PACS and SPIRE data) to create the
160~$\mu$m image of NGC~4254 and the 160-350~$\mu$m images of NGC~4303
and 4535.  For all other galaxies, we used standard PACS and SPIRE
scan map data.  Details on the observations are given by
\citet{bendo12a} for the VNGS galaxies, \citet{auld13} for HeViCS,
\citet{boselli10} and \citet{cortese14} for HRS galaxies not in the
HeViCS fields, and \citet{kennicutt11} for KINGFISH galaxies not in
the HRS sample.

The individual PACS data frames were all uniformly reprocessed using
the standard PACS pipeline within the {\it Herschel} Interactive
Processing Environment \citep[{\small HIPE}; ][]{ott10} version 12.0,
which includes cosmic ray removal and crosstalk correction.  The data
were then mosaicked using {\small SCANAMORPHOS} version 23
\citep{roussel13}, which also removes variations in the instrumental
background signal over time.  The final pixel scales are 1.7~arcsec at
100~$\mu$m and 2.85~arcsec and 160~$\mu$m.  We then applied colour
correction factors of $1.03 \pm 0.06$ to the 100 micron data and $1.01
\pm 0.07$ to the 160~$\mu$m data based on what would be expected for
emission from a modified blackbody with a temperature between 15 and
40~K and a $\beta$ value between 1 and 2
\citep{muller11}\footnote{http://herschel.esac.esa.int/twiki/pub/Public/PacsCalibrationWeb\\ /cc\_report\_v1.pdf}.
The FWHM of the PSF is approximately 7~arcsec at 100~$\mu$m and
12~arcsec at 160~$\mu$m \citep{lutz12}\footnote{
  https://herschel.esac.esa.int/twiki/pub/Public/PacsCalibrationWeb\\ /bolopsf\_20.pdf}.
The flux calibration uncertainty is 5\%
\citep{pacs13}\footnote{http://herschel.esac.esa.int/Docs/PACS/pdf/pacs\_om.pdf}, and along with the colour correction, the overall uncertainties in the
scaling of the flux densities is $\sim$9\%.

The SPIRE data were all uniformly reprocessed using HIPE developer
build version 12.0.  The timeline data were processed through the
standard pipeline, which includes electrical crosstalk correction,
cosmic ray removal, signal jump removal, and response corrections.  In
place of the temperature drift removal and baseline signal removal,
however, we used the BRIght Galaxy ADaptive Element method
\citep[][Smith et al., in preparation]{smith12, auld13}, which uses
the SPIRE thermistors to characterise the drift in the background
signal.  Relative gain corrections were applied to correct for the
responsivities of the bolometers to extended sources.  The data were
then mapped using the naive mapmaker within {\small HIPE} using final
pixel scales of 6~arcsec at 250~$\mu$m and 8~arcsec and 350~$\mu$m.
To optimise the data for extended source photometry, we applied the
$K_{\mbox{PtoE}}$ values of 91.289 and 51.799 MJy sr$^{-1}$ (Jy
beam$^{-1}$)$^{-1}$ as suggested by the SPIRE Handbook Version 2.5
\citep{spire14}\footnote{http://herschel.esac.esa.int/Docs/SPIRE/spire\_handbook.pdf}.
We also applied colour corrections of $0.997 \pm 0.029$ to the
250~$\mu$m data and $0.991 \pm 0.030$ to the 350~$\mu$m data, which
should be appropriate for extended modified blackbody emission with
temperatures between 10 and 40 K and $\beta$ of 1.5 or 2
\citep{spire14}.  The FWHM is approximately 18~arcsec at 250~$\mu$m
and 25~arcsec at 350~$\mu$m \citep{spire14}.  The overall
uncertainties in the scaling of the flux densities, which encompass
the uncertainties from the flux calibration from \citet{bendo13}, the
correction factors for the older flux calibration, the beam area
uncertainties, and the colour correction uncertainties, is $\sim$5\%.

Note that the uncertainties in the flux densities and the correction
factors applied to the data do not generally affect most of analysis,
which is dependent upon the relative variations in surface brightness
ratios and therefore independent of the exact scale of the surface
brightnesses.  The only part of the analysis that is significantly
affected is the analysis on the SEDs in
Section~\ref{s_discussion_sed}, where the absolute values affect the
resulting dust masses and temperatures.

\subsection{Star formation tracers}
\label{s_sftracer}

To trace the energy from star forming regions that is heating the
dust, we have a few options.  Ultraviolet, H$\alpha$, and mid-infrared
(22 or 24~$\mu$m) images are available for most or all of the sample
galaxies.  However, these three star formation tracers have different
strengths and weaknesses.

Uniform ultraviolet images from the {\it Galaxy Evolution Explorer}
are available for 22 of the galaxies, but the emission is very
strongly affected by dust extinction.  Also, while H$\alpha$ and
24~$\mu$m emission tend to trace O stars with $<$3~Myr lifespans,
ultraviolet light can trace B and A stars with lifespans of tens of
Myr.  Consequently, ultraviolet emission traces different structures
than H$\alpha$ or 24~$\mu$m emission \citep[e.g.][]{calzetti05}.
Given these complications, we did not work with ultraviolet data
for our analysis.

H$\alpha$ images are very useful to work with because they trace not
only emission from photoionised gas within star forming regions
themselves but also the photoionising light from the star forming
regions that is absorbed by the diffuse ISM.  If the diffuse dust
outside star forming regions are partially heated by these photons,
then we should find a better correlation between the far-infrared
colours and H$\alpha$ emission.  Additionally, the H$\alpha$ band will
be relatively unaffected by older stellar populations since stars with
ages of $<$3~Myr produce relatively few photoionising photons (as can
be seen, for example, in the simulations from \citet{leitherer99}).
H$\alpha$ light is affected by dust extinction, but most star forming
regions are still visible in optical light; only $\sim$4\% of all star
forming regions are completely obscured in the H$\alpha$ band
\citep{prescott07}.  Although H$\alpha$ is theoretically a good star
formation tracer, the H$\alpha$ images available for our analysis are
sometimes difficult to deal with in practice.  The data were acquired
from multiple telescopes, have multiple pixel scales and PSF sizes,
and cover varying amounts of area around the targets.  Moreover,
because H$\alpha$ images are produced by subtracting continuum
emission from the data, the images often contain artefacts such as
residual background structures or incompletely subtracted starlight
from either foreground stars or the bulge stars within the target
galaxies.  Furthermore, H$\alpha$ images often contain additional
[N{\small II}] emission that cannot be easily removed on a
pixel-by-pixel basis.

Mid-infrared hot dust emission as observed at 24~$\mu$m by the
Multiband Imaging Photometer for {\it Spitzer} \citep{rieke04} on the
{\it Spitzer} Space Telescope \citep{werner04} or at 22~$\mu$m by the
{\it Wide-field Infrared Survey Explorer} \citep[{\it WISE};
][]{wright10} is available for all of the galaxies in the sample.
Mid-infrared bands are as effective as hydrogen recombination lines in
measuring star formation within compact sources
\citep[e.g.][]{calzetti05, calzetti07, prescott07}, and the emission
is unaffected by dust extinction except in extreme environments.
However, the band may also include emission from diffuse dust heated
by older stars \citep[e.g. ][]{kennicutt09}. The mid-infrared data are
relatively straightforward to work with in practice.  While the data
from the two telescopes differ in format, the images from each
telescope all have a uniform format, even though the {\it Spitzer}
data may have been processed by different groups using slightly
different data reduction pipelines.

Several authors have also published equations in which a tracer of
unobscured star formation and a tracer of obscured star formation are
combined together to produce a more robust star formation tracer.  We
decided to use a combined H$\alpha$ and mid-infrared (22 or 24~$\mu$m)
metric for most of our analysis because both trace very similar
structures \citep{calzetti07}, which means that they trace regions
affected by the same populations of photoionising stars.  For
combining H$\alpha$ emission with mid-infrared emission, we used
\begin{equation}
f(\mbox{H}\alpha)_{\mbox{corrected}} = 
    f(\mbox{H}\alpha)_{\mbox{observed}} + 0.020 \nu f_\nu(24 \mu\mbox{m})
\label{e_hacorr}
\end{equation}
from \citet{kennicutt09}, which we will henceforth refer to as the
extinction-corrected H$\alpha$ emission.  We decided to use this
relation because it was derived to work with emission from both
compact and diffuse emission (as opposed to the calibration derived by
\citet{calzetti07}, which is better suited for compact regions after
the application of local background subtraction).  We lack {\it
  Spitzer} 24~$\mu$m data for NGC~3631 and NGC~5364 and used {\it
  WISE} data as a substitute (although, when we refer to the {\it
  Spitzer} and {\it WISE} data together, we will use the term
24~$\mu$m data for simplicity).  Although separate calibrations have
been derived for combining H$\alpha$ and {\it WISE} data by
\citet{lee13}, the coefficients in the equivalent equation differs by
a factor of 1.5 from the coefficients for {\it Spitzer} 24~$\mu$m data
by \citet{kennicutt09} or other authors \citep[e.g.][]{zhu08} even
though the {\it Spitzer} 24~$\mu$m and {\it WISE} 22~$\mu$m flux
densities do not differ this much.  Hence, we also use
Equation~\ref{e_hacorr} for the {\it WISE} data.  We did not have
access to suitable H$\alpha$ data for NGC~4736 or 6946 that cover the
full optical discs of these galaxies, so we used the {\it Spitzer}
24~$\mu$m data by itself as a star formation tracer for these
galaxies.

The combined H$\alpha$ and mid-infrared metric is still not a perfect
star formation tracer, mainly because it still potentially includes
diffuse 24~$\mu$m emission heated by the ISRF from older stars.  It is
also possible that the dust heated by star forming regions may
correlate better with one of the star formation tracers instead of the
other.  Hence, in Appendix~\ref{a_ha24comp}, we examined how using
H$\alpha$ or 24~$\mu$m emission by themselves affected the results for
the quantitative analysis in Section~\ref{s_singlesourceanalysis}.  We
found five galaxies where using one tracer instead of the other could
have a significant impact on identifying the dust heating sources, but
for most galaxies, the results do not depend upon the selected star
formation tracer.  It is also possible that other star formation
tracers could be better related to the infrared surface brightness
variations because they more accurately trace the population that is
heating the dust, but it was beyond the scope of this paper to
investigate tracers other than H$\alpha$ and 24~$\mu$m emission.

\begin{table*}
\centering
\begin{minipage}{163mm}
\caption{Information on H$\alpha$ data.}
\label{t_hadata}
\begin{tabular}{@{}lcccccccc@{}}
\hline
Galaxy &   
           H$\alpha$ &     
           PSF &        
           Pixel &
           $1\sigma$ Sensitivity &           
           Calibration &
           Foreground &            
           [N {\small II}]/H$\alpha$ &
           [N {\small II}]/H$\alpha$ \\
& 
           Source &                       
           FWHM &
           Size &              
           ($10^{-17}$  erg cm$^{-2}$  &
           Uncertainty &
           Extinction$^b$ &                       
           Ratio &
           Reference$^c$\\          
& 
           & 
           (arcsec) &
           (arcsec) &              
           s$^{-1}$ arcsec$^{-2}$)$^a$ &
           &
           &                       
           &
           \\          
\hline
NGC 628 &  
           SINGS &
           2 &
           0.43 &
           0.9 &
           10\% &
           0.152 &
           0$^d$ & 
           \\
NGC 925 &  
           \citet{boselli02} &
           5 &
           0.69 &
           4.5 & 
           5\% &
           0.165 &
           0.20 &
           K09 \\
NGC 2403 & 
           \citet{boselli02} &
           3 &
           0.69 &
           2.1 &
           5\% &
           0.087 &
           0.22 &
           K09 \\
NGC 3031 & 
           \citet{boselli02} &
           3 &
           0.69 &
           5.6 &
           5\% &
           0.174 &
           0.55 &
           K09 \\
NGC 3184 & 
           SINGS &
           2 &
           0.30 &
           3.7 &
           10\% &
           0.036 &
           0.52 &
           K09 \\
NGC 3621 & 
           SINGS$^e$ &
           1 &
           0.27 &
           27.9 &
           10\% &
           0.046 &
           0.40 &
           K09 \\
NGC 3631 & 
           Boselli et al. (in prep) &
           2 &
           0.31 &
           6.8 &
           5\% &
           0.036 &
           0.11 &
           B13 \\
NGC 3938 & 
           SINGS & 
           2 &
           0.30 &
           1.4 &
           10\% &
           0.046 &
           0.42 &
           K09 \\
NGC 3953 & 
           Boselli et al. (in prep) & 
           2 &
           0.31 &
           2.3 &
           5\% &
           0.065 &
           0.52 &
           B13 \\
NGC 4254 & 
           \citet{boselli02} &
           3 &
           0.69 &
           1.9 &
           5\% &
           0.084 &
           0.42 &
           B13 \\
NGC 4303 & 
           \citet{boselli02} &
           2 &
           0.41 &
           2.5 &
           5\% &
           0.048 &
           0.48 &
           B13 \\
NGC 4321 & 
           \citet{boselli02} &
           3 &
           0.69 &
           1.7 &
           5\% &
           0.057 &
           0.48 &
           B13 \\
NGC 4501 & 
           \citet{boselli02} &
           3 &
           0.69 &
           2.0 &
           5\% &
           0.082 &
           0.47 &
           B13 \\
NGC 4535 & 
           \citet{boselli02} &
           3 &
           0.69 &
           2.8 &
           5\% &
           0.042 &
           0.49 &
           B13 \\
NGC 4548 & 
           Boselli et al. (in prep) & 
           2 &
           0.41 &
           1.3 &
           5\% &
           0.083 &
           0.41 &
           B13 \\
NGC 4579 & 
           Boselli et al. (in prep) &
           2 &
           0.41 &
           6.3 &
           5\% &
           0.088 &
           0.92 &
           B13 \\
NGC 4725 & 
           SINGS &
           2 &
           0.31 & 
           1.6 &
           10\% &
           0.026 &
           0.49$^g$ &
           \\
NGC 4736 & 
           $^f$ &
           &
           &
           &
           &
           &
           \\
NGC 5055 & 
           SINGS$^e$ &
           2 &
           0.38 &
           6.9 &
           10\% &
           0.038 & 
           0.49 &
           K09 \\
NGC 5236 & 
           \citet{meurer06} &
           2 &
           0.43 &
           5.3 &
           4\% &
           0.144 &
           0.40$^h$ &
           K08 \\
NGC 5364 & 
           Boselli et al. (in prep) & 
           2 &
           0.31 &
           5.3 &
           5\% &
           0.059 &
           0.46$^g$ &
           \\
NGC 5457 & 
           \citet{hoopes01}$^e$ &  
           6 &
           2.03 &
           0.7 &
           15\% &
           0.019 &
           0.54 &
           K08 \\
NGC 6946 & 
           $^f$ &
           &
           &
           &
           &
           &
           \\
NGC 7793 & 
           SINGS &
           1 &
           0.43 &
           2.6 &
           10\% &
           0.042 &
           0$^d$ &
           \\
\hline
\end{tabular}
$^a$ These $1\sigma$ sensitivities correspond to measurements in
  the original data (before any convolution or rebinning steps are
  applied).  It does include corrections that remove foreground extinction and
  emission from the [N {\small II}] line. \\
$^b$ The foreground extinction is based on the R-band values calculated by 
  NED using the results from \citet{schlafly11}.\\
$^c$ The following abbreviations correspond to the following references: 
  B13: \citet{boselli13}; K08: \citet{kennicutt08}; K09: \citet{kennicutt09}.\\
$^d$ The H$\alpha$ filter used in these observations did not include any
  [N {\small II}] emission, so no correction was needed.\\
$^e$ Not enough information was provided with these H$\alpha$ image files 
  to allow us to convert the data to intensity units, so we rescaled the
  integrated signal within the optical discs of these galaxies using
  H$\alpha$ fluxes from the literature. The correction for [N {\small II}]
  emission was applied during this step.\\
$^f$ The H$\alpha$ data available for these galaxies were not suitable
  for analysis.  See Section~\ref{s_sftracer} for details.\\
$^g$ We were unable to find any references with data on the 
  [N~{\small II}]/H$\alpha$ ratios for these galaxies, so we instead
  estimated the ratio using the relation between the 
  [N~{\small II}]/H$\alpha$ ratio and B-band emission given by
  \citet{kennicutt08}.  We used B-band absolute magnitudes of -20.77
  for NGC~4725 \citep[based on the photometry from ][]{dale07} and
  -20.60 for NGC~5364 \citep[based on the photometry 
  from ][]{devaucouleurs91}.   Both absolute magnitudes were 
  calculated using the distances given in Table~\ref{t_sample} and
  include foreground extinction corrections calculated by NED using the
  results from  \citet{schlafly11}.\\
$^h$ The filters used for these H$\alpha$ observations only included
  emission from the 6584 \AA \ line, so we adjusted the [N~{\small
      II}]/H$\alpha$ ratio by 0.75 based on the spectroscopy data from
  \citet{storey00}.
\end{minipage}
\end{table*}

Table~\ref{t_hadata} lists the sources of the H$\alpha$ data that we
used in this analysis as well as some of the characteristics of the
data and corrections applied to the data.  We generally gave
preference to data that covered the entire target and where the
background was close to flat.  For NGC~3621 and 3938, we use images
that have not been converted into physical units (e.g. erg cm$^{-2}$
s$^{-1}$) and that we could not determine how to convert into physical
units using the available reference information for the images.  We
therefore rescaled the total fluxes within the optical disc of each
galaxy using the global H$\alpha$ fluxes from \citet{kennicutt09}.
The image of NGC~5457 from \citet{hoopes01} was rescaled in a similar
way using the photometry published within that paper.  For all images,
we also applied a foreground extinction correction based on the R-band
extinction calculated by the NASA/IPAC Extragalactic Database (NED)
using the results from \citet{schlafly11}.  These corrections change
the H$\alpha$ emission by $<$10\%.  Many but not all of the H$\alpha$
images include emission from one or both [N {\small II}] lines at 6548
and 6583 \AA.  We include corrections for these lines based on [N
  {\small II}]/H$\alpha$ ratios from several references in the
literature.  Since the ratios are either global or nuclear
measurements, we are only able to use these ratios to uniformly
rescale the H$\alpha$ emission, although the ratio may vary within the
galaxies.

Table~\ref{t_infraredsource} lists the sources of the 24~$\mu$m data
used in the analysis.  For VNGS and HRS galaxies, we used the {\it
  Spitzer} 24~$\mu$m images from \citet{bendo12b} except for NGC~3631
and NGC~5364, where we used the {\it WISE} 22~$\mu$m images.  For the
KINGFISH galaxies, we used the SINGS images from \citet{dale07} except
in the case of NGC 5457, where we used the {\it Spitzer} Local Volume
Legacy (LVL) Survey data from \citet{dale09}.  All MIPS 24~$\mu$m data
have plate scales of 1.5~arcsec, PSF FWHM of 6~arcsec
\citep{engelbracht07}, and calibration uncertainties of 4\%
\citep{engelbracht07}.  The {\it WISE} data are taken from the All-Sky Data
Release (14 March 2012) and have plate scales of 1.375~arcsec, PSF
FWHM of 12~arcsec \citep{wright10}, and calibration uncertainties of
3\%
\citep{cutri13}\footnote{http://wise2.ipac.caltech.edu/docs/release/allsky/expsup/}.
We examined the effects of applying a correction, such as the one
given by \citet{helou04}, to remove the stellar continuum emission
within the the 24~$\mu$m band, but we found that this is not necessary
for our analysis.  Typically, $<$5\% of the emission in the 24~$\mu$m
band is from the stellar population seen at 3.6~$\mu$m, so applying
the correction has a $<$1\% effect on the logarithm of the 24~$\mu$m
surface brightness and therefore does not have a significant effect on
our overall results.  We discuss this further in
Appendix~\ref{a_24stellarrem}.

\subsection{Stellar population tracers}
\label{s_totstartracer}

Near-infrared bands sample the Rayleigh-Jeans side of the starlight
SED and therefore trace the evolved stellar populations very
effectively.  We have access to uniformly-processed Two-Micron All-Sky
Survey (2MASS) J-, H-, and K-band data from \cite{jarrett03}.  We also
have {\it Spitzer} 3.6~$\mu$m data for all of the sample galaxies, and
while the data were processed by several different groups, the groups
used similar processing pipelines to produce images with a uniform
format.  From a physical standpoint, each band has different
advantages and disadvantages.  The shorter bands are more affected by
dust extinction, while the 3.6~$\mu$m band may sometimes include
thermal emission from dust or polycyclic aromatic hydrocarbons
\citep{lu03, mentuch09, mentuch10}, although the contribution of these
sources of emission to the 3.6~$\mu$m band may be $\ltsim$10\%
\citep[e.g. ][]{meidt12}.  Technically, the 3.6~$\mu$m data are better
than the 2MASS data.  The 3.6~$\mu$m data have a smaller PSF, which
aids in detecting and removing foreground stars, and the
signal-to-noise ratios are much higher in the 3.6~$\mu$m data.

Hence, we opted to use the 3.6~$\mu$m data as a tracer of the evolved
stellar populations in our analysis because of its better data
quality.  However, we did examine the relation between H-band and
3.6~$\mu$m data for our sample galaxies to demonstrate that the
selection of the band has a negligible impact on the overall analysis.
This is shown in Appendix~\ref{a_hiraccomp}.  We also investigated the
possibility of using the extinction-corrected H$\alpha$ emission to
remove the contribution of the star forming regions to the 3.6~$\mu$m
band, which would make the band a more effective tracer of
intermediate-aged and evolved stars.  When we applied such a
correction to the 3.6~$\mu$m band, we found that it has an effect of
$\ltsim$2\% for most regions within the sample galaxies.  This is
equivalent to a change of $<$1\% in the logarithm of the 3.6~$\mu$m
surface brightness, so we decided not to apply such corrections.  This
is discussed further in Appendix~\ref{a_iracsfrrem}.

\begin{table}
\caption{Sources of mid-infrared data.}
\label{t_infraredsource}
\begin{center}
\begin{tabular}{@{}lcc@{}}
\hline
Galaxy &   \multicolumn{2}{c}{Data Source}\\
&  
           3.6~$\mu$m &            24~$\mu$m  \\
\hline
NGC 628 &  
           SINGS &                 SINGS \\
NGC 925 &  
           SINGS &                 SINGS \\
NGC 2403 & 
           SINGS &                 \citet{bendo12b} \\
NGC 3031 & 
           SINGS &                 \citet{bendo12b} \\
NGC 3184 & 
           SINGS &                 SINGS \\
NGC 3621 & 
           SINGS &                 SINGS \\
NGC 3631 & 
           S$^4$G &                {\it WISE}$^a$ \\
NGC 3938 & 
           SINGS &                 SINGS \\
NGC 3953 & 
           S$^4$G &                \citet{bendo12b} \\
NGC 4254 & 
           SINGS &                 \citet{bendo12b} \\
NGC 4303 & 
           S$^4$G &                \citet{bendo12b} \\
NGC 4321 & 
           SINGS &                 \citet{bendo12b} \\
NGC 4501 & 
           S$^4$G &                \citet{bendo12b} \\
NGC 4535 & 
           S$^4$G &                \citet{bendo12b} \\
NGC 4548 & 
           S$^4$G &                \citet{bendo12b} \\
NGC 4579 & 
           SINGS &                 \citet{bendo12b} \\
NGC 4725 & 
           SINGS &                 \citet{bendo12b} \\
NGC 4736 & 
           SINGS &                 SINGS \\
NGC 5055 & 
           SINGS &                 SINGS \\
NGC 5236 & 
           LVL &                   \citet{bendo12b} \\
NGC 5364 & 
           S$^4$G &                {\it WISE}$^a$ \\
NGC 5457 & 
           LVL &                   LVL \\
NGC 6946 & 
           SINGS &                 SINGS \\
NGC 7793 & 
           SINGS &                 SINGS \\
\hline
\end{tabular}
\end{center}
\footnotesize
$^a$ {\it Spitzer} 24~$\mu$m data were not available for these galaxies.  We
  used WISE 22~$\mu$m data as a substitute. 
\end{table}

Table~\ref{t_infraredsource} lists the sources of the 3.6~$\mu$m data
used in this analysis.  Most of the images are from SINGS, the LVL
Survey, or the Spitzer Survey of Stellar Structure in Galaxies
\citep[S$^4$G; ][]{sheth10}. All data are in maps with pixel sizes of
1.5~arcsec.  The FWHM of the PSF is 1.7~arcsec, and the calibration
uncertainty is 3\%
\citep{irac13}\footnote{http://irsa.ipac.caltech.edu/data/SPITZER/docs/irac\\ /iracinstrumenthandbook/IRAC\_Instrument\_Handbook.pdf}.
We multiplied the 3.6~$\mu$m data by the surface brightness correction
factor 0.91 as recommended by the \citet{irac13}.  The 2MASS data are
obtained from \citet{jarrett03}.  These data have pixel sizes of
1~arcsec, PSF FWHMs of 2-3~arcsec, and calibration uncertainties of
3\%.

\subsection{Data preparation}
\label{s_data_preparation}

The data preparation steps are almost identical to the steps applied
by \citet{bendo12a}.  We began with identifying foreground stars in
the H$\alpha$, 1.6~$\mu$m and 3.6~$\mu$m images, which appear as
unresolved sources with very blue colours (with 3.6/24~$\mu$m flux
density ratios $\gtrsim$10) or as sources with positive and negative
counterparts in the H$\alpha$ data.  We then removed these sources by
interpolating over them.  We also removed residual continuum emission
from the bulges in the H$\alpha$ images of NGC~3031, 3953, 4548 and
4725, which we could identify as potential stellar emission because it
was unusually diffuse compared to other sources within the galaxies or
because we could see artefacts in the central regions (e.g. negative
sources or asymmetric PSFs) that appeared similar to the artefacts
associated with foreground stars.  We also removed a couple of very
bright stars that are visible in the 24~$\mu$m image on NGC~3621.

After this, we applied convolution kernels from \citet{aniano11} to
match the PSFs of all of the data to the PSF of the 350~$\mu$m data,
which has the largest FWHM.  These kernels not only change the FWHM of
the PSFs but also match the extended structure outside the central
peak.  Specific kernels were created for matching the IRAC 3.6~$\mu$m,
{\it WISE} 22~$\mu$m, MIPS 24~$\mu$m, PACS 160~$\mu$m, and SPIRE
250~$\mu$m PSF to the SPIRE 350~$\mu$m PSF.  For the other images, we
assumed that the PSFs are approximately Gaussian and used kernels that
match Gaussian functions to the SPIRE 350~$\mu$m PSF.  

Following this, we matched the coordinate systems of all images to the
coordinates of the 350~$\mu$m data (which has 8~arcsec pixels).  Next,
we subtracted the background from the data using locations outside the
optical discs of the galaxies.  These data were used to produce
160/250 and 250/350~$\mu$m ratio maps that can be used in the
qualitative analyses in Section~\ref{s_qualanalysis}.

For quantitative analyses, we rebinned the data into 24~arcsec bins,
which is both an integer multiple of the 250 or 350~$\mu$m map pixels
and approximately equivalent to the FWHM of the data after
convolution.  The rebinning is performed in such a way that the centre
of each galaxy will correspond to the centre of a bin.  Since we are
using data where the bins are equivalent to the FWHM of the PSF, these
binned data will generally sample individual resolution elements in
the data and will therefore be statistically independent.  

NGC~3031 contains a low-luminosity active galactic nucleus that
potentially produces nonthermal emission in these wavebands.  We
excluded the central $3\times3$ bins from the quantitative analyses,
although we still show these data in the plots of the 24~arcsec binned
data.

\section{Qualitative analysis of far-infrared surface brightness ratios}
\label{s_qualanalysis}

\begin{figure*}
\includegraphics[width=0.95\textwidth]{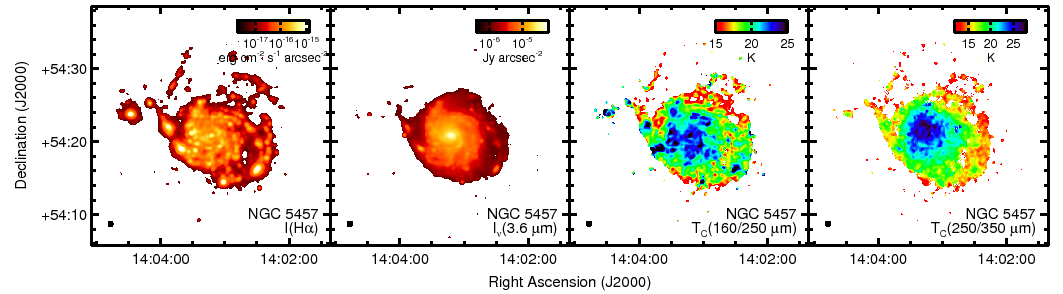}
\includegraphics[width=0.95\textwidth]{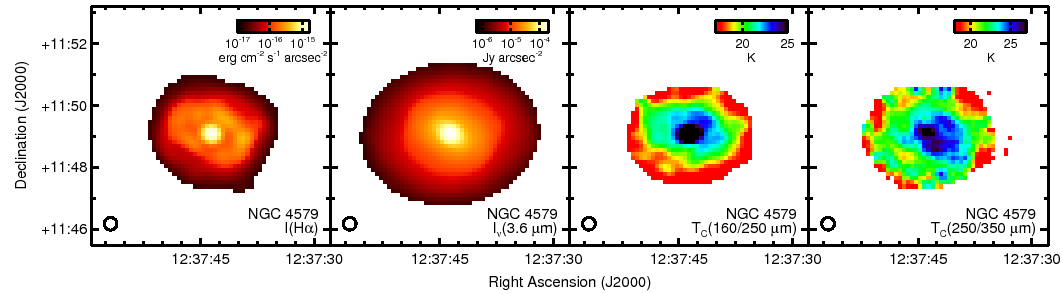}
\caption{H$\alpha$, 3.6~$\mu$m, 160/250~$\mu$m, and 250/350~$\mu$m
  images of two example galaxies.  NGC~5457 is an example of a galaxy
  where the smooth structures in the 250/350~$\mu$m ratio is more
  strongly related to those in the 3.6~$\mu$m image than in the
  H$\alpha$ image, although the locally-enhanced colour temperatures
  in the 160/250~$\mu$m map qualitatively appear more like the
  H$\alpha$ image.  NGC~4579 is an example of a galaxy where the
  160/250 and 250/350~$\mu$m colour variations resemble structures in
  both the H$\alpha$ and 3.6~$\mu$m images.  Equivalent images for the
  other galaxies in the sample are shown in Figure~\ref{f_maptempall}
  in Appendix~\ref{a_fig}.  All images are based on data where the PSF
  has been matched to the PSF of the 350~$\mu$m image, which has a
  FWHM of 25~arcsec.  The 160/250 and 250/350~$\mu$m ratios are shown
  as colour temperatures (based on modified blackbodies with
  $\lambda^{-2}$ emissivities) so that the colours can be related to
  the SED shape, although the actual temperatures may vary.  We only
  show data within approximately the optical disc of each galaxy
  detected at the $5\sigma$ level in the H$\alpha$ or 3.6~$\mu$m maps
  or above the $3\sigma$ level in the colour temperature maps; other
  parts of the image are left blank.}
\label{f_maptemp}
\end{figure*}

Figure~\ref{f_maptemp} shows the extinction-corrected H$\alpha$ images
(as a tracer of star formation), the 3.6~$\mu$m images (as a tracer of
the evolved stellar population), and the 160/250 and 250/350~$\mu$m
surface brightness ratios (shown as colour temperatures) for all
galaxies in the sample.  If dust is heated locally by star forming
regions, multiple unresolved hot regions will appear in the colour
temperature maps in locations with strong star formation activity as
seen in the H$\alpha$ images, and the colour temperatures will also
appear higher in structures with enhanced star formation activity such
as spiral arms.  If dust is heated locally by the evolved stellar
population, then the colour temperatures will appear to decrease more
smoothly with radius and will only appear locally enhanced in broader
structures with enhanced 3.6~$\mu$m emission.  This ignores the issue
that the scale height of the stars, particularly the older stars, may
be much larger than that of the dust, but given that the stellar
emission drops exponentially (or more steeply) with height above the
disc of the galaxy, the projected stellar emission should still
approximate the radiation field of the stars intermixed with the dust.

The ratios are shown as colour temperatures as a useful reference for
relating the ratios to SED shape.  The actual temperatures of the dust
seen at these wavelengths may differ depending on the dust emissivity
and may have a broad range of temperatures, although the colour
temperatures should be approximately similar to the mean temperature
of dust emitting at any pair of wavelengths.

These maps demonstrate that the heating sources for dust vary
significantly among nearby galaxies.  The 160/250~$\mu$m colour
temperature maps for about half the sample (NGC 628, 925, 2403, 3184,
3938, 4254, 4303, 4501, 4535, 4725, 5055, 5236, 5364, and 5457) appear more
strongly associated with star forming regions than with the structures
traced by the evolved stellar population.   Additionally, the
250/350~$\mu$m colour temperatures for NGC~925, 4254, 4303, 4501, and
5364 also look more closely related to star forming regions.  In some
cases, the colours clearly appear enhanced either in individual star
forming regions or in star forming structures such as the spiral arms
in NGC~5236.  In NGC~4254 and 5055, the relation is also seen in the
match between the asymmetric distribution of star formation and the
asymmetric colour temperatures.  In NGC~4254, 4303, and 4501, the relation of
the colour temperature variations to star formation appears stronger
than the relation to the evolved stellar population because the colour
temperatures do not peak in the centre like the evolved stellar
population does but instead appear enhanced in a broader region
tracing the locations of enhanced star formation.

In contrast, the 250/350~$\mu$m colour temperatures in several
galaxies (NGC~628, 2403, 3031, 3621, 3938, 3953, 5055, and 5457) trace
structures that are similar to the emission seen at 3.6~$\mu$m.  We
can clearly see that the colour temperatures are not enhanced in star
forming regions that appear in the H$\alpha$ maps in many of these
galaxies.  Instead, the colour temperatures tend to vary smoothly with
radius.  In the case of NGC~3031, 3621, and 3953, this applies to the
160/250~$\mu$m ratios as well.  In NGC 628, 2403, 3938, 5055, and
5457, however, the 160/250~$\mu$m colours are strongly affected by
star forming regions; this result along with the result for the
250/350~$\mu$m ratios indicates that the transition from dust heated
by star forming regions to dust heated by evolved stars occurs
between 160 and 250~$\mu$m for these galaxies.

In most of the rest of the galaxies, the interpretation of both of the
colour temperature maps is ambiguous.  The corrected H$\alpha$ (or
24~$\mu$m) and 3.6~$\mu$m maps for NGC 3631, 4321, 4579,
4736, and 6946 trace structures that are not visually distinct enough
to allow us to visually see differences between them or to see that
the colour temperature maps are better related to one heating source
instead of another.  Meanwhile, the colour temperature maps of
NGC~4548 and 7793 appear intermediate between the H$\alpha$ and
3.6~$\mu$m maps.

Many of the colour temperature maps are affected by artefacts that
make them more difficult to interpret.  Artefacts from
imperfectly-matched PSFs appear around some very bright sources.
Noise in the shorter-wavelength data tend to create hot spots,
especially near the edges of the regions detected at the $3\sigma$
level.  

Sometimes the 250/350~$\mu$m color temperatures appear enhanced in
lopsided regions that do not correspond to enhancements in either the
H$\alpha$ or 3.6~$\mu$m surface brightness, most notably in NGC~4725,
5457, 6946 and 7793.  It is unclear whether these structures are
intrinsic to the galaxies or a result of either foreground cirrus
structures or image processing artefacts.  NGC~6946 is at a galactic
latitude of $\sim$12 degrees, so the enhanced 250/350~$\mu$m colour
temperatures on the east side of the galaxy could be the result of
foreground cirrus structure.  To investigate this further, we examined
the dust optical depths at 353 GHz ($\tau_{353\mbox{GHz}}$) from
version R1.20 of the Planck thermal dust model map\footnote{Available
  from
  http://irsa.ipac.caltech.edu/data/Planck/release\_1/all-\\ sky-maps/previews/HFI\_CompMap\_ThermalDustModel\_2048\_R1.20/\\ index.html
  .}  \citep{planck11}.  For NGC~6946, the mean and standard deviation
in the background values of $\tau_{353\mbox{GHz}}$ (measured in an
annulus 1.5-2.0 times the optical diameter of the galaxy specified in
Table~\ref{t_sample}) were $(1.9 \pm 0.2)\times10^{-5}$.  For the
other sample galaxies, the mean background values of
$\tau_{353\mbox{GHz}}$ were $0.6-5.1 \times 10^{-6}$, a factor of
$\sim$10 lower than for NGC~6946, and the standard deviations in
$\tau_{353\mbox{GHz}}$ for the other sample galaxies
($0.2-1.3\times10^{-6}$) were also lower than the standard deviation
for NGC~6946.  These results imply that foreground cirrus structures
could be responsible for the structures in the NGC~6946 250/350~$\mu$m
colour temperature maps.  NGC~4725, 5457, and 7793, however, are at
higher latitudes where cirrus structures should not has as much of an
effect on the data. Moreover, the mean and standard deviation in the
background $\tau_{353\mbox{GHz}}$ values for these galaxies are
actually below the median values for the sample overall, so it is
unlikely that lopsided structures in the colour temperature maps of
these galaxies are related to foreground cirrus emission.

The origin of the elongated structure in the 250/350~$\mu$m colour
temperature map of NGC~5236 is unclear.  It could be caused by actual
enhancements in the colour temperatures, an artefact in one of the
scans (some of which are aligned with the galaxy's bar), or a
combination of these effects.

In the 160/250~$\mu$m colour temperature map for a few galaxies,
particularly NGC~5236 and 6946, we can see small offsets between the
spiral arm structures traced by the star forming regions and the
structures traced by the 160/250~$\mu$m ratios.  This could
potentially be the result of either the non-axisymmetric propagation
of light from star forming regions within the spiral arms in these
galaxies or dust heating by a population of B and A stars that have
migrated out of the star forming regions in the spiral arms.  This is
discussed further in Appendix~\ref{a_offsetlinecut}.

\section{One-on-one comparison of far-infrared surface brightness ratios to 
  heating sources}
\label{s_singlesourceanalysis}

We use the binned data to examine the quantitative relationship
between the 160/250 and 250/350~$\mu$m ratios and potential dust
heating sources.  This analysis relies upon three assumptions.  First,
we assume that projection effects do not significantly affect the
results; the remarkable correlation between the infrared surface
brightness ratios and the stellar surface brightness found previously
in the early-type spiral galaxy M81 by \citet{bendo12a} would suggest
that this is not a significant issue.  Second, we assume that the dust
observed within the 24~arcsec bins is heated by the stars observed
within those bins.  As discussed in Appendix~\ref{a_offsetlinecut},
the physical sizes of the bins are usually larger than the mean free
path of light.  In the closest galaxies in the sample, the bins may be
smaller than the mean free path of light, and if the light propagates
asymmetrically from star forming regions, as may be the case in some
nearby grand-design spiral galaxies (see
Appendix~\ref{a_offsetlinecut}), this assumption may not work well.
Third, we assume that the dust properties themselves are not variable.
Some authors have also suggested that $\beta$ is variable and also
affects the far-infrared colours, although these variations may be
related to a degeneracy with temperature in SED fitting
\citep{shetty09, galametz12, kirkpatrick14}.
 
When comparing a ratio to a heating source we only selected data that
are detected at the $3\sigma$ level in the ratio that we wish to
compare, the 3.6~$\mu$m band, and the star formation tracer ( either
the extinction-corrected H$\alpha$ emission calculated using
Equation~\ref{e_hacorr} or, when H$\alpha$ data were not available,
the 24~$\mu$m band)\footnote{For clarity, when both H$\alpha$ and
  24~$\mu$m data were available for a galaxy, we did not require that
  both the uncorrected H$\alpha$ and 24~$\mu$m emission were detected
  at the $3\sigma$ level when selecting data for the analysis in
  Section~\ref{s_singlesourceanalysis}.  We only required that the
  extinction corrected H$\alpha$ emission calculated using
  Equation~\ref{e_hacorr} was measured at the $3\sigma$ level.  Hence,
  our analysis includes data from both strongly obscured and
  unobscured star forming regions.} As an additional step to filter
out low signal-to-noise data, we only use data from locations where
the ratio is measured at the $3\sigma$ level both before and after
rebinning the data into 24~arcsec bins.  We often have more data
points in the comparisons including the 250/350~$\mu$m ratios because
those data have higher signal-to-noise levels in general.

For the analysis in this section, we use extinction-corrected
H$\alpha$ emission as a star formation tracer when it is available or
the 24~$\mu$m emission otherwise.  In Appendix~\ref{a_ha24comp}, we
show how the results change when using either the uncorrected
H$\alpha$ or 24~$\mu$m surface brightnesses as alternative star
formation tracers.  In galaxies where we do not identify significant
central 24~$\mu$m emission without corresponding H$\alpha$ emission,
the selection of star formation tracers only has a marginal effect on
the results except for NGC~628, 3184, 3938, 5364, and 7793.  We
will mention these exceptions in the discussion and otherwise rely
upon the results with the corrected H$\alpha$ intensities.

Because we are working with galaxies at different distances, we also
include an analysis in Appendix~\ref{a_reseffect} that demonstrates
how the results change for NGC~3031 and 5457 (the two galaxies in the
sample with the largest angular sizes) when data are measured in bins
that are 120~arcsec in size.  Using these data, we would reach the
same conclusions regarding the dust heating sources for these two
galaxies as we would using the 24~arcsec bins, demonstrating that the
analysis is relatively robust against resolution effects.

\begin{figure*}
\includegraphics{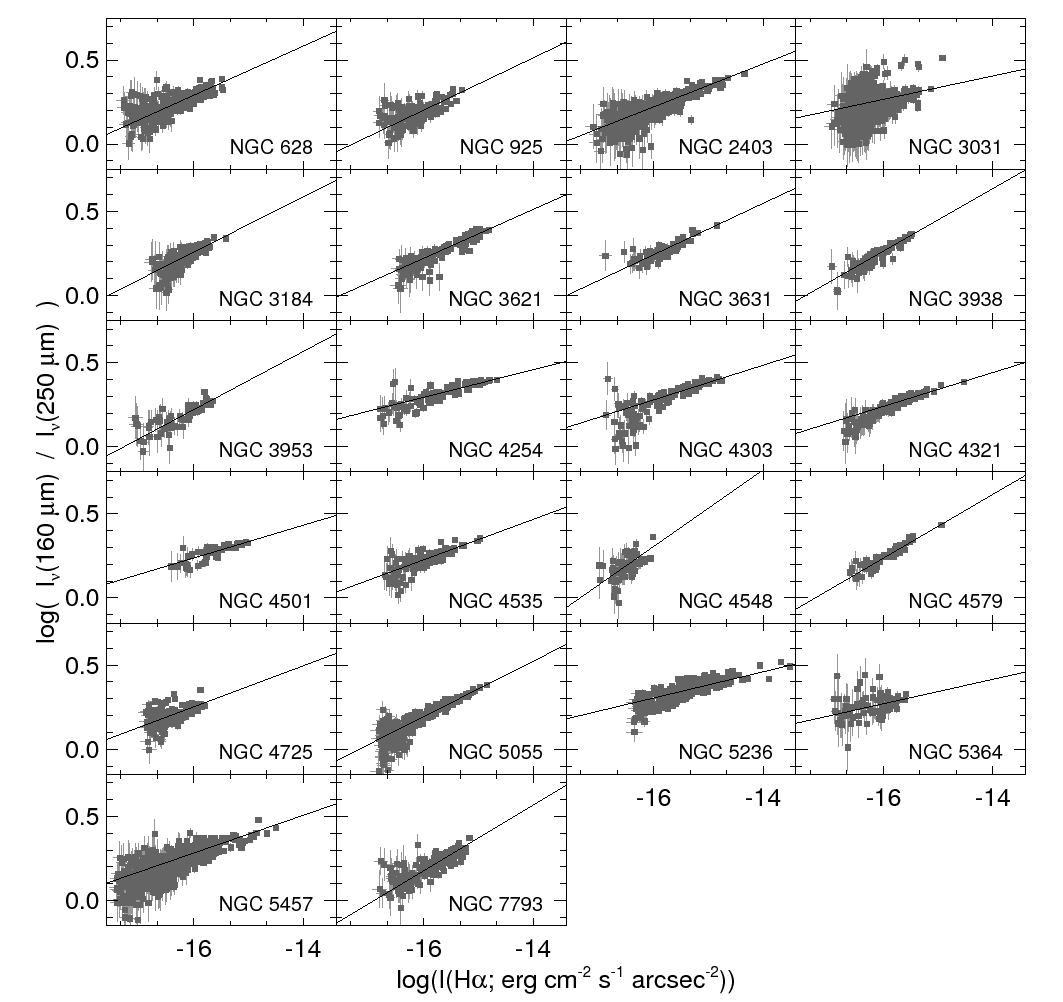}
\includegraphics{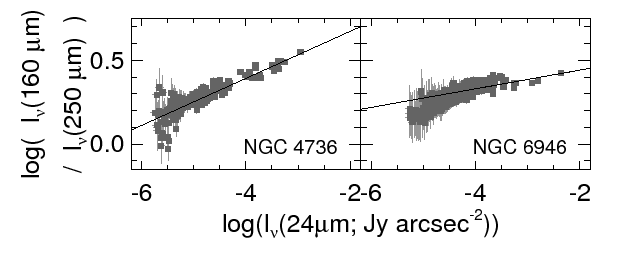}
\caption{Plots of 160/250~$\mu$m surface brightness ratios versus
  extinction-corrected H$\alpha$ intensities (top) and, for galaxies
  with no available H$\alpha$ data, 24~$\mu$m surface brightnesses
  (bottom) measured in the 24~arcsec binned data.  Only data detected
  at above the $3\sigma$ level in the plotted ratio, the 3.6~$\mu$m
  emission, and the star formation tracer (either the
  extinction-corrected H$\alpha$ emission or, when H$\alpha$ data were
  not available, the 24~$\mu$m emission) are shown; see the text of
  Section~\ref{s_singlesourceanalysis} for additional details.  The
  solid lines show the best fit lines.  Table~\ref{t_singlesource}
  gives the weighted Pearson correlation coefficients for the
  relations in these plots.}
\label{f_160250vsha}
\end{figure*}

\begin{figure*}
\includegraphics{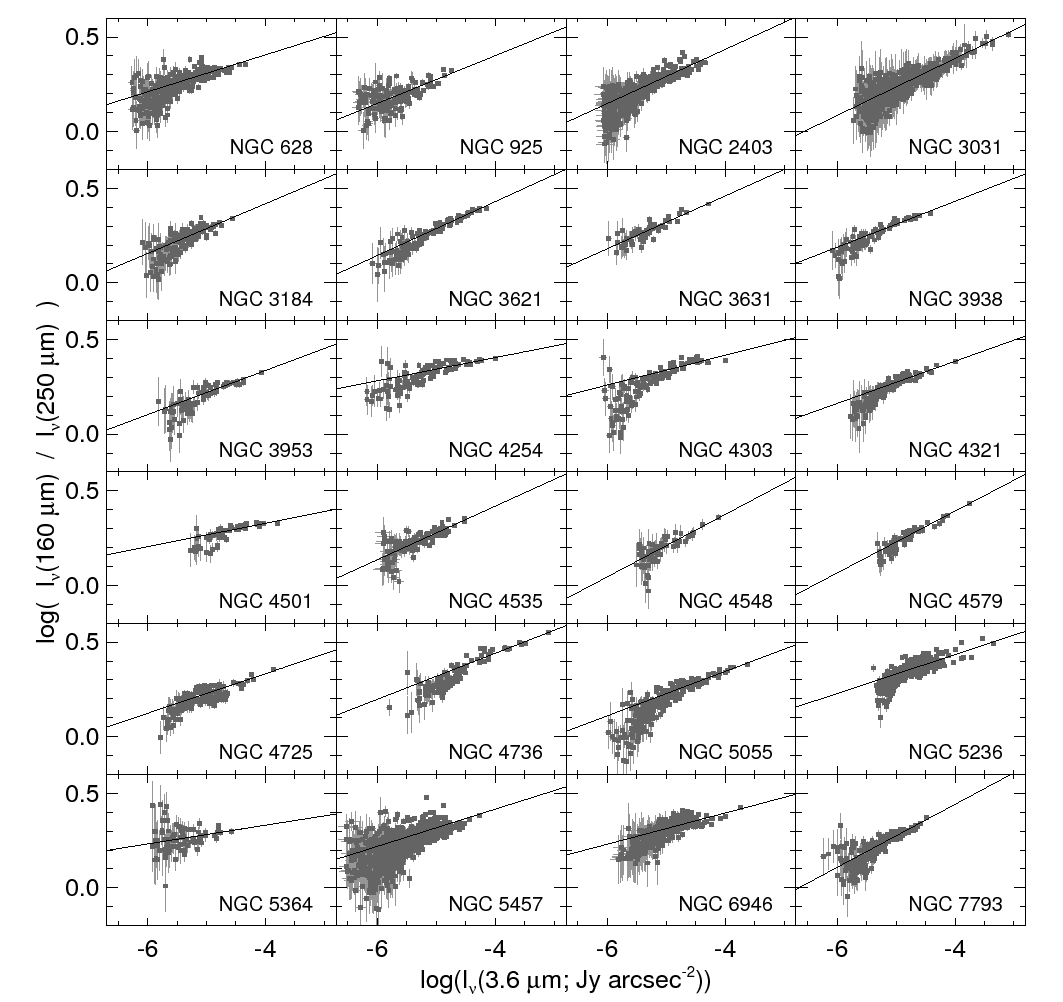}
\caption{Plots of 160/250~$\mu$m surface brightness ratios versus
  3.6~$\mu$m surface brightnesses measured in the 24~arcsec binned
  data.  See the caption for Figure~\ref{f_160250vsha} for additional
  details regarding this figure.}
\label{f_160250vsirac}
\end{figure*}

\begin{figure*}
\includegraphics{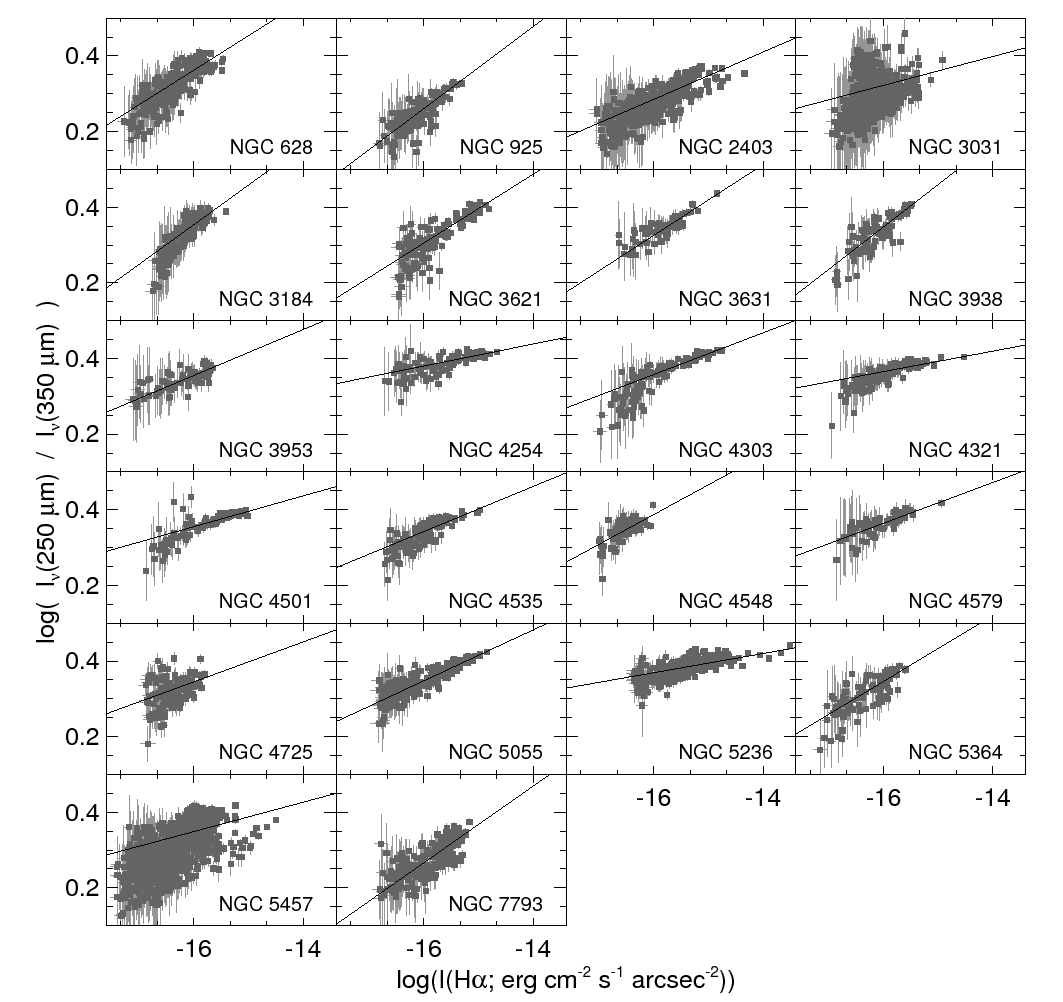}
\includegraphics{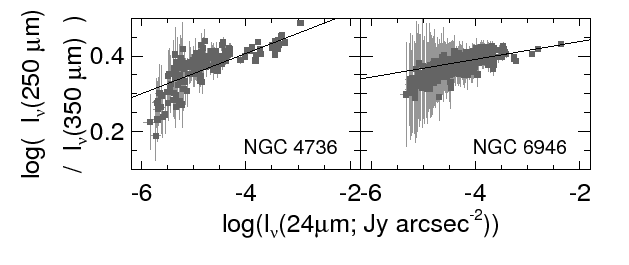}
\caption{Plots of 250/350~$\mu$m surface brightness ratios versus
  extinction-corrected H$\alpha$ intensities (top) and, for galaxies
  with no available H$\alpha$ data, 24~$\mu$m surface brightnesses
  (bottom) measured in the 24~arcsec binned data.  See the caption for
  Figure~\ref{f_160250vsha} for additional details regarding this
  figure.}
\label{f_250350vsha}
\end{figure*}

\begin{figure*}
\includegraphics{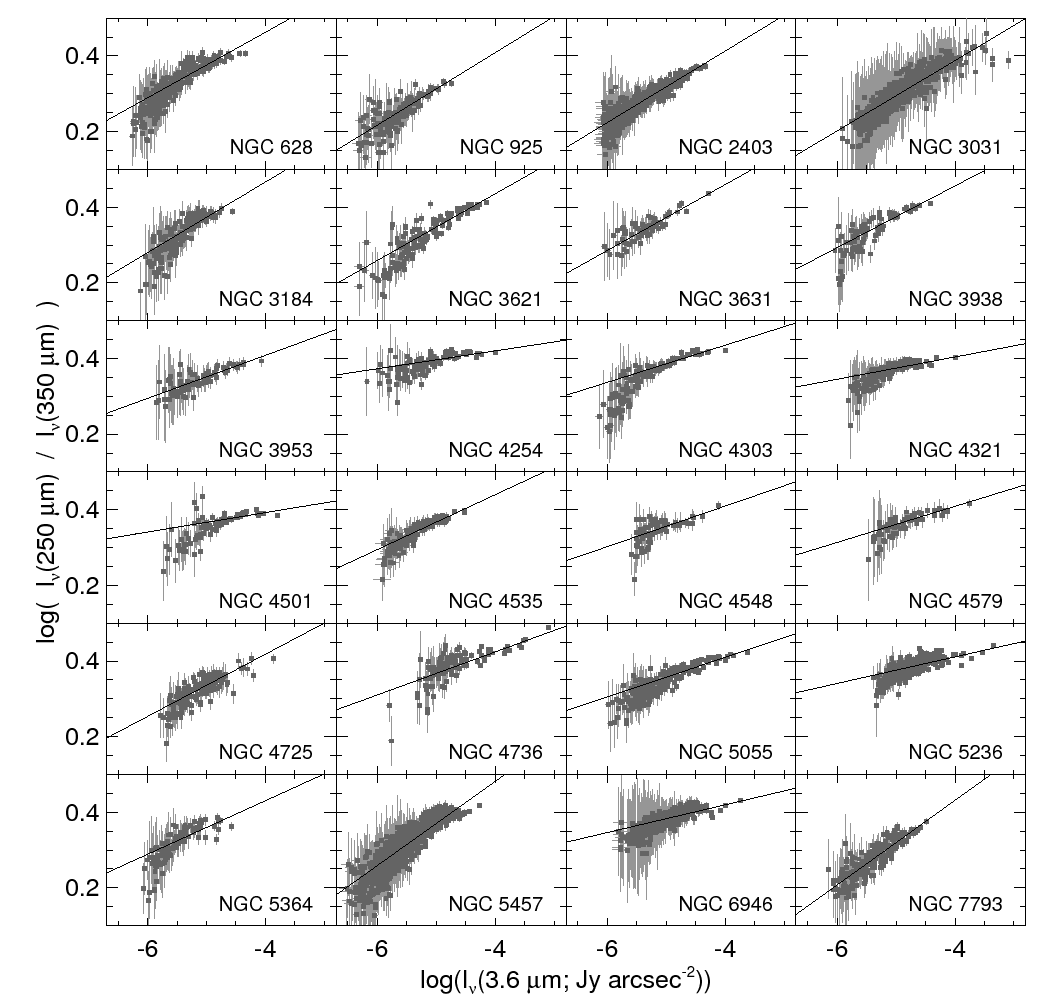}
\caption{Plots of 250/350~$\mu$m surface brightness ratios versus
  3.6~$\mu$m surface brightnesses measured in the 24~arcsec binned
  data.  See the caption for Figure~\ref{f_160250vsha} for additional
  details regarding this figure.}
\label{f_250350vsirac}
\end{figure*}

\begin{table*}
\begin{minipage}{138mm}
\caption{Weighted correlation coefficients for relations between
  160/250 and 250/350~$\mu$m ratios and tracers of heating sources.}
\label{t_singlesource}
\begin{tabular}{@{}lccccc@{}}
\hline
Galaxy &   
    \multicolumn{5}{c}{Weighted Correlation Coefficients} \\
&
    \multicolumn{2}{c}{$\log(I_\nu(160\mu\mbox{m})/I_\nu(250\mu\mbox{m}))$} &
    \multicolumn{2}{c}{$\log(I_\nu(250\mu\mbox{m})/I_\nu(350\mu\mbox{m}))$} &
    $\log(I_\nu(3.6\mu\mbox{m}))$ \\
&
    vs. $\log(I(\mbox{H}\alpha))^a$ &
    vs. $\log(I_\nu(3.6\mu\mbox{m}))$ &
    vs. $\log(I(\mbox{H}\alpha))^a$ &
    vs. $\log(I_\nu(3.6\mu\mbox{m}))$ &
    vs. $\log(I(\mbox{H}\alpha))^a$ \\
\hline
NGC 628 &
            0.82 &     0.71 &     0.75 &     0.87 &     0.53 \\
NGC 925 &
            0.82 &     0.77 &     0.88 &     0.88 &     0.73 \\
NGC 2403 &
            0.92 &     0.81 &     0.75 &     0.94 &     0.48 \\
NGC 3031 &
            0.17 &     0.84 &     0.16 &     0.87 &     0.14 \\
NGC 3184 &
            0.81 &     0.79 &     0.76 &     0.81 &     0.85 \\
NGC 3621 &
            0.88 &     0.96 &     0.89 &     0.90 &     0.80 \\
NGC 3631 &
            0.95 &     0.96 &     0.95 &     0.96 &     0.99 \\
NGC 3938 &
            0.96 &     0.93 &     0.88 &     0.89 &     0.84 \\
NGC 3953 &
            0.83 &     0.84 &     0.71 &     0.86 &     0.63 \\
NGC 4254 &
            0.88 &     0.81 &     0.69 &     0.68 &     0.96 \\
NGC 4303 &
            0.88 &     0.73 &     0.90 &     0.84 &     0.78 \\
NGC 4321 &
            0.97 &     0.96 &     0.85 &     0.82 &     0.99 \\
NGC 4501 &
            0.86 &     0.75 &     0.82 &     0.64 &     0.90 \\
NGC 4535 &
            0.91 &     0.89 &     0.87 &     0.91 &     0.92 \\
NGC 4548 &
            0.67 &     0.91 &     0.69 &     0.73 &     0.70 \\
NGC 4579 &
            0.96 &     0.95 &     0.84 &     0.81 &     0.97 \\
NGC 4725 &
            0.48 &     0.75 &     0.38 &     0.77 &     0.35 \\
NGC 4736 &
            0.95$^b$ &     0.94 &     0.81$^b$ &     0.91 &     0.92$^b$ \\
NGC 5055 &
            0.98 &     0.95 &     0.94 &     0.92 &     0.96 \\
NGC 5236 &
            0.89 &     0.87 &     0.82 &     0.83 &     0.98 \\
NGC 5364 &
            0.43 &     0.37 &     0.79 &     0.74 &     0.75 \\
NGC 5457 &
            0.89 &     0.57 &     0.40 &     0.89 &     0.40 \\
NGC 6946 &
            0.89$^b$ &     0.86 &     0.81$^b$ &     0.83 &     0.99$^b$ \\
NGC 7793 &
            0.85 &     0.90 &     0.72 &     0.87 &     0.68 \\
\hline
\end{tabular}
$^a$ The H$\alpha$ data used here were corrected for extinction using 24~$\mu$m
    data.\\
$^b$ No H$\alpha$ data were available for these galaxies.  These coefficients
    are based on relations between the infrared surface brightness ratios and
    24~$\mu$m emission.
\end{minipage}
\end{table*}

The relations between the infrared surface brightness ratios and
either 3.6~$\mu$m emission or the best available star formation tracer
are shown in Figures~\ref{f_160250vsha}-\ref{f_250350vsirac}.  For
modified blackbody emission from dust with a temperature range similar
to what is seen in the far-infrared in these galaxies, the relation
between far-infrared surface brightness ratios and energy absorbed and
re-emitted can be approximated by a power law to within a few percent.
Hence, we frequently see linear relations between the surface
brightness ratios and the dust heating sources plotted in these
figures.  However, more scatter is seen at the low surface brightness
end of the plots.  To assess the strength of these relations (by
examining which relations for each surface brightness ratio show less
scatter), we do not necessarily need to use a nonparametric
correlation coefficient (e.g. the Spearman correlation
coefficient) because the relations in logarithm space are typically
linear, but we do need to compensate for the variation in the
signal-to-noise of the data in the relations.  We therefore used the
weighted Pearson correlation coefficients \citep{pozzi12}, where the
uncertainties in the logarithm of the surface brightness ratios are
used to weight the data (as the uncertainties in the 3.6~$\mu$m
emission or star formation tracer in comparison to the dynamical range
of the data are generally less significant).  The weighted correlation
coefficients for the relations in
Figures~\ref{f_160250vsha}-\ref{f_250350vsirac} are given in
Table~\ref{t_singlesource}.  We interpret differences of 0.05 in the
correlation coefficients as indicating that one heating source was
more strongly linked to dust heating than the other.  This is greater
than the variation in the coefficients we have seen when using
different versions of the {\it Herschel} data, when making adjustments
to the data selection criteria, or when making adjustments to the
astrometry.

Based on these criteria, we can clearly identify the 160/250~$\mu$m
ratios as being better correlated with star formation tracers than
with 3.6~$\mu$m emission in 8 out of 24 galaxies.  Additionally, we
see that the 250/350~$\mu$m ratios are also better correlated with
star formation tracers in NGC~4303, 4501, and 5364 which would
indicate that dust heated by star formation is the dominant source of
emission at 250~$\mu$m and possibly even at longer wavelengths,
although the results for NGC~4303 and 5364 depend on the star
formation tracer used as discussed in Appendix~\ref{a_ha24comp}.
NGC~628, 2403, and 5457 are cases where the 160/250~$\mu$m ratios are
clearly better correlated with star formation while the 250/350~$\mu$m
ratios are clearly more strongly correlated with the evolved stellar
population, which demonstrates a clear switch in dust heating sources
between 160 and 250~$\mu$m.  In the other galaxies, we can clearly
identify star forming regions as having the stronger influence on the
160/250~$\mu$m ratios, but the identification of which source more
strongly affects the 250/350~$\mu$m ratios is more ambiguous or
dependent on the star formation tracer used.

In 9 of the 24 galaxies within the sample, the 250/350~$\mu$m ratios
are more strongly correlated with the 3.6~$\mu$m emission than with
star formation, indicating that in about half of the galaxies, the
dust seen at $\geq 250$~$\mu$m is heated by the evolved stellar
population.  NGC~3621 and 4548 are odd cases where the two correlation
coefficients in Table~\ref{t_singlesource} for the 250/350~$\mu$m
ratio differ only by $<$0.05 but where the data for the 160/250~$\mu$m
ratio show that the ratio is more strongly correlated with 3.6~$\mu$m
emission, implying that the emission at $\leq 160$~$\mu$m is from dust
heated by the evolved stellar population.  (We at least do not see any
case where the 250/350~$\mu$m ratios are clearly related to star
formation while the 160/250~$\mu$m ratio is more strongly related to
the 3.6~$\mu$m band tracing the evolved stellar population.)  Aside from
NGC~3621 and 4548, we also find that the 160/250~$\mu$m ratios for
NGC~3031, 4725, and 7793 are also well-correlated with the 3.6~$\mu$m
emission.  In the other galaxies, we find either that the
160/250~$\mu$m ratios are more clearly correlated with star formation
tracers or that we cannot clearly relate the 160/250~$\mu$m ratios to
a heating source.

In 8 galaxies, we see no clear difference in the way both the 160/250
and 250/350~$\mu$m ratios are related to either dust heating source
based on this analysis alone.  In many of these galaxies, this is
because the extinction-corrected H$\alpha$ emission is very well
correlated with the 3.6~$\mu$m emission as measured within the
24~arcsec binned data.  The rightmost column in
Table~\ref{t_singlesource} shows that the correlation coefficients for
the relation between the logairthms of the H$\alpha$ and 3.6~$\mu$m
emission for 7 of these 8 galaxies is $>$0.90; the lowest value is
0.84 for NGC~3938.  Hence, surface brightness ratios measured in the
24~arcsec binned data for these galaxies will tend to look equally
correlated with both tracers of different dust heating sources in
these galaxies.  In constrast, the correlation coefficients for the
relations between H$\alpha$ and 3.6~$\mu$m emission are $\leq0.90$ for
all galaxies where we can definitely identify a stronger relation
between one of the dust heating sources and the surface brightness
ratios.

In the 8 galaxies where we have difficulty statistically identifying a
dominant dust heating source, the qualitative analysis on the colour
temperature maps in Section~\ref{s_qualanalysis} implies that the
160/250~$\mu$m ratios for NGC~3938, 4535, 5055, and 5236 are more
closely related to star formation while the 250/350~$\mu$m ratios for
NGC~3938 and 5055 are more closely related to the evolved stellar
population.  However, given the issues encountered in the analysis
here, it may be that these qualitative interpretations are overly
subjective. The qualitative results based on the 250/350~$\mu$m maps
for NGC~4535, and 5236 as well as the results for both ratio maps for
NGC~3631, 4321, 4579, and 6946 are as ambiguous as the results from
the statistical approach.

The quantitative approach used in this section generally implies that
the contribution of the evolved stellar populations to dust heating is
higher than what is indicated by the qualitative analysis of the
colour temperature maps in Section~\ref{s_qualanalysis}.  We may have
obtained this result for a few reasons.  First, we may be visually
sensitive to picking out compact locations with hotter colours in the
maps, which would tend to bias us towards identifying the surface
brightness variations as being more strongly related qualitatively to
star formation.  Second, the evolved stellar populations will have a
smoother distribution than the star forming regions and will tend to
decrease smoothly with radius.  This type of gradient can be difficult
to pick out visually, but correlations between radially-varying
quantities are easy to detect statistically.

\section{Decomposition of far-infrared surface brightness ratio variations 
into separate components}
\label{s_decompositionanalysis}

As an alternative approach for identifying the heating source for the
dust, we fit the 160/250 and 250/350~$\mu$m surface brightness ratios
to a combination of the 3.6~$\mu$m emission and one of the star
formation tracers using
\begin{equation}
\ln \left( \frac{I_\nu(\lambda_1)}{I_\nu(\lambda_2)} \right)
    = \alpha \ln(I(\mbox{SFR}) + A_1 I_\nu (3.6 \mu \mbox{m}))
    + A_2
\label{e_multfit}
\end{equation}
derived by \citet{bendo12a} from the Stefan-Boltzmann law upon the
assumptions that the dust temperatures in the pairs of bands used here
can be approximated as a single temperature and that the relation
between dust temperature and either far-infrared ratio can be
approximated by a power law.  $I_\nu(\lambda_1)/I_\nu(\lambda_2)$ is
the far-infrared surface brightness ratio.  $I(\mbox{SFR})$ is the
emission from the star formation tracer, which is related to the
energy $E_{SF}$ from the star forming regions, and $I_\nu (3.6 \mu
\mbox{m}))$ is the surface brightness at 3.6~$\mu$m, which is related
to the energy $E_{ES}$ from the evolved stellar population.  The terms
$\alpha$, $A_1$, and $A_2$ are derived in the fit.  In fitting this
equation to the far-infrared ratios, we effectively decompose the
ratios into star formation and evolved stellar population components.
As a result, we can calculate the contribution of evolved stellar
population to the total dust heating $E_{Tot}$ using
\begin{equation}
\eta(\lambda_1/\lambda_2) = \frac{E_{ES}}{E_{Tot}} = \frac{A_1 I_\nu (3.6 \mu 
    \mbox{m})}{I(\mbox{H} \alpha) + A_1 I_\nu (3.6 \mu \mbox{m})}
\label{e_efrac}
\end{equation}
In practice, $\eta(\lambda_1/\lambda_2)$ also indicates the relative
fraction of the emission in the $\lambda_1$ band originates from dust
heated by the evolved stellar population, at least as long as
$\lambda_2$ is located on the Rayleigh-Jeans side of the dust SED.
This is because $I_\nu(\lambda_1)/I_\nu(\lambda_2)$ is equivalent to
$I_\nu(\lambda_1)$ normalised for dust mass, in which case it only
depends on dust temperature.  In Appendix~\ref{a_efractest}, we
perform some tests with simplified models of dust heating in galaxies
in which we test the effectiveness of using
$\eta(\lambda_1/\lambda_2)$ to rescale the 160 and 250~$\mu$m
emission.  We generally find that the numbers are accurate to within
0.15, although the derived $\eta$ may decrease with distance.

\begin{table*}
\centering
\begin{minipage}{166mm}
\caption{Results from fitting Equation~\ref{e_multfit} to the 160/250 and 
  250/350~$\mu$m surface brightness ratios.}
\label{t_multfit}
\begin{tabular}{@{}lcccccccc@{}}
\hline
Galaxy &
\multicolumn{4}{c}{Results for fits to 
    $I_\nu(160\mu\mbox{m})/I_\nu(250\mu\mbox{m})$} &
\multicolumn{4}{c}{Results for fits to 
    $I_\nu(250\mu\mbox{m})/I_\nu(350\mu\mbox{m})$} \\
&
    $A_1^a$ &
    $A_2$ &
    $\alpha$ &
    $\eta$ &
    $A_1^a$ &
    $A_2$ &
    $\alpha$ &
    $\eta$ \\
\hline
NGC 628 &
    $4.7 \pm   0.6$ &
    $5.8 \pm 0.1$ &
    $0.141 \pm 0.003$ &
    $0.25 \pm 0.02$ &
    $53.7 \pm  17.0$ &
    $4.0 \pm 0.1$ &
    $0.089 \pm 0.003$ &
    $0.77 \pm 0.05$ \\
NGC 925 &
    $22.0 \pm   4.8$ &
    $6.0 \pm 0.2$ &
    $0.152 \pm 0.006$ &
    $0.40 \pm 0.05$ &
    $41.6 \pm  15.8$ &
    $4.4 \pm 0.2$ &
    $0.106 \pm 0.006$ &
    $0.54 \pm 0.08$ \\
NGC 2403 &
    $29.0 \pm   2.5$ &
    $5.8 \pm 0.1$ &
    $0.147 \pm 0.002$ &
    $0.48 \pm 0.02$ &
    $634.9 \pm 225.4$ &
    $3.7 \pm 0.1$ &
    $0.090 \pm 0.002$ &
    $0.94 \pm 0.02$\\
NGC 3031 &
    $31.0 \pm   4.6$ &
    $6.5 \pm 0.2$ &
    $0.167 \pm 0.005$ &
    $0.79 \pm 0.02$ &
    $186.9 \pm 134.7$ &
    $4.2 \pm 0.2$ &
    $0.103 \pm 0.006$ &
    $0.95 \pm 0.03$ \\
NGC 3184 &
    $11.5 \pm   3.2$ &
    $6.1 \pm 0.3$ &
    $0.153 \pm 0.007$ &
    $0.41 \pm 0.06$ &
    $31.7 \pm  19.0$ &
    $4.2 \pm 0.3$ &
    $0.094 \pm 0.007$ &
    $0.65 \pm 0.12$\\
NGC 3621 &
    $77.2 \pm   9.9$ &
    $5.9 \pm 0.1$ &
    $0.151 \pm 0.002$ &
    $0.73 \pm 0.02$ &
    $36.2 \pm   6.8$ &
    $4.1 \pm 0.1$ &
    $0.093 \pm 0.002$ &
    $0.56 \pm 0.04$ \\
NGC 3631 &
    $67.4 \pm  20.0$ &
    $5.7 \pm 0.1$ &
    $0.143 \pm 0.004$ &
    $0.67 \pm 0.08$ &
    $40.1 \pm  25.6$ &
    $3.9 \pm 0.2$ &
    $0.088 \pm 0.005$ &
    $0.55 \pm 0.17$ \\
NGC 3938 &
    $6.1 \pm   2.2$ &
    $6.6 \pm 0.3$ &
    $0.165 \pm 0.007$ &
    $0.25 \pm 0.06$ &
    $16.1 \pm  14.2$ &
    $4.2 \pm 0.3$ &
    $0.094 \pm 0.008$ &
    $0.46 \pm 0.17$ \\
NGC 3953 &
    $5.1 \pm   1.7$ &
    $6.2 \pm 0.5$ &
    $0.157 \pm 0.013$ &
    $0.38 \pm 0.07$ &
    $23.4 \pm  23.2$ &
    $2.8 \pm 0.3$ &
    $0.056 \pm 0.009$ &
    $0.72 \pm 0.17$ \\
NGC 4254 &
    $0.0^b$ &
    $4.5 \pm 0.3$ &
    $0.103 \pm 0.008$ &
    0.00 &
    $0.0^b$ &
    $2.3 \pm 0.3$ &
    $0.040 \pm 0.008$ &
    0.00 \\
NGC 4303 &
    $0.0^b$ &
    $6.1 \pm 0.3$ &
    $0.149 \pm 0.008$ &
    0.00 &
    $7.4 \pm   4.3$ &
    $2.8 \pm 0.1$ &
    $0.054 \pm 0.002$ &
    $0.21 \pm 0.09$ \\
NGC 4321 &
    $5.3 \pm   3.3$ &
    $4.3 \pm 0.1$ &
    $0.102 \pm 0.002$ &
    $0.20 \pm 0.09$ &
    $0.0^b$ &
    $3.0 \pm 0.3$ &
    $0.059 \pm 0.008$ &
    0.00 \\
NGC 4501 &
    $0.0^b$ &
    $4.7 \pm 0.7$ &
    $0.115 \pm 0.018$ &
    0.00 &
    $0.0^b$ &
    $3.2 \pm 0.3$ &
    $0.064 \pm 0.009$ &
    0.00 \\
NGC 4535 &
    $16.6 \pm   5.0$ &
    $5.2 \pm 0.1$ &
    $0.129 \pm 0.004$ &
    $0.40 \pm 0.07$ &
    $66.2 \pm  19.5$ &
    $3.2 \pm 0.1$ &
    $0.068 \pm 0.004$ &
    $0.72 \pm 0.07$ \\
NGC 4548 &
    $20.6 \pm  16.6$ &
    $6.6 \pm 0.6$ &
    $0.170 \pm 0.016$ &
    $0.82 \pm 0.09$ &
    $3.1 \pm   1.9$ &
    $3.4 \pm 0.4$ &
    $0.069 \pm 0.009$ &
    $0.43 \pm 0.12$ \\
NGC 4579 &
    $5.5 \pm   1.7$ &
    $7.2 \pm 0.2$ &
    $0.183 \pm 0.006$ &
    $0.42 \pm 0.07$ &
    $5.4 \pm   6.1$ &
    $2.6 \pm 0.3$ &
    $0.049 \pm 0.008$ &
    $0.42 \pm 0.20$ \\
NGC 4725 &
    $5.0 \pm   0.5$ &
    $6.5 \pm 0.2$ &
    $0.164 \pm 0.006$ &
    $0.46 \pm 0.02$ &
    $12.8 \pm   3.8$ &
    $4.3 \pm 0.2$ &
    $0.098 \pm 0.006$ &
    $0.66 \pm 0.06$ \\
NGC 4736$^c$ &
    $1.8 \pm 0.1$ &
    $2.1 \pm 0.1$ &
    $0.137 \pm 0.001$ &
    $0.50 \pm 0.02$  &
    $^d$ &
    $1.8 \pm 0.1$ &
    $0.084 \pm 0.007$ &
    1.00 \\
NGC 5055 &
    $0.0^b$ &
    $7.3 \pm 0.2$ &
    $0.188 \pm 0.006$ &
    0.00 &
    $3.3 \pm   1.5$ &
    $3.1 \pm 0.1$ &
    $0.063 \pm 0.002$ &
    $0.24 \pm 0.08$ \\
NGC 5236 &
    $4.1 \pm   0.5$ &
    $3.6 \pm 0.0$ &
    $0.080 \pm 0.000$ &
    $0.14 \pm 0.01$ &
    $373.5 \pm 202.5$ &
    $2.0 \pm 0.0$ &
    $0.034 \pm 0.001$ &
    $0.92 \pm 0.04$ \\
NGC 5364 &
    $5.4 \pm   6.1$ &
    $2.5 \pm 0.6$ &
    $0.051 \pm 0.016$ &
    $0.26 \pm 0.16$ &
    $4.0 \pm   4.3$ &
    $3.7 \pm 0.3$ &
    $0.079 \pm 0.008$ &
    $0.21 \pm 0.14$ \\
NGC 5457 &
    $10.0 \pm   0.5$ &
    $5.3 \pm 0.1$ &
    $0.129 \pm 0.001$ &
    $0.37 \pm 0.01$ &
    $^d$ &
    $2.5 \pm 0.1$ &
    $0.139 \pm 0.004$ &
    1.00 \\
NGC 6946$^{c}$ &
    $0.0^b$ &
    $1.9 \pm 0.1$ &
    $0.123 \pm 0.008$ &
    0.00 &
    $73 \pm 96$ &
    $1.3 \pm 0.0$ &
    $0.054 \pm 0.005$ &
    $0.92 \pm 0.07^e$ \\
NGC 7793 &
    $50.7 \pm   7.1$ &
    $7.3 \pm 0.2$ &
    $0.192 \pm 0.005$ &
    $0.60 \pm 0.03$ &
    $401.2 \pm 210.0$ &
    $4.4 \pm 0.2$ &
    $0.111 \pm 0.004$ &
    $0.92 \pm 0.04$ \\
\hline
\end{tabular}
$^a$ $A_1$ has units of $10^{-12}$ erg cm$^{-2}$ s$^{-1}$ Jy$^{-1}$ unless 
    otherwise noted.\\ 
$^b$ For these fits, $A_1 I_\nu (3.6 \mu \mbox{m})$ in
    Equation~\ref{e_multfit} was found to be negligible compared to $I
    (\mbox{SFR})$.  We therefore report $A_1$ as 0.  The $A_2$ and $\alpha$
    values are based on fitting the star formation tracer to the infrared
    surface brightness ratio using a power law.\\
$^c$ H$\alpha$ data were not for these galaxies, so we used 24~$\mu$m data
    as a substitute star formation tracer.  $A_1$ in these cases is 
    dimensionless.\\
$^d$ For these fits, $I (\mbox{SFR})$ in Equation~\ref{e_multfit} was
    found to be negligible compared to $A_1 I_\nu (3.6 \mu \mbox{m})$.
    We therefore fit $\ln(I_\nu(250 \mu \mbox{m})/I_\nu(350 \mu
    \mbox{m}))$ to $\ln(I_\nu (3.6 \mu \mbox{m}))$ using a linear
    function.  The $\alpha$ listed here is the slope, and $A_2$ is the
    constant.\\
$^e$ The $\eta$(250/350~$\mu$m) values for this galaxy are
    suspect.  See the text in Section~\ref{s_decompositionanalysis} for details.
\end{minipage}
\end{table*}

\begin{figure*}
\includegraphics[width=0.3\textwidth]{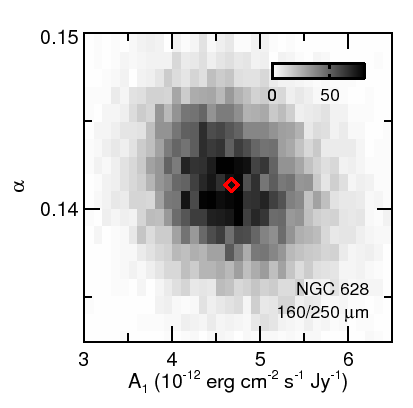}
\includegraphics[width=0.3\textwidth]{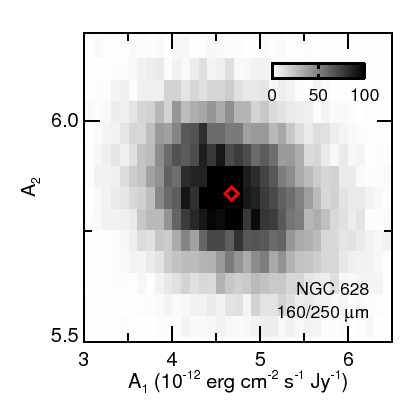}
\includegraphics[width=0.3\textwidth]{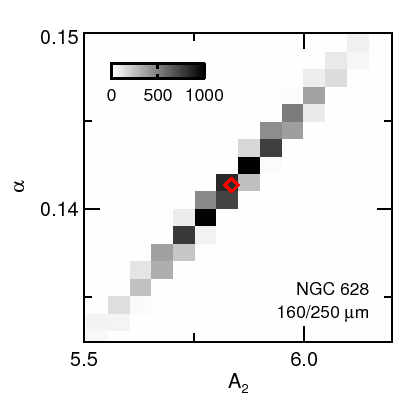}
\caption{Plots showing the distributions of best fitting parameters
  when using a Monte Carlo approach to fit Equation~\ref{e_multfit} to
  the 160/250~$\mu$m data for NGC~628.  For illustration purposes, the
  parameter space has been divided into multiple bins, and the shades
  of grey indicate the number of iterations that produce parameters
  that fall within each of these bins. The results are based on 10000
  iterations.  The red diamond shows the median parameters.}
\label{f_multfitunc}
\end{figure*}

Table~\ref{t_multfit} shows the best-fitting parameters to
Equation~\ref{e_multfit} for each galaxy for each surface brightness
ratio as well as the resulting $\eta$ values.  The values and
uncertainties are derived using a Monte Carlo approach. We perform a
least-squares fit to Equation~\ref{e_multfit} in 10000 iterations in
which Gaussian noise (scaled based on the measured standard deviations
in the binned data) has been added to both the ratios and the tracers
of dust heating sources, and then we determine the median and standard
deviation for each of the parameters and $\eta$ after removing
$5\sigma$ outliers.  For fitting each ratio, we selected 24~arcsec
bins within the optical disk of each galaxy where either the
3.6~$\mu$m or star formation tracer was detected at the $3\sigma$
level and where the far-infrared ratio was measured at the $3\sigma$
level.  The extinction-corrected H$\alpha$ data are used to trace
heating by star-forming regions except for NGC~4736 and 6946, where we
used 24~$\mu$m data because suitable H$\alpha$ data were not
available.  The uncertainties for $A_1$ may sometimes be very large,
but $\eta$ is frequently better constrained because it is also
dependent upon the magnitude of $A_1 I_\nu (3.6 \mu \mbox{m})$
relative to $I(\mbox{SFR})$.  We did find a strong degeneracy between
$A_2$ and $\alpha$ as well as much weaker degeneracies between $A_1$
and the other parameters.  This is illustrated by
Figure~\ref{f_multfitunc}, which shows the range of the best fitting
parameters for the 160/250~$\mu$m data for NGC~628.  However, these
plots also show that the fitting procedure will converge upon a single
best fitting value rather than converging upon a local minimum in
$\chi^2$.

We had a few galaxies where the fits produced results where one of the
heating source terms in Equation~\ref{e_multfit} (either
$I(\mbox{SFR})$ or $A_1 I_\nu (1.6 \mu \mbox{m})$) was negligible
compared to the other term.  In these situations, we performed a
linear fit between $\ln (I_\nu(\lambda_1)/I_\nu(\lambda_2)$ and the
logarithm of the heating source.  In most cases, we used the heating
source which was better correlated with the surface brightness ratio
according to the results in Table~\ref{t_singlesource}.  For the
250/350~$\mu$m ratio for NGC~4254 and the 160/250~$\mu$m ratio for
NGC~6946, however, the correlation coefficients in
Table~\ref{t_singlesource} do not clearly demonstrate that one heating
source is dominant.  In these cases, the rms residuals from fits
between the ratios and the star formation tracer are smaller than for
the corresponding fits with the 3.6~$\mu$m emission, so we quote
results in Table~\ref{t_multfit} based on the fit to the star
formation tracer, although these particular results should be treated
somewhat cautiously.

\begin{figure*}
\includegraphics[width=0.8\textwidth]{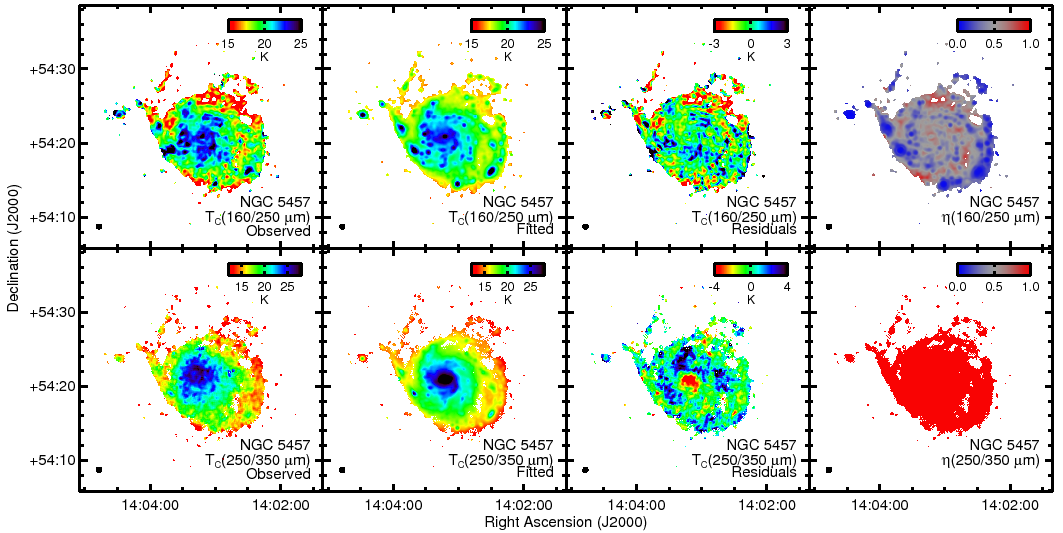}
\includegraphics[width=0.8\textwidth]{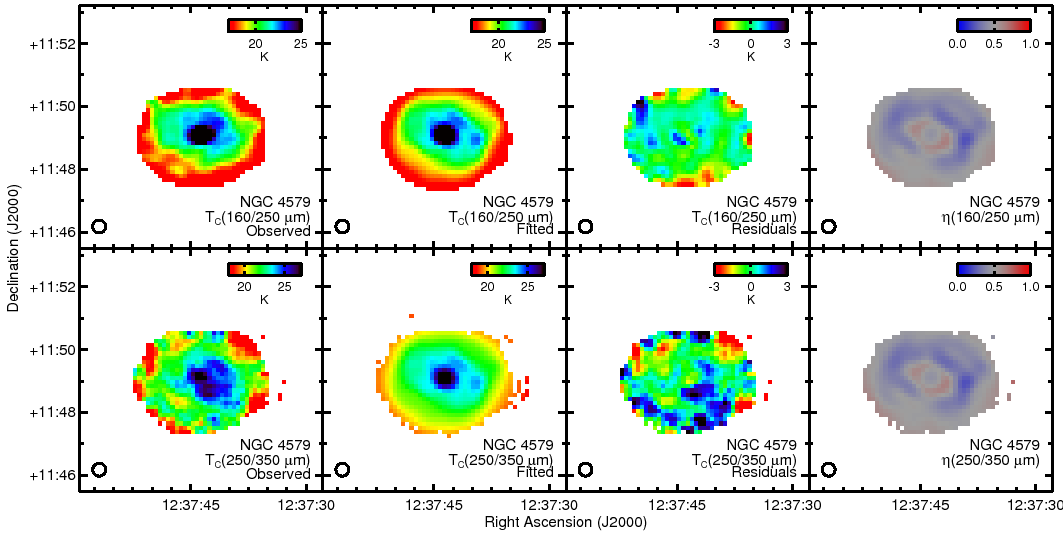}
\caption{Examples of the results from using Equation~\ref{e_multfit}
  to fit the 160/250 and 250/350~$\mu$m surface brightness ratios as a
  function of both the H$\alpha$ and 3.6~$\mu$m data.  The first
  columns of images on the left show the observed colour temperatures.
  The second column of images show the colour temperatures produced by
  adding together the H$\alpha$ and 3.6~$\mu$m images using
  Equation~\ref{e_multfit} and the best fitting parameters from
  Table~\ref{t_multfit}.  The third column shows the residuals from
  subtracting the fitted colour temperature map from the observed
  colour temperature map.  The fourth column shows the fraction of
  dust heating by the evolved stellar populations as predicted by the
  parameters in Table~\ref{t_multfit} and Equation~\ref{e_efrac}.  The
  same galaxies are used as were shown in Figure~\ref{f_maptemp}.
  Equivalent images for the other galaxies in the sample are shown in
  Figure~\ref{f_multfitall} in Appendix~\ref{a_fig}.  The images are
  formatted in the same way as the colour temperature images in
  Figure~\ref{f_maptemp}.}
\label{f_multfit}
\end{figure*}

In addition to the tabular data, Figure~\ref{f_multfit} shows the
observed surface brightness ratio maps for each galaxy, the surface
brightness ratios from best fitting parameters in
Table~\ref{t_multfit}, the residuals from the fit (shown as the
difference between the observed and fitted colour temperature maps),
and maps of $\eta$ based on the fits.  The levels in the residual
images are scaled roughly by the total dynamic range in the colour
temperatures of the observed data and can be used to assess the
quality of the fits.  Residual images for good fits will either look
like noise or reveal imaging artefacts.

The results show a broad spread in the relative contributions of star
forming regions and the evolved stellar population to dust heating.  We
found 10 galaxies where $\eta$(160/250~$\mu$m) $<$0.30, and 6 of these
galaxies have $\eta$(250/350~$\mu$m) $<$0.30.  In contrast, we found
11 galaxies (excluding NGC~6946 for reasons discussed below) with
either $\eta$(160/250~$\mu$m) $>$0.70 or $\eta$(250/350~$\mu$m)
$>$0.70, which would indicate that the evolved stellar population is the
dominant heating source for dust seen at least at $\geq250$~$\mu$m.
For 5 galaxies, we find that $0.30<\eta<0.70$ for both the 160/250 and
250/350~$\mu$m ratios, which would indicate that star forming regions
and the evolved stellar populations equally contribute to heating the
dust seen at 160-350~$\mu$m.  However, just because this ratio is near
0.50 for an individual galaxy does not necessarily mean that star
forming regions and the evolved stellar population contribute equally to
dust heating everywhere within that galaxy.  Several of the $\eta$
maps, such as the 160/250~$\mu$m maps for NGC~925, 2403, and 5457 show
that star forming regions may be dominant local dust heating sources
but that dust seen in diffuse regions or near the centres of the
galaxies at these wavelengths is heated more by light from the evolved
stellar population.  We also see in the 160/250~$\mu$m maps for
NGC~4725 and 4736 that star forming regions may contribute more to
heating dust within ring-like structures seen at these wavelengths but
that the evolved stellar population contributes more to the dust heating
in other locations.

For the majority of galaxies in this analysis, $\eta$(160/250~$\mu$m)
is smaller than $\eta$(250/350~$\mu$m).  This matches our
expectations, as we would expect to see dust heated by star forming
regions at shorter wavelengths and dust heated by evolved stars
at longer wavelengths.  The result also gives us more confidence in
the results from using this multiple component fit to identify heating
sources.  In NGC~3621, 4321, and 4548, however, $\eta$(250/350~$\mu$m)
was significantly smaller ($\gtrsim$0.15) than $\eta$(160/250~$\mu$m).
The 250/350~$\mu$m residual images from the fits applied to both
NGC~3621 and 4548 show some broad asymmetric structures that were not
fit using Equation~\ref{e_multfit} properly and could potentially have
caused issues with the fits in these two specific cases.  The reason
for the inconsistencies in NGC~4321 are probably the result of the
$I_\nu(3.6\mu\mbox{m})$ term being relatively small compared to the
$I(\mbox{SFR})$ term.

The 250/350~$\mu$m data of NGC~6946 stand out as a special case where
we had difficulty accurately reproducing the observed colour
temperature structures.  Neither the star forming regions (traced by
the 24~$\mu$m data in this case) nor the evolved stellar populations
trace the structures seen in the 250/350~$\mu$m map, so we would
naturally not be able to fit this as the function of the sum of the
3.6 and 24~$\mu$m data.  We therefore flag the $\eta$(250/350~$\mu$m)
value as suspect in Table~\ref{t_multfit} and disregard it in the rest
of the analysis.

As a consistency check, we can compare the $\eta$ values for NGC~2403,
3031, and 5236 to the $\eta$ values from \citet{bendo12a}.  They
expressed their results as $E_{SF}/E_{Tot}$, which represents the
relative fraction of temperature variations linked to heating by star
forming regions; $\eta$ is equivalent to $1-E_{SF}/E_{Tot}$.  Most of
the $\eta$ values from both studies are within $\sim0.15$, which is
within the range of uncertainty of our methodology stated in
Appendix~\ref{a_efractest}.  The $\eta$(250/350~$\mu$m) measured in
our new analysis for NGC~5236 is $\sim0.30$ higher than the value from
\citet{bendo12a}, which affects our interpretation of the dust
heating.  The 250/350~$\mu$m colour temperature map also includes
significant structure (which may be either artefacts of the data
processing or structures in the galaxy) that were not accurately
fitted in the analysis, and both the extinction-corrected H$\alpha$
and 3.6~$\mu$m maps trace similar structures in NGC~5236, so it is
possible that the uncertainties in the $\eta$ values for this galaxy
are much higher than our assumed uncertainty of 0.15.  The results are
also potentially affected by multiple other changes in the analysis,
including the use of updated 160-350~$\mu$m data, the use of different
tracers of dust heating sources, and changes in the angular resolution
used in the analysis.

In another check, we compared the $\eta$ values obtained in 24 and
120~arcsec binned data in NGC~3031 and 5457.  In this comparison,
which is described in Appendix~\ref{a_reseffect}, we found that the
$\eta$ values decrease somewhat in the 120~arcsec binned data but that
the decrease was consistent with the uncertainties from the Monte
Carlo analysis and the uncertainty of 0.15 in the methodology as
described in Appendix~\ref{a_efractest}.  We recommend treating the
comparison of nearby galaxies to distant galaxies with some caution,
but the derived $\eta$ values should not be so severely biased that
they are unusable.

\begin{figure}
\includegraphics[width=0.45\textwidth]{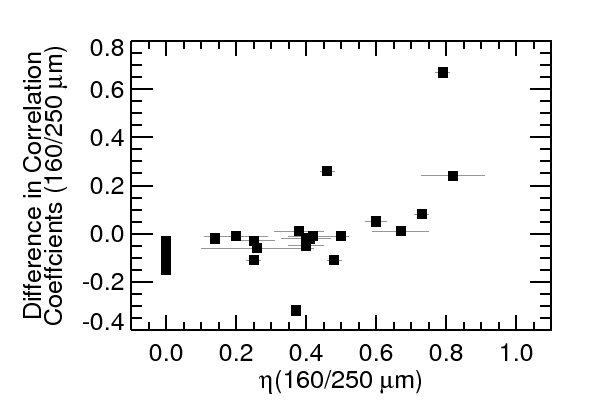}
\includegraphics[width=0.45\textwidth]{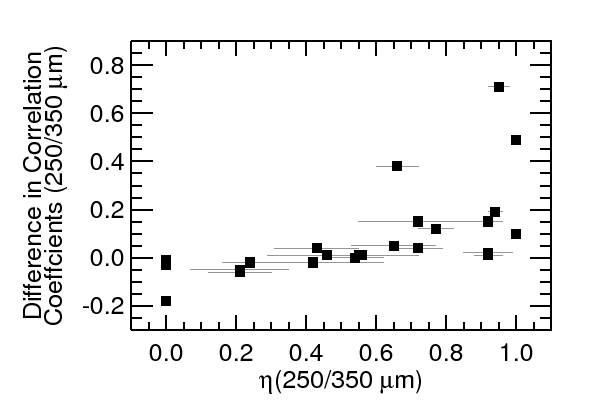}
\caption{The y-axes show the difference between correlation
  coefficients in Table~\ref{t_singlesource}.  For a given infrared
  surface brightness ratio, the correlation coefficient for the
  relation to H$\alpha$ emission is subtracted from the coefficient
  for the relation to 3.6~$\mu$m emission.  Positive values in the
  difference correspond to galaxies where variations in the infrared
  surface brightness ratio being more dependent on emission from the
  evolved stellar population, while negative values correspond to
  galaxies where the ratios primarily depend on emission from star
  forming regions.  The x-axes show the $\eta$ values from
  Table~\ref{t_multfit} for the corresponding infrared surface
  brightness ratio, with higher values corresponding to more dust
  heating by older stars at the shorter wavelength in the ratio.}
\label{f_competacoeff}
\end{figure}

In yet another check of the analysis, we also found that the $\eta$
values from Table~\ref{t_multfit} are consistent with the results from
comparing the 160/250 and 250/350~$\mu$m ratios to the emission from
H$\alpha$ (or 24~$\mu$m emission) and 3.6~$\mu$m emission in
Section~\ref{s_singlesourceanalysis}.  When $\eta>0.70$, we typically
find that the corresponding ratio is significantly better correlated
with 3.6~$\mu$m emission, and when $\eta>0.30$, we typically find that
the corresponding ratio is better correlated with star formation.  We
do frequently find individual cases where one of the analyses produces
less definitive results than the other, and the analyses sometimes
produce correspondingly indeterminate results, but we do not have any
example where the $\eta$ values and the
Section~\ref{s_singlesourceanalysis} analysis identify different
dominant heating sources.  Figure~\ref{f_competacoeff} shows a
comparison between $\eta$ and the difference between the correlation
coefficients from Table~\ref{t_singlesource} for a given infrared
surface brightness ratio.  The correlation coefficient differences are
based on subtracting the coefficient for the relation based on
H$\alpha$ emission from the coefficients for the relations based on
3.6~$\mu$m emission; higher values correspond to a higher fraction of
the dust heating being attributed to the evolved stellar
population. The relations are nonlinear, but the data do show a trend.
The Spearman correlation coefficients (which are more useful than
Pearson correlation coefficients when assessing whether a nonlinear
trend is present) are 0.70 for the relation in the 160/250~$\mu$m data
in Figure~\ref{f_competacoeff} and 0.85 for the relation in the
250/350~$\mu$m data, indicating a good correspondence between $\eta$
and the difference between the coefficients in
Table~\ref{t_singlesource}.  Overall, these results demonstrate that
the dust heating sources identified using the $\eta$ values derived in
this section should generally be consistent with the analysis in
Section~\ref{s_singlesourceanalysis}.

\subsection{Residual structures}
\label{s_decompositionresudial}

The residual temperature structure maps can reveal issues with the
quality of the fits.  Some of the fits are quite good; the replicated
colour temperatures match the observed colour temperatures to within
2-3~K.  In quite a few galaxies, however, we see structures in the
residual maps that are potentially imaging artefacts that would not be
readily apparent when looking at the individual infrared images.  Two
of the most egregious examples of this are the 250/350~$\mu$m images
for NGC~4548 and 5236, which, as also discussed above, have residual
asymmetric structures that align with the scan direction of the SPIRE
instrument.  We also see a few regions where the residual images,
particularly the 250/350~$\mu$m images, exhibit asymmetric hot or cold
structures that cannot be easily replicated using
Equation~\ref{e_multfit}.  As mentioned previously, the 250/350~$\mu$m
data for NGC~6946 exhibit the most severe artefacts, but we see
similar notable artefacts in the 250/350~$\mu$m data for NGC~628,
3184, 4254, and 4736.  These may be imaging artefacts, colour
variations related to foreground cirrus structures, or real
temperature structures produced by non-local dust heating effects, but
in any case, the $\eta$(250/350~$\mu$m) derived here may not
be reliable for some of these galaxies.  Additionally, some of the
residual temperature structure may look particularly pronounced
because of the high dynamic range of the 160/250 or 250/350~$\mu$m
data, which could be the case for NGC~3031 and 4736.

NGC 628, 3031, 4321, 4725, 4736, 5457, and 6946 all exhibit nuclei that
appear unusually cold in the 250/350~$\mu$m residual maps, and the
nucleus of NGC~6946 also appears cold in the map of the 160/250~$\mu$m
data.  This could arise for a few reasons.  The fits to the NGC~628,
4736, and 6946 250/350~$\mu$m data were already flagged as
problematic, and asymmetric dust temperature structures may have
contributed to the appearance of the cold nuclei in these galaxies.
NGC~3031, 4321, and 4725 have AGN \citep{ho97, moustakas10} that can
potentially produce synchrotron emission at submillimetre wavelengths.
The synchrotron emission could appear relatively bright compared to
the thermal emission from the dust because the surface density of the
dust near the centres of these galaxies is relatively low compared to
areas in the outer disc.  The data in Figure~\ref{f_multfit} clearly
show that the 250/350~$\mu$m ratio is unusually suppressed for
NGC~3031, but the results are not as clear for NGC~4725.  It would
also be unlikely that nonthermal AGN emission would explain the cold
residuals in NGC~628, 5457, and 6946, which have optical nuclei
spectra classified as star forming \citep{ho97, moustakas10}.  Another
possibility is that, when we use the 3.6~$\mu$m emission as a tracer
of starlight from the evolved stellar population, the cold residuals
appear because we are not removing emission from the bulge stars
outside the planes of the target galaxies that is not heating the dust
within the plane of the galaxies.  As a result, when we recreate the
colour temperature maps using the 3.6~$\mu$m, H$\alpha$, and 24~$\mu$m
data, the bulge appears warmer in the fitted map than in the observed
map.  This would satisfactorally explain the appearance of the cold
nuclear residuals in galaxies with large bulges like NGC~4725 and
4736, but it is less clear that it explains why these residuals
nuclear features appear in galaxies with small bulges like NGC~628,
5457, and 6946.  A final possibility is that the relation of dust to
its heating sources differs between the nuclei and discs in some
galaxies, which is related to the hypothesis from \citet{sauvage10}
stating that the infrared emission from the nuclei and discs of
galaxies may be decoupled \citep[see also ][]{roussel01}.  Ultimately,
some combination of the above physical and analytical issues may be
responsible for the appearance of the cold nuclear regions in the
residual maps.

We also sometimes see hot spiral structures offset from spiral arms in
a few galaxies, most notably in the 160/250~$\mu$m maps for NGC~628,
5236, and 6946.  As discussed in Appendix~\ref{a_offsetlinecut}, this
potentially appears because the dust around the star forming regions
in these structures is distributed asymmetrically, which then leads to
asymmetries in how light from star forming regions propagates into the
ISM and heats the dust.

\subsection{Relations between $\eta$ and other galaxy properties}

We looked at the relations between the $\eta$ values and various other
galaxy properties including morphological type, distance, inclination,
luminosities, surface brightnesses, and luminosity ratios based on
H$\alpha$, 3.6~$\mu$m, 24~$\mu$m, 160~$\mu$m, 250~$\mu$m, 350~$\mu$m,
and 500~$\mu$m data (and including metrics related to star formation
rates, specific star formation rates, stellar mass, dust mass, dust
temperature, and dust obscuration), but we found no clear dependency
of $\eta$ on any global galaxy property.  At best, we can only make a
couple of tentative statements about the possible influence of various
galaxy properties on $\eta$.

We did find that $\eta$ was relatively high in NGC~3031 and 4736, two
of the three Sab galaxies in this analaysis.  This could be because of
the increased role that the large bulges play in dust heating in these
galaxies, as also found by \citet{sauvage92} and
\citet{engelbracht10}.  However, the $\eta$ values for NGC~4725 are
not as high as for the other two Sab galaxies, and some of the Sb-Sd
galaxies have $\eta$ values that are just as high as the values for
NGC~3031 and 4736.  Second, we found weak trends in which
$\eta$(160/250~$\mu$m) decreases as the monochromatic luminosities
measured in each of the 24-350~$\mu$m bands increases.  However, the
strongest relation (between $\eta$(160/250~$\mu$m) and the 160~$\mu$m
luminosity) had a Spearman correlation coefficient with an absolute
value of only 0.75, and the relations with infrared luminosities in
other bands had coefficients of $\sim 0.70$, which would imply that
$\ltsim$50\% of the variance in $\eta$(160/250~$\mu$m) depends upon
infrared luminosity.  It is also unclear why the correlation
coefficients for corresponding $\eta$(250/350~$\mu$m) are
$<$0.50, which would indicate that $\eta$(250/350~$\mu$m) shows no
significant relation to infrared luminosity.  

We also found a slight bias in $\eta$(250/350~$\mu$m) with distance.
The Spearman correlation coefficient for this relation is -0.63, which
is relatively weak.  However, $\eta$(250/350~$\mu$m)$>$0.90 for most
galaxies within distances of 7 Mpc but not for galaxies at larger
distances.  As explained in Appendix~\ref{a_reseffect}, this could be
because diffuse dust is more easily separated from star forming
regions in data with higher angular resolutions \citep[see also
][]{galliano11}, but the results from Appendix~\ref{a_reseffect} show
that we will still obtain $\eta$(250/350~$\mu$m)$>$0.90 for sources
like NGC~5457 at distances grater than 15~Mpc.  It is also unclear why
this effect is not seen for $\eta$(160/250~$\mu$m), where the
corresponding correlation coefficient is -0.15.  The other possibility
is that, because of selection effects, we are seeing different forms
of dust heating in nearby and distant galaxies, which is quite likely
given the inhomogeneity of the sample.  Many of the galaxies within
5~Mpc would not be selected for this analysis if they were at 20~Mpc
simply because their angular sizes would be too small; it is plausible
that dust heating in these galaxies may be different than the dust
heating seen in the physically larger galaxies at larger distances.

Given the weakness of these trends, the relatively small and
inhomogeneous nature of our sample, and some of the issues with our
derivation of $\eta$ (including how the $\eta$ values depend on the
star formation tracer used in the computation and potential biases in
$\eta$ with distance), we suggest being extremely cautious regarding
any of these trends in $\eta$.  Further analyses with larger,
homogeneous samples of galaxies or with better measurements of the
fractions of dust heated by different sources may produce more
reliable results.

\section{Discussion}
\label{s_discussion}

\subsection{Implications for dust modelling and SED fitting}
\label{s_discussion_sed}

We have identified far-infrared emission from multiple nearby galaxies
produced by dust heated by intermediate-aged and older
stars.  What is surprising, however, is that we also find some
galaxies where the far-infrared emission at $\leq 250$~$\mu$m and
possibly at longer wavelengths is mainly from dust heated by star
forming regions.  Previously published results had implied that the
transition between emission from warmer dust heated by star formation
and colder dust heated by evolved stars should fall within a
relatively narrow wavelength range.  \citet{bendo10},
\citet{boquien11}, and \citet{bendo12a}, who had used the same
techniques applied here but who had studied only a limited number of
galaxies, had suggested that the transition was 160-250~$\mu$m.
\citet{hughes14} presented NGC~891 as an example of a spiral galaxy
where, using the same techniques, star forming regions could be
identified as the heating source for the dust seen at wavelengths as
long as 350~$\mu$m.  However, it was unclear whether the results for
NGC~891 were just a consequence of issues with applying these analysis
techniques to an edge-on galaxy where emission is integrated along the
line of sight.  It is clear now that NGC~891 is not the only galaxy
where dust at $\geq250$~$\mu$m may be heated by star forming regions.
The results from the \citet{draine07a} models applied to multiple
galaxies \citep[e.g. by ][]{draine07b, aniano12, dale12,
  mentuchcooper12, ciesla14} had suggested that the transition was at
30-100~$\mu$m.  Our results, which include several of the galaxies
contained in these studies, show that this transition point is
typically at longer wavelengths.  Indeed, the results from
some radiative transfer models \citep[e.g.][]{law11, delooze12} have
placed this transition point in a wavelength range that more closely
matches our emipirical results.

Many existing dust emission and radiative transfer models
\citep[e.g. such as those published by ][]{silva98, draine07a,
  bianchi08, dacunha08, baes11, popescu11, domingueztenreiro14} can
accurately reproduce either globally-integrated infrared galaxy SEDs
or the SEDs of individual subregions within galaxies.  To accurately
characterise the dust emission, however, these models should also
account for the broad variation in the transition wavelength between
the two different dust components that we have identified.  If the
transition is at a wavelength that is too short, it could lead to dust
temperatures that are too high and dust masses that are too low, while
the converse would occur if the transition is at too long a
wavelength.

In future research, we will examine either using existing models or
developing new models to replicate not only the global SEDs of these
galaxies but also the observed infrared surface brightness ratio
variations within them.  For now, we can at least examine the
implications of using single modified blackbody functions to estimate
global dust temperature and mass.

\citet{bendo10} had proposed a method of fitting a single modified
blackbody to the SED of M81 in which the function was fit to the
160-500~$\mu$m data points and the 70~$\mu$m data point was treated as
an upper limit.  This is still appropriate for M81 because the
emission in the full 160-500~$\mu$m range mostly originates from dust
heated by the radiation field from the evolved stellar population.  It
may also be appropriate for galaxies where the dust is primarily
heated by a single source, which could be either the evolved stellar
population or star forming regions (as in NGC~4254, 4303, 4501, and
5364 in this paper or as in NGC~891 as reported by \citet{hughes14}).
Based on the analysis from \citet{bianchi13}, the dust mass estimated
from the modified blackbody fit will be accurate as long as the fitted
data do not include emission from stochastically-heated dust grains
and do not sample regions with widely-varying large grain dust
temperatures (but also note the potential issues with
resolution-related effects discussed by \citet{galliano11}) or major
variations in other dust grain properties.  However, our results show
that, in a significant fraction of galaxies, the emission in the
160-350~$\mu$m range originates from dust heated by both star forming
regions and the evolved stellar population.  In a few of these
galaxies, the star forming regions and the evolved stellar population
appear to contribute equally to heating the emission at all
wavelengths in this range; the single thermal components may still
produce reasonable dust mass estimates in these cases, although this
would depend on whether the large dust grains are close to the same
temperature.  In other galaxies, the dust observed at 160~$\mu$m is
heated much more by star forming regions than the dust observed at
350~$\mu$m.  It is these cases where the single modified blackbodies
fit to the 160-500~$\mu$m data will yield the most inaccurate dust
masses.

To examine how the dust temperatures and masses change if we divide
the dust SED into different components based on the dust heating
sources, we performed SED fits to the 160-500~$\mu$m data for NGC~628,
2403, and 5457, three galaxies where, based on our results, star
forming regions were responsible for $\geq50$\% of the heating for the
dust seen at $\leq160$~$\mu$m but the evolved stellar populations were
the dominant source of heating for the dust seen at $\geq250$~$\mu$m.
Among the galaxies in the sample, the dust temperatures and masses
from fits to the measured 160-500~$\mu$m data for these three galaxies
should diverge the most from the quantities that would be found if we
separated the SED into separate components based on the results from
Section~\ref{s_decompositionanalysis}.

\begin{table*}
\centering
\begin{minipage}{101mm}
\caption{Measured and rescales SEDs for NGC~628, 2403, and NGC~5457.}
\label{t_seddata}
\begin{tabular}{@{}lccccc@{}}
\hline
Galaxy &
    Wave- &
    Measured &
    Scaling Factor &
    \multicolumn{2}{c}{Rescaled Flux Densities (Jy)} \\
&
    length &
    Flux Density &
    (Cold &
    Warm &
    Cold \\
&
    ($\mu$m) &
    (Jy)$^a$ &
    Component)$^b$ &
    Component &
    Component \\
\hline
NGC 628 & 
    70 & 
    $34 \pm 4$ & 
    0 & 
    $34 \pm 4$ &
    0 \\
&
    100 &
    $83 \pm 7$ &
    0 &
    $83 \pm 7$ &
    0 \\
&
    160 &
    $115 \pm 10$ &
    $0.25 \pm 0.15$ &
    $86 \pm 19$ &
    $29 \pm 17$ \\
&
    250 &
    $65 \pm 3$ &
    $0.77 \pm 0.15$ &
    $15 \pm 10$ &
    $50 \pm 10$ \\
&
    350 &
    $31.3 \pm 1.3$ &
    1 &
    0 &
    $31.3 \pm 1.3$ \\
&
    500 &
    $12.8 \pm 0.5$ &
    1 &
    0 &
    $12.8 \pm 0.5$ \\
NGC 2403 & 
    70 & 
    $86. \pm 10$ & 
    0 & 
    $86. \pm 10$ & 
    0 \\
&
    100 &
    $150 \pm 20$ &
    0 &
    $150 \pm 20$ &
    0 \\
&
    160 &
    $210 \pm 20$ &
    $0.48 \pm 0.15$ &
    $110 \pm 30$ &
    $100 \pm 30$ \\
&
    250 &
    $129 \pm 5$ &
    $0.94 \pm 0.15$ &
    $8 \pm 20$ &
    $120 \pm 20$ \\
&
    350 &
    $68 \pm 3$ &
    1 &
    0 &
    $68 \pm 3$ \\
&
    500 &
    $29.3 \pm 1.2$ &
    1 &
    0 &
    $29.3 \pm 1.2$ \\
NGC 5457 & 
    70 & 
    $118 \pm 14$ & 
    0 & 
    $118 \pm 14$ & 
    0 \\
&
    100 &
    $250 \pm 20$ &
    0 &
    $250 \pm 20$ &
    0 \\
&
    160 &
    $370 \pm 30$ &
    $0.37 \pm 0.15$ &
    $220 \pm 60$ &
    $150 \pm 60$ \\
&
    250 &
    $208 \pm 8$ &
    1 &
    0 &
    $208 \pm 8$ \\
&
    350 &
    $102 \pm 4$ &
    1 &
    0 &
    $102 \pm 4$ \\
&
    500 &
    $42.4 \pm 1.7$ &
    1 &
    0 &
    $42.4 \pm 1.7$ \\
\hline
\end{tabular}
$^a$ All flux density measurements except the 60~$\mu$m measurements
    and the 100~$\mu$m measurement for NGC~2403 include color
    corrections.\\
$^b$ See the text in Section~\ref{s_discussion_sed} for how these numbers 
    are set.
\end{minipage}
\end{table*}

The global 70-500~$\mu$m flux densities that we use are given in
Table~\ref{t_sedfit}.  Because the PACS and SPIRE flux calibration
have been recently updated, we needed to re-measure the flux
densities; the measurements were made within apertures $1.5\times$ the
size of the optical disc given by \citet{devaucouleurs91} (although
this crosses the edges of the PACS images for NGC~5457).  For
NGC~2403, which was not observed at 100~$\mu$m with PACS, we used the
100~$\mu$m flux density from the Infrared Astronomical Satellite
(IRAS) given by \citet{rice88}.  We also included MIPS 70~$\mu$m flux
densities from \citet{dale09}.  Even though PACS 70~$\mu$m data were
available, the MIPS 70~$\mu$m data have better signal-to-noise ratios
and therefore seemed more trustworthy.  Colour corrections and other
photometric corrections have been applied to all PACS and SPIRE flux
densities; colour corrections for IRAS and MIPS data are smaller than
the uncertainties in the data and are therefore not applied.

To create SEDs representing separate warmer and colder components, we
assumed for these three specific galaxies that all $\leq100$~$\mu$m
emission originated from dust heated by star forming regions and all
$\geq350$~$\mu$m emission originated from dust heated by the evolved
stellar population (although these are oversimplifications).  The
160~$\mu$m data were multiplied by $\eta$(160/250~$\mu$m)
and the 250~$\mu$m by $\eta$(250/350~$\mu$m) to get the
emission at those wavelengths for the cold component, with the
remainder assigned to the warm component.  The uncertainties in the
scaling terms is conservatively set at 0.15 based on the results from
Appendix~\ref{a_efractest}.

\begin{table*}
\centering
\begin{minipage}{159mm}
\caption{Dust temperatures and masses for NGC~628, 2403, and 5457.}
\label{t_sedfit}
\begin{tabular}{@{}lcccccc@{}}
\hline
Galaxy &
    \multicolumn{2}{c}{Unscaled Data} &
    \multicolumn{4}{c}{Rescaled Data} \\
&
    &
    &
    \multicolumn{2}{c}{Warm Component} &
    \multicolumn{2}{c}{Cold Component} \\
&
    Dust &
    Dust &
    Dust &
    Dust \\
&
    Temperature (K) &
    Mass (M$_\odot$) &
    Temperature (K) &
    Mass (M$_\odot$) &
    Temperature (K) &
    Mass (M$_\odot$) \\
\hline
NGC 628 &
    $18.2 \pm 0.6$ &
    $(7.2 \pm 0.7)\times10^7$ &
    $22.5 \pm 1.3$ &
    $(2.1 \pm 0.8)\times10^7$ &
    $14.5 \pm 1.8$ &
    $(1.3 \pm 0.5)\times10^8$ \\
NGC 2403 &
    $16.9 \pm 0.5$ &
    $(2.0 \pm 0.2)\times10^7$ &
    $26.1 \pm 1.1$ &
    $(1.6 \pm 0.4)\times10^6$ &
    $13.7 \pm 1.0$ &
    $(3.5 \pm 1.0)\times10^7$ \\
NGC 5457 &
    $18.1 \pm 0.6$ &
    $(1.1 \pm 0.1)\times10^8$ &
    $23.8 \pm 1.7$ &
    $(2.0 \pm 0.8)\times10^7$ &
    $16.8 \pm 0.6$ &
    $(1.3 \pm 0.1)\times10^8$ \\
\hline
\end{tabular}
\end{minipage} 
\end{table*}

\begin{figure}
\includegraphics[width=0.45\textwidth]{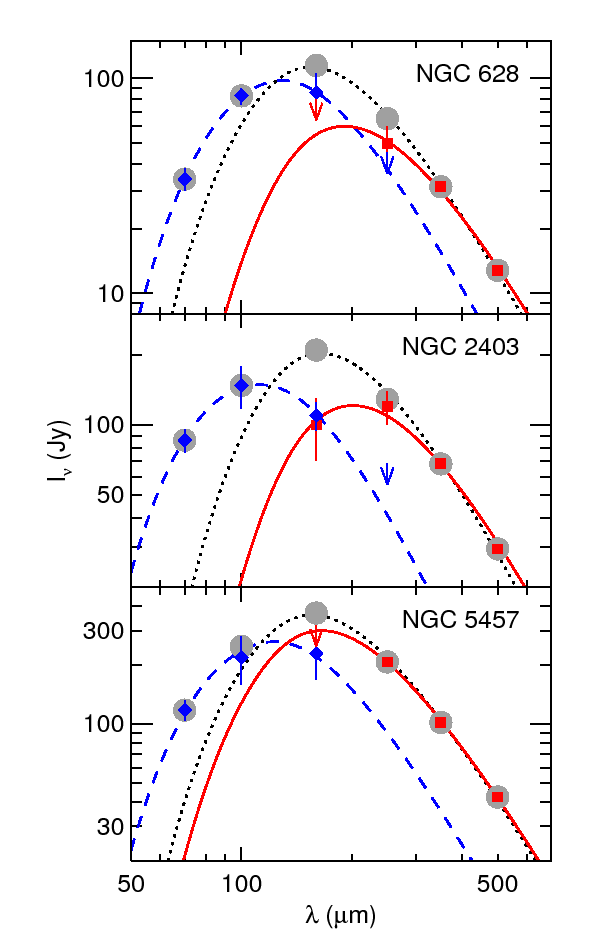}
\caption{70-500~$\mu$m SEDs for NGC~628, 2403, and 5457.  The grey
  circles show the measured unscaled global flux densities for these
  galaxies.  The blue diamonds show the rescaled data for the
  component of the SED from dust heated by star forming regions, and
  the red squares show the component from dust heated by the evolved
  stellar population, both of which were found by rescaling the data
  using the results from Section~\ref{s_decompositionanalysis}. Flux
  densities which are not measured at the $3\sigma$ level are
  represented as arrows showing the $3\sigma$ upper limits; the upper
  limits were not used in the fits.  These data and the rescaling
  factors are listed in Table~\ref{t_seddata}.  The black dotted line
  shows the best fitting modified blackbody function for the measured
  global flux densities, and the blue dashed and red solid lines show
  the best fitting function for the rescaled data.  The emissivity
  function for the modified blackbody function is equivalent to the
  function for the tabulated $\kappa_\nu$ values given by
  \citet{draine03}, where $\beta$ is approximately equal to 2.  The
  dust temperatures and masses for the best fitting functions are
  listed in Table~\ref{t_sedfit}.}
\label{f_sedfit}
\end{figure}

We fit the data using the function
\begin{equation}
M_{dust}=\frac{f_\nu D^2}{\kappa_\nu B_\nu (T)},
\label{e_dustmass}
\end{equation}
which is a variant of the equation originally derived by
\citet{hildebrand83}.  In this equation, $f_\nu$ is the observed flux
density, $D$ is the distance, $\kappa_\nu$ is the dust opacity, and
$B_\nu (T)$ is a blackbody function based on the temperature from the
fits.  We used $\kappa_\nu$ values based on interpolating in logarithm
space between the tabulated values given by \citet{draine03}; for
these $\kappa_\nu$, $\beta$ is equal to approximately (but not
exactly) 2 in the 70-500~$\mu$m range.  Although many {\it Herschel}
results indicate that $\beta$ may be between 1.5 and 2
\citep[e.g.][]{boselli12}, it is unclear how to rescale the
\citet{draine03} opacities if $\beta \neq 2$, so we will not attempt using
alternate values of $\beta$.  We first perform fits to both the global
160-500~$\mu$m data in Table~\ref{t_seddata} and used the
70-100~$\mu$m flux densities as upper limits.  We then fit the
70-250~$\mu$m data for the warm component and the 160-500~$\mu$m data
for the cold components separately (excluding rescaled data where the
measurements are $<$3$\sigma$). Note that the warm component is
potentially emission from a sum of multiple thermal components and may
not be accurately physically represented as a single modified
blackbody, although the cold component can be physically approximated
as a single modified blackbody \citep{bianchi13}.  The global SEDs for
these galaxies as well as the separate rescaled components and
best-fitting modified blackbodies functions are shown in
Figure~\ref{f_sedfit}.  The resulting temperatures and masses are
listed in Table~\ref{t_sedfit}.

The resulting temperatures change by 2-4~K between the unscaled data
and rescaled cold component, which is higher than implied by the
uncertainties. This change in temperature is small compared to the
dust temperature variations seen in modified blackbodies fitted to
160-500~$\mu$m data (or similar wavelength ranges) for large samples
of nearby galaxies, such as those studied by \citet{galametz12} and
\citet{cortese14}, but the resulting dust temperatures for the
rescaled cold components are lower than what is obtained by these
authors, even when they used the same emissivity functions.  While the
temperatures obtained by these authors are characteristic of the
colours, they probably do not represent the actual dust temperatures
of the coldest dust heated by the evolved stellar population in
galaxies like NGC~628, 2403, and 5457.  The actual variance in dust
temperatures in nearby galaxies may therefore be higher than implied
by the fits to 100-500~$\mu$m data.

The estimated masses for the rescaled cold component are not well
defined and the estimates are $<$3$\sigma$ for NGC~628.  None the less,
the results show that the dust masses may change by up to a factor 2
between the unscaled and rescaled data.  Gas-to-dust ratios obtained
using dust masses from unscaled 160-500~$\mu$m data will tend to be
high but still close enough to assess whether the gas-to-dust ratios
are similar to the nominally-expected values of 100-200 or whether
dust masses appear unusually low relative to gas masses, as would be
expected in low-metallicity regions.  The data may exhibit more
scatter if dust masses in some but not all of the galaxies are
underestimated through fits to unscaled 160-500~$\mu$m data, although
this increase scatter may not be significant in logarithm space.

In addition to the issues described above, dust masses can potentially
be underestimated from fitting modified blackbodies to SEDs in other
ways.  Emission at $<$100~$\mu$m potentially includes stochastically
heated dust that will appear warmer than the larger dust grains that
are in thermodynamic equilibrium with the illuminating radiation
field; fitting a single modified blackbody to data from 60, 70, or possibly
even 100~$\mu$m to 500~$\mu$m could produce temperatures that are
warmer than the bulk of the dust grains and dust masses that are too
low.  Additionally, many authors allow $\beta$ to vary as a free
parameter in SED fits, but degeneracies between $\beta$ and
temperature \citep[e.g. ][]{shetty09, galametz12, kirkpatrick14} could
cause dust temperatures to generally appear warmer than the actual
dust temperatures and thus make the dust masses appear lower.
Moreover, $<$15~K dust visible only on the Rayleigh-Jeans side of the
dust SED could make $\beta$ appear lower than it actually is
\citep{kirkpatrick14}, which not only gives the false appearance that
$\beta$ is variable among or within galaxies but also potentially
leads to even lower dust masses.

As our analysis shows, it is very difficult to use the
$\eta$ values from Section~\ref{s_decompositionanalysis} to
accurately rescale dust SEDs.  Our SED fitting approach oversimplifies
the physics of the dust emission and makes some relatively crude
assumptions about the division between warm and cold thermal
components.  To measure more accurate dust masses within individual
galaxies, we recommend not only using dust models with more realistic
modelling of the dust heating and radiation but also performing fits
with these models that replicate not only the observed flux densities
but also the observed colour temperature variations within the
galaxies.  See De Looze et al. (2014, in preparation) for an example
of this approach.

\subsection{Implications for measuring star formation rates}

The results here also have implications for using infrared emission to
measure star formation rates.  A commonly held classical perspective
has been that dust emission traces star formation because dust is
primarily absorbing light from star forming regions.  This has
generally been true for mid-infrared wavebands tracing hot dust, such
as the {\it Spitzer} 24~$\mu$m or WISE 22~$\mu$m band.  However, prior
to our analysis, several different authors had presented multiple
observational results implying that this did not apply to the
infrared, including the demonstration that the ratio of far-infrared
to H$\alpha$ emission was higher in spiral galaxies with large bulges
than in spiral galaxies with small bulges \citep{sauvage92}; the
observation that the ratio of infrared to H$\alpha$ emission for
subregions within galaxies was higher for longer infrared wavelengths
than shorter wavelengths \citep{calzetti10}; and the prior {\it
  Herschel}-based analyses relating variations in the far-infrared
surface brightness ratios to the evolved stellar population.
Additionally, \citet{boquien11} and \citet{bendo12a} had demonstrated
that far-infrared emission could still be correlated with other star
formation tracers, such as H$\alpha$ and 24~$\mu$m emission, even when
variations in the infrared surface brightness ratios demonstrated that
the dust was heated by the evolved stellar populations.  Both authors
suggested that these seemingly-contradictory results could be valid if
the dust heated by the evolved stellar population was tracing the gas
that fuels star formation, which would be related to star formation
through the Kennicutt-Schmidt law \citep{schmidt59, kennicutt98}.
They also proposed a scenario where emission at shorter wavelengths
originated from dust near young stars within star forming regions that
was optically thick to ultraviolet light while emission at longer
wavelengths came from dust in outer shells that were shielded from the
ultraviolet photons from the star forming regions and that were
therefore mainly heated by light from the older stellar population.

Our new results show that this scenario describing the relation of
far-infrared dust emission to star formation is oversimplistic.  In
some cases, the far-infrared emission still appears related to star
formation through the Kennicutt-Schmidt law, as the dust is primarily
heated by the evolved stellar population, and the \citet{boquien11}
and \citet{bendo12a} scenario of the structure of dust around star
forming regions still applies.  In other galaxies, all of the dust is
heated by the star forming regions.  Using far-infrared emission to
trace star formation therefore becomes problematic, as the functional
relation between star formation rate and infrared emission changes in
these two scenarios.  In the case where far-infrared emission
originates from dust heated directly by the star forming regions, the
relation between the emission and star formation rate should be
linear.  In the case where the emission is from dust not heated by
star formation but tracing the fuel for star formation, the relation
could be expected to follow a power law with an index of 1-2.  If
far-infrared emission is sometimes related to star formation rate
through direct dust heating by star forming regions and sometimes
related through the Kennicutt-Schmidt law, then we can expect the
relation between the two quantities to exhibit significant scatter,
especially when looking at longer wavelengths where more inconsistency
is expected.  In cases where dust emission is related to star
formation through the Kennicutt-Schmidt law, it should also be
possible to see dust emission from locations that have not achieved
the critical density for star formation described by
\citet{kennicutt98}, and this will add scatter to the empirical
relation between far-infrared emission and star formation.
\citet{kennicutt09} pointed out that such diffuse dust emission can be
seen at wavelengths as short as 24~$\mu$m, so dust emission
unassociated with star formation would naturally be expected at longer
wavelengths.

Several authors, including \citet{calzetti07}, \citet{leroy08},
\citet{zhu08}, \citet{kennicutt09}, \citet{hao11}, and
\citet{lee13}, have developed star formation metrics based on
combining measurements of ultraviolet or H$\alpha$ emission (tracing
the fraction of unextinguished light from star forming regions) with
infrared light (tracing the fraction of extinguished light from star
forming regions).  This naturally works if the re-radiated light is
from the dust heated by the star forming regions.  However, our
analysis shows that far-infrared emission sometimes originates from
dust heated by the intermediate and older stars.  Because of this,
using far-infrared emission to correct ultraviolet or optical star
formation tracers may potentially be inaccurate, although the
far-infrared emission may still correlate well with the column density
of dust obscuring star forming regions and may therefore still be
useful for extinction corrections.  Most previously-published
extinction-corrected star formation metrics, however, rely either on
mid-infrared dust emission or infrared emission integrated over a
large wavelength range such as 3-1100~$\mu$m.  Mid-infrared emission
should mainly contain light from dust heated directly by star forming
regions, so that waveband should be relatively accurate for correcting
for dust extinction.  However, integrated infrared emission may
include substantial emission from dust heated by older stars,
especially in galaxies with low star formation rates or large
spheroids of evolved stars, and therefore may provide an innacurate
extinction correction for ultraviolet and optical star formation
tracers \citep[see also ][]{delooze12}.

\section{Conclusions}
\label{s_conclusions}

In our analysis of variations in the 160/250 and 250/350~$\mu$m
surface brightness ratios observed within 24 nearby galaxies, we have
identified a broad variation in the heating of the dust observed in
the 160-350~$\mu$m range.  At one extreme, we definitively identified
star forming regions as the primary heating source for dust seen at
$\leq250$~$\mu$m and possibly at longer wavelengths in at least four
galaxies, and a fifth galaxy may also fall into this category.  At the
other extreme, we find strong evidence that emission at
$\geq160$~$\mu$m and possibly longer wavelengths originates from dust
heated by the evolved stellar population in at least three galaxies, and
weaker evidence suggests that two other galaxies may fall into this
category.  Among the rest of the galaxies, we either see a transition
from observing dust heated by star formation at 160~$\mu$m to
observing dust heated by the evolved stellar population at 350~$\mu$m,
or we see 160-350~$\mu$m emission from dust heated by a mixture of
young and old stars.

\begin{table}
\caption{Summary of identified dust heating sources related to surface
  brightness ratios.}
\label{t_summary}
\begin{center}
\begin{tabular}{@{}lccc@{}}
\hline
Galaxy &         \multicolumn{2}{c}{Heating Source$^a$}\\
&                160/250~$\mu$m &          250/350~$\mu$m \\
\hline
NGC 628 &        SF &                      ES \\
NGC 925 &        SF &                      mixed \\
NGC 2403 &       SF &                      ES \\
NGC 3031 &       ES &                      ES \\
NGC 3184 &       mixed &                   ES? \\
NGC 3621 &       ES &                      ES$^b$ \\
NGC 3631 &       mixed &                   mixed \\
NGC 3938 &       SF &                      mixed \\
NGC 3953 &       mixed &                   ES \\
NGC 4254 &       SF &                      SF \\
NGC 4303 &       SF &                      SF \\
NGC 4321 &       SF? &                     SF? \\
NGC 4501 &       SF &                      SF \\
NGC 4535 &       mixed &                   ES? \\
NGC 4548 &       ES  &                     ES$^b$ \\
NGC 4579 &       mixed &                   mixed \\
NGC 4725 &       ES? &                     ES? \\
NGC 4736 &       mixed &                   ES \\
NGC 5055 &       SF &                      mixed$^c$ \\
NGC 5236 &       SF &                      ES? \\
NGC 5364 &       SF &                      SF \\
NGC 5457 &       SF &                      ES \\
NGC 6946 &       SF? &                     $^d$ \\
NGC 7793 &       ES? &                     ES \\
\hline
\end{tabular}
\end{center} 
$^a$ SF stands for star formation.  ES stands for the evolved stellar
  population.  Mixed indicates that both of these sources may heat the
  dust equally or that the results are ambiguous.  Question marks are
  placed next to cases where only one of the quantitative analyses 
  identified this as the heating source and the other analyses produced
  ambiguous results.\\
$^b$ The results for the 160/250~$\mu$m data strongly indicated dust
  heating by the evolved stellar population at those wavelengths, but
  the results from the 250/350~$\mu$m data were more indeterminate.
  In these cases, it seems more likely in these cases that the
  250/350~$\mu$m is also linked to the evolved stellar population.\\
$^c$ The qualitative appearance of the 250/350~$\mu$m colour
  temperature map does not match what is predicted by some of the
  quantitative results, so we labelled this as mixed.\\
$^d$ The 250/350~$\mu$m colour temperature structures for this galaxy cannot be
  straightforwardly related to any dust heating source.
\end{table}

In Table~\ref{t_summary}, we list summaries of what we have identified
as the heating sources that influence the 160/250 and 250/350~$\mu$m
ratios for each galaxy based on the results from the analyses in
Sections~\ref{s_qualanalysis}-\ref{s_decompositionanalysis}.  We
identify definite heating sources for galaxies and ratios where the
analyses from at least two of the three sections show that the ratios
depend primarily on one heating source.  We mark cases as questionable
when one of the quantitative analyses suggests a predominant heating
source influencing a ratio but when the results from the other
quantitative analysis and the qualitative analysis of the colour
temperature maps provide ambiguous results.  Other galaxies and ratio
are marked as “mixed”.  These mixed cases are often situations where
the 160/250 or 250/350~$\mu$m ratios are influenced equally by star
forming regions and the evolved stellar population or cases where the
structures traced by both heating sources are so similar that the
colour temperatures appear to be equally related to both.  We also
highlight a number of cases where the analysis produced confusing or
illogical results as well as the confusing results for NGC~6946, where
the 250/350~$\mu$m colour temperature map did not look related to any
dust heating source.

We were unable to link any individual galaxy property to the dust
heating mechanisms.  We found that dust tended to be heated more by the
evolved stellar population in the Sab galaxies than in the Sb-Sd
galaxies but that dust tended to be heated by star forming regions
more in galaxies with high infrared surface brightnesses.  However,
the evidence for all of this is statistically weak because of the
limited sample size used and because of the limited work that could be
done with our analysis results.

None the less, our results demonstrate a substantial variation in the
relative magnitudes of the emission from warmer dust heated by star
forming regions and colder dust heated by the evolved stellar
populations within galaxies and a substantial variation in the
wavelength marking the transition point between emission from warmer
and colder thermal components.  SED fitting performed with either
modified blackbodies or more complex dust emission or radiative
transfer models should be able to account for or replicate these
variations to more accurately measure dust temperatures and masses.
These functions and models are typically fitted to observed flux
densities or surface brightnesses and often match the magnitudes of
the flux densities very well.  It would be interesting to test whether
these fits can accurately replicate the colours of these objects
(effectively the derivative of the SED curve), including the
variations in infrared surface brightness ratios that we displayed in
Figure~\ref{f_maptemp}.  Additionally, it would be appropriate to
either build new dust emission or radiative transfer models or adjust
existing models so that they can replicate these colour variations as
well as the magnitudes of the global and local SEDs.

\section*{Acknowledgments}

We thank the reviewer for the helpful comments on this paper.
GJB is funded by the STFC.  IDL is a postdoctoral researcher of the
FWO-Vlaanderen (Belgium).  The {\it Herschel} spacecraft was designed,
built, tested, and launched under a contract to ESA managed by the
{\it Herschel}/{\it Planck} Project team by an industrial consortium
under the overall responsibility of the prime contractor Thales Alenia
Space (Cannes), and including Astrium (Friedrichshafen) responsible
for the payload module and for system testing at spacecraft level,
Thales Alenia Space (Turin) responsible for the service module, and
Astrium (Toulouse) responsible for the telescope, with in excess of a
hundred subcontractors.  SPIRE has been developed by a consortium of
institutes led by Cardiff University (UK) and including
Univ. Lethbridge (Canada); NAOC (China); CEA, LAM (France); IFSI,
Univ. Padua (Italy); IAC (Spain); Stockholm Observatory (Sweden);
Imperial College London, RAL, UCL-MSSL, UKATC, Univ. Sussex (UK); and
Caltech, JPL, NHSC, Univ. Colorado (USA). This development has been
supported by national funding agencies: CSA (Canada); NAOC (China);
CEA, CNES, CNRS (France); ASI (Italy); MCINN (Spain); SNSB (Sweden);
STFC, UKSA (UK); and NASA (USA).  HIPE is a joint development (are
joint developments) by the {\it Herschel} Science Ground Segment
Consortium, consisting of ESA, the NASA {\it Herschel} Science Center,
and the HIFI, PACS and SPIRE consortia.  This publication makes use of
data products from the Wide-field Infrared Survey Explorer, which is a
joint project of the University of California, Los Angeles, and the
Jet Propulsion Laboratory/California Institute of Technology, funded
by the National Aeronautics and Space Administration.  This research
has made use of the NASA/IPAC Extragalactic Database (NED) which is
operated by the Jet Propulsion Laboratory, California Institute of
Technology, under contract with the National Aeronautics and Space
Administration.

{}

\appendix

\section{Comparisons of H$\alpha$ and 24~$\mu$m emission}
\label{a_ha24comp}

\begin{table*}
\begin{minipage}{160mm}
\caption{Results of using different star formation tracers in comparison to
    infrared surface brightness ratios}
\label{t_ha24comp}
\begin{tabular}{@{}lccccccc@{}}
\hline
Galaxy &    \multicolumn{7}{c}{Weighted Correlation Coefficients}\\
&           log($I_\nu$(24~$\mu$)) &
            \multicolumn{3}{c}{log($I_\nu$(160~$\mu$m)/$I_\nu$(250~$\mu$m))} &
            \multicolumn{3}{c}{log($I_\nu$(250~$\mu$m)/$I_\nu$(350~$\mu$m))} \\
&           vs. log($I$(H$\alpha$)) &
            vs. log($I$(H$\alpha$)&
            vs. log($I$(H$\alpha$)&
            vs. log($I_\nu$(24~$\mu$))&
            vs. log($I$(H$\alpha$)&
            vs. log($I$(H$\alpha$)&
            vs. log($I_\nu$(24~$\mu$))\\
&           (uncorrected) &
            (uncorrected) &
            (corrected) &
            &
            (uncorrected) &
            (corrected) &
            \\
\hline
NGC 628 &
            0.86 &
            0.73 &     0.82 &     0.83 &
            0.57 &     0.72 &     0.76 \\
NGC 925 &
            0.92 &
            0.81 &     0.82 &     0.84 &
            0.86 &     0.88 &     0.91 \\
NGC 2403 &
            0.97 &
            0.91 &     0.92 &     0.91 &
            0.74 &     0.75 &     0.77 \\
NGC 3031 &
            0.93 &
            0.40 &     0.47 &     0.55 &
            0.29 &     0.36 &     0.48 \\
NGC 3184 &
            0.80 &
            0.67 &     0.81 &     0.82 &
            0.59 &     0.75 &     0.80 \\
NGC 3621 &
            0.94 &
            0.84 &     0.87 &     0.90 &
            0.87 &     0.89 &     0.90 \\
NGC 3631 &
            0.99 &
            0.94 &     0.95 &     0.96 &
            0.93 &     0.95 &     0.96 \\
NGC 3938 &
            0.95 &
            0.92 &     0.96 &     0.97 &
            0.84 &     0.88 &     0.90 \\
NGC 3953 &
            0.91 &
            0.76 &     0.83 &     0.84 &
            0.62 &     0.68 &     0.71 \\
NGC 4254 &
            0.98 &
            0.89 &     0.88 &     0.87 &
            0.68 &     0.69 &     0.70 \\
NGC 4303 &
            0.83 &
            0.87 &     0.88 &     0.85 &
            0.86 &     0.91 &     0.91 \\
NGC 4321 &
            1.00 &
            0.97 &     0.97 &     0.97 &
            0.85 &     0.85 &     0.85 \\
NGC 4501 &
            0.90 &
            0.82 &     0.86 &     0.86 &
            0.74 &     0.82 &     0.87 \\
NGC 4535 &
            0.89 &
            0.85 &     0.91 &     0.92 &
            0.85 &     0.87 &     0.87 \\
NGC 4548 &
            0.68 &
           -0.44 &     0.67 &     0.74 &
            0.07 &     0.69 &     0.72 \\
NGC 4579 &
            0.99 &
            0.97 &     0.96 &     0.94 &
            0.84 &     0.83 &     0.83 \\
NGC 4725 &
            0.82 &
            0.72 &     0.77 &     0.76 &
            0.44 &     0.64 &     0.78 \\
NGC 5055 &
            0.97 &
            0.97 &     0.98 &     0.98 &
            0.92 &     0.93 &     0.92 \\
NGC 5236 &
            0.97 &
            0.89 &     0.89 &     0.88 &
            0.81 &     0.81 &     0.80 \\
NGC 5364 &
            0.87 &
            0.47 &     0.46 &     0.41 &
            0.71 &     0.75 &     0.79 \\
NGC 5457 &
            0.90 &
            0.80 &     0.89 &     0.90 &
            0.21 &     0.38 &     0.50 \\
NGC 7793 &
            0.80 &
            0.78 &     0.84 &     0.91 &
            0.62 &     0.69 &     0.80 \\
\hline
\end{tabular}
\end{minipage}
\end{table*}

Multiple authors \citep[e.g. ][]{calzetti05, calzetti07, prescott07}
have found a good correspondence between H$\alpha$ and 24~$\mu$m
emission from compact sources within nearby galaxies.  However, it is
possible for diffuse dust to produce 24~$\mu$m emission, and this
diffuse emission will not necessarily correspond to H$\alpha$ emission
\citep{kennicutt09}.  Additionally, the ratio of H$\alpha$ to
24~$\mu$m emission may vary with metallicity as well.  To examine
whether this could affect our analysis, we compared uncorrected
H$\alpha$ and 24~$\mu$m emission measured within the 24~arcsec binned
data in our analysis, as we also compared the correlations of the
160/250 and 250/350~$\mu$m ratios to the uncorrected H$\alpha$,
corrected H$\alpha$, and 24~$\mu$m emission.  We selected data for
24~arcsec bins where both the uncorrected H$\alpha$ and 24~$\mu$m
emission were measured at the $3\sigma$ level and where the data
otherwise met the criteria for use in the analysis in
Section~\ref{s_singlesourceanalysis}; the data meeting the criteria
for analysis on the 250/350~$\mu$m ratio were also used for
calculating correlation coefficients for the relations between the
uncorrected H$\alpha$ and 24~$\mu$m emission.  These selection
criteria are necessary for directly comparing H$\alpha$ and 24~$\mu$m
emission but may cause the resulting coefficients to differ slightly
($<$0.05) from the coefficients in Table~\ref{t_singlesource} listed
for the same relations.

The weighted correlation coefficients from this analysis are presented
in Table~\ref{t_ha24comp}.  First, we can see that while the
uncorrected H$\alpha$ and 24~$\mu$m data are often strongly
correlated, the correlation coefficients do not always equal 1, and
some of the values drop to $<$0.80.  Although the H$\alpha$ and
24~$\mu$m emission is clearly correlated, enough scatter exists that
swapping one for the other could potentially change the relations we
see when comparing these star formation tracers to the 160/250 and
250/350~$\mu$m surface brightness ratios.  Further analysis to
understand these variations and, in particular, to identify whether
they are associated with other properties of the ISM (such as the
gas-to-dust ratio or metallicity), would be useful, but this is beyond
the scope of this paper.

In the comparison of the different star formation tracers to the
infrared surface brightness ratios, we generally found that the
correlations with the uncorrected H$\alpha$ emission were the weakest
and the correlations with the 24~$\mu$m emission were strongest.  The
correlation coefficients for the relations using the
extinction-corrected H$\alpha$ emission usually fell between the other
two, which is expected given that the corrected H$\alpha$ emission is
based on a combination of the uncorrected H$\alpha$ and 24~$\mu$m
emission.  The coefficients for the relations with the corrected
H$\alpha$ emission are sometimes higher than for the corresponding
relations with the 24~$\mu$m emission, but with one exception (the
relations for the 160/250~$\mu$m ratio for NGC~5364, where all
coefficients are $<$0.50), the difference never exceeds 0.05.

For 15 of the 22 galaxies, the weighted correlation coefficients were
$\geq 0.05$ higher for at least one (but usually two) of the relations
between the far-infrared ratios and the 24~$\mu$m surface brightness
than for the corresponding relations between the ratios and the
uncorrected H$\alpha$ intensity.  NGC~3031, 3953, 4548 and 4725 are
all cases where we masked emission in the centres of the H$\alpha$
image that we determined was incompletely-subtracted continuum
emission (based on the diffuse appearance of the emission and the
presence of artefacts similar to what was seen for foreground stars).
We did not remove any 24~$\mu$m emission from these regions, which
probably originates from dust heated in part or completely by the
evolved stellar population, particularly the bulge stars.  If the
emission at 160-350~$\mu$m also originates from dust heated by the
evolved stellar population, then the 160/250 and 250/350~$\mu$m ratios
may naturally correlate with the 24~$\mu$m band very well within these
central regions, and the correlation coefficients for the overall
relations between the far-infrared ratios and 24~$\mu$m emission will
be higher than the relations between the ratios and the H$\alpha$
emission.  We therefore should disregard the ``improved'' relation
found between the ratios and the 24~$\mu$m emission in these four
galaxies.  Of the remaining 11 galaxies where a significantly stronger
correlation is found between either ratio and 24~$\mu$m emission, the
major question is whether changing the star formation tracer would
alter our interpretation of whether the far-infrared ratios were more
strongly affected by star forming regions or the evolved stellar
population as presented in Section~\ref{s_singlesourceanalysis}.
NGC~628, 925, 3184, 3938, 4303, 5364, and 7793 are the only remaining
cases where this could be an issue.  In these six galaxies, we would
be more likely to infer that dust heating by star forming regions is
less significant when using the uncorrected H$\alpha$ emission or more
significant when using the 24~$\mu$m emission.  This may still be
caused in part by 24~$\mu$m emission from dust heated by the diffuse
ISRF.

For most galaxies, though, the selection of the star formation tracer
does not affect the analysis in Section~\ref{s_singlesourceanalysis}.
Given these results, we will proceed with using the
extinction-corrected H$\alpha$ data as the primary star formation
tracer when comparing to the 3.6~$\mu$m emission.  However, we will
still use the 24~$\mu$m data as a star formation tracer in the cases
where H$\alpha$ data is not available, and we will also note any cases
where the choice of star formation tracer would affect the
identification of the heating source.

\section{Tests on the removal of stellar emission from the 24~$\mu$m band}
\label{a_24stellarrem}

To tests the effects of removing the stellar photospheric emission from
the 24~$\mu$m band, we applied the correction
\begin{equation}
\begin{split}
I_\nu(24\mu\mbox{m (corrected)})=I_\nu(24\mu\mbox{m
  (observed)})\\ -0.032I_\nu(3.6\mu\mbox{m}),
\end{split}
\label{e_corr24}
\end{equation}
from \citet{helou04}.  As we are primarily focused on how the
logarithm of the 24~$\mu$m emission relates to the logarithms of the
160/250 and 250/350~$\mu$m surface brightness ratios in our primary
analysis, we will focus on how the logarithm of the 24~$\mu$m changes
when the correction is applied.  We selected data from the 24~arcsec
bins within the optical discs of the galaxies which met the criteia
for the analysis on the 250/350~$\mu$m ratio in
Section~\ref{s_singlesourceanalysis}.

\begin{table}
\caption{Statistics from tests of subtracting stellar continuum from 
    24~$\mu$m data.}
\label{t_24corrtest}
\begin{center}
\begin{tabular}{@{}lccccc@{}}
\hline
Galaxy &   
           \multicolumn{2}{c}{Change in $I_\nu(24\mu\mbox{m})$} &
           \multicolumn{2}{c}{Change in $\log(I_\nu(24\mu\mbox{m}))$} \\
&
           Median &            Max &
           Median &            Max \\
\hline
NGC 628 &
             0.8 \% &          2.2 \% &
             0.1 \% &          0.2 \% \\
NGC 925 &
             1.1 \% &          3.9 \% &
             0.1 \% &          0.3 \% \\
NGC 2403 &
             1.1 \% &          5.0 \% &
             0.1 \% &          0.4 \% \\
NGC 3031 &
             4.0 \% &         22.0 \% &
             0.3 \% &          2.3 \% \\
NGC 3184 &
             1.2 \% &          2.2 \% &
             0.1 \% &          0.2 \% \\
NGC 3621 &
             0.9 \% &          2.3 \% &
             0.1 \% &          0.2 \% \\
NGC 3631 &
             0.7 \% &          1.4 \% &
             0.1 \% &          0.1 \% \\
NGC 3938 &
             0.9 \% &          2.1 \% &
             0.1 \% &          0.2 \% \\
NGC 3953 &
             1.5 \% &          6.3 \% &
             0.1 \% &          0.6 \% \\
NGC 4254 &
             0.5 \% &          1.2 \% &
             0.1 \% &          0.1 \% \\
NGC 4303 &
             0.8 \% &          2.5 \% &
             0.1 \% &          0.2 \% \\
NGC 4321 &
             1.0 \% &          3.0 \% &
             0.1 \% &          0.2 \% \\
NGC 4501 &
             1.7 \% &          4.5 \% &
             0.2 \% &          0.4 \% \\
NGC 4535 &
             1.1 \% &          1.9 \% &
             0.1 \% &          0.2 \% \\
NGC 4548 &
             2.6 \% &          8.1 \% &
             0.2 \% &          0.8 \% \\
NGC 4579 &
             3.2 \% &          7.2 \% &
             0.3 \% &          0.6 \% \\
NGC 4725 &
             2.7 \% &          13.4 \% &
             0.2 \% &          1.2 \% \\
NGC 4736 &
             2.7 \% &          4.9 \% &
             0.3 \% &          0.4 \% \\
NGC 5055 &
             1.3 \% &          2.7 \% &
             0.1 \% &          0.2 \% \\
NGC 5236 &
             0.7 \% &          2.5 \% &
             0.1 \% &          0.2 \% \\
NGC 5364 &
             1.3 \% &          3.8 \% &
             0.1 \% &          0.3 \% \\
NGC 5457 &
             1.0 \% &          3.6 \% &
             0.1 \% &          0.3 \% \\
NGC 6946 &
             0.7 \% &          2.6 \% &
             0.1 \% &          0.2 \% \\
NGC 7793 &
             1.2 \% &          3.7 \% &
             0.1 \% &          0.3 \% \\
\hline
\end{tabular}
\end{center}
\end{table}

Statistics on comparisons of the 24~$\mu$m data with and without the
correction are given in Table~\ref{t_24corrtest}.  The corrections
have up to a 5\% effect on most of the 24~$\mu$m surface brightnesses.
In a few situations, though, the corrections are significantly
stronger.  The most notable cases are the centres of NGC~3031 and
4725, where the 24~$\mu$m surface brighnesses change by up to 25\%.
However, our analysis is more strongly affected by the logarithms of
the 24~$\mu$m surface brightnesses, and these values do not change
significantly.  In the case of NGC~3031, the maximum change in the logarithm
of the surface brightness is 2.5\%, in the case of NGC~4725, it is
1.2\%, and for all other galaxies, it is $<$1\%.  Moreover, the
logarithms of the 24~$\mu$m surface brightnesses with and without the
corrections are very well correlated; the weighted correlation
coefficients for the relations is equivalent to 1.00 to two decimal
places.

Given these results, subtracting the stellar continuum emission from
the 24~$\mu$m band is not critical for our analysis.  In NGC~3031 and
NGC~4725, where the corrections have a significant effect, we already
mention in Section~\ref{s_sftracer} that the 24~$\mu$m emission
includes stellar emission that could affect its relation to the
160/250 and 250/350~$\mu$m ratios.  In other galaxies, the effects of
stellar emission on the 24~$\mu$m is minor or even negligible.  We
also have the issue that the 3.6~$\mu$m data may also include emission
from hot dust associated with star formation \citep{lu03, mentuch09,
  mentuch10}, and while the analysis in Appendix~\ref{a_hiraccomp}
suggests that the effects of this hot dust are minimal, it does
complicate the stellar continuum subtraction.  We will therefore use
the 24~$\mu$m data without applying the correction given by
Equation~\ref{e_corr24}.

\section{Comparisons of 1.6 and 3.6~$\mu$m emission}
\label{a_hiraccomp}

As stated in Section~\ref{s_totstartracer}, mid-infrared data from the
1.6-3.6~$\mu$m range will generally trace starlight from the
Rayleigh-Jeans side of the stellar SED.  However, shorter wavelengths will be
more strongly affected by dust extinction than longer wavelengths,
while longer wavebands might include emission from hot dust.  Therefore, we
may expect to see minor variations in the shape of the 1.6-3.6~$\mu$m SED that
could potentially affect the relation between near-infrared emission and 
the 160/250 or 250/350~$\mu$m ratios.

To investigate how the use of different near-infrared data may affect
our results, we examined the relations between the logarithms of the
1.6 and 3.6~$\mu$m surface brightnesses for all galaxies in our
sample.  We used data from 24~arcsec bins within the optical discs of
the galaxies that met the criteria for the analysis on the
250/350~$\mu$m surface brightness ratios in
Section~\ref{s_singlesourceanalysis}.

\begin{table}
\caption{Data from comparison of 1.6 and 3.6~$\mu$m emission.}
\label{t_hiraccomp}
\begin{center}
\begin{tabular}{@{}lc@{}}
\hline
Galaxy &   Mean Fractional Change in \\
&          $\log (I_\nu (1.6\mu\mbox{m})/\log (I_\nu (3.6\mu\mbox{m})$ \\
&          over Full Range of $\log (I_\nu (3.6\mu\mbox{m}))$\\
\hline
NGC 628 &
           $-0.036 \pm 0.003$ \\
NGC 925 &
           $0.003 \pm 0.019$ \\
NGC 1097 &
           $0.001 \pm 0.001$ \\
NGC 2403 &
           $-0.039 \pm 0.003$ \\
NGC 3031 &
           $-0.033 \pm 0.001$ \\
NGC 3184 &
           $-0.035 \pm 0.004$ \\
NGC 3351 &
           $-0.025 \pm 0.006$ \\
NGC 3621 &
           $-0.014 \pm 0.002$ \\
NGC 3631 &
           $-0.009 \pm 0.003$ \\
NGC 3938 &
           $-0.043 \pm 0.006$ \\
NGC 3953 &
           $-0.036 \pm 0.004$ \\
NGC 4254 &
           $-0.029 \pm 0.002$ \\
NGC 4303 &
           $-0.026 \pm 0.011$ \\
NGC 4321 &
           $-0.015 \pm 0.002$ \\
NGC 4501 &
           $-0.031 \pm 0.002$ \\
NGC 4535 &
           $-0.040 \pm 0.007$ \\
NGC 4548 &
           $-0.020 \pm 0.001$ \\
NGC 4579 &
           $-0.014 \pm 0.002$ \\
NGC 4725 &
           $-0.010 \pm 0.001$ \\
NGC 4736 &
           $-0.033 \pm 0.001$ \\
NGC 5055 &
           $-0.021 \pm 0.001$ \\
NGC 5236 &
           $-0.005 \pm 0.001$ \\
NGC 5364 &
           $-0.044 \pm 0.016$ \\
NGC 5457 &
           $-0.026 \pm 0.003$ \\
NGC 6946 &
           $0.013 \pm 0.002$ \\
NGC 7793 &
           $-0.032 \pm 0.003$ \\
\hline
\end{tabular}
\end{center}
\end{table}

\begin{figure}
\includegraphics{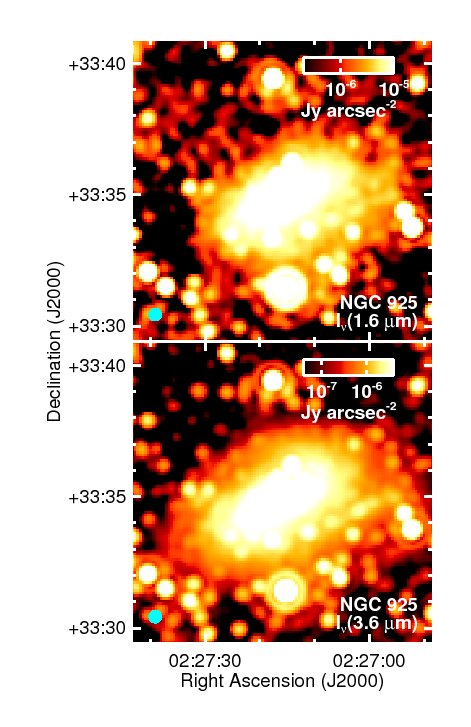}
\caption{Example 1.6 and 3.6~$\mu$m images of NGC~925, with the color
  scale adjusted to show the differences in the radial profiles of the
  observed 1.6 and 3.6~$\mu$m emission as well as low-level background
  artefacts in the 1.6~$\mu$m map.  The PSFs of the data have been
  matched to the PSF of the 350~$\mu$m data, which has a FWHM of
  25~arcsec.  The cyan circles in each panel show the size of the
  PSF.}
\label{f_hiraccompexample}
\end{figure}

We found a very strong correlation between the 1.6 and 3.6~$\mu$m
surface brightnesses.  The weighted correlation coefficients for all
of these relations are $\sim1.00$; the lowest value is 0.987 for
NGC~5457 and 6946.  This indicates that swapping one waveband for the
other in the analyses in Sections~\ref{s_singlesourceanalysis} and
\ref{s_decompositionanalysis} should have a minimal impact on the
results.  Table~\ref{t_hiraccomp} gives data showing the mean
fractional change in the $\log (I_\nu (1.6\mu\mbox{m})/\log (I_\nu
(3.6\mu\mbox{m})$ over the full range of $\log (I_\nu
(3.6\mu\mbox{m}))$, which was found by fitting a line to the two
quantities.  The variations in the
$\log(1.6\mu\mbox{m})/\log(3.6\mu\mbox{m})$ ratio are typically
$<$5\%, which suggests that we will not see any major differences
related to near- infrared colour variations if we use one of these
wavebands instead of the other.  It is very likely that the systematic
variations that we do see in the
$\log(1.6\mu\mbox{m})/\log(3.6\mu\mbox{m})$ ratio are caused by
background artefacts, usually in the 2MASS 1.6~$\mu$m image, where
often spot alternating dark and bright rectangular regions.
Figure~\ref{f_hiraccompexample} shows NGC~925 as an example in which
the low surface brightness structure in the 1.6~$\mu$m image appears
much more asymmetrical than the background in the 3.6~$\mu$m image,
probably because the background was oversubtracted in a data frame
covering the southeast side of the galaxy.

Aside from these background artefacts, we detect only a small number
of high signal-to-noise sources where the
$\log(1.6\mu\mbox{m})/\log(3.6\mu\mbox{m})$ ratios deviate
significantly (more than 0.2) from the mean values found in each
galaxy. These are typically bright star forming regions, such as the
brightest star forming region in NGC 2403, one of the bright regions
in NGC 5457, and a few of the regions in NGC~6946.  Some of the
galactic nuclei with strong star formation activity, such as the
nuclei of NGC~5236 and NGC~6946, have ratios that are suppressed
relative to the rest of the data from the galaxies.  In these
situations, it is possible that we are either seeing
higher-than-average dust attenuation that is reddening the stellar
emission, or we are seeing significant hot dust emission in the
3.6~$\mu$m bands, as suggested by \citet{mentuch09} and
\citet{mentuch10}.

Overall, though, we simply do not see many regions where the
near-infrared colours look significantly redder than average.  This
may be because we are averaging over relatively large areas (areas
that are $4\times$ larger than what \citet{mentuch10} used).  Reddening
caused by extinction or hot dust emission in extreme environments may
be diluted by regions with more typical colours.  Given these results, 
we feel confident that the 3.6~$\mu$m data in our analysis are generally 
unaffected by hot dust emission, although we do acknowledge that we may 
encounter some problems with a few extreme cases.  Moreover, it is more 
advantageous to use the 3.6~$\mu$m images because of the superior signal-
to-noise and better background subtraction in the 3.6~$\mu$m images.  We 
therefore rely on the 3.6~$\mu$m band as a tracer of emission from the 
evolved stellar population in our analysis.

\section{Tests of the removal of emission from star forming regions from 
    the 3.6~$\mu$m band}
\label{a_iracsfrrem}

It would potentially be desirable to treat the 3.6~$\mu$m band as a
tracer of the intermediate and older stellar population by removing
the contributing emission from star forming regions.  To derive such a
correction, we used the Starburst99 model version 6.0.3
\citep{leitherer99}\footnote{Accessed at
  http://www.stsci.edu/science/starburst99/docs/default.htm .} to
model the H$\alpha$ and 3.6~$\mu$m from stars formed either in an
instantaneous burst of star formation or in continuous star formation.
We primarily used a Kroupa initial mass function with $\Phi(M) \propto
M^{2.3}$ for stars with masses $>$0.5~M$_\odot$ and $\Phi(M) \propto
M^{1.3}$ for stars with masses $<$0.5~M$_\odot$ \citep{kroupa01}.  We
performed simulations only for solar metallicities (which should
approximate the metallicities of the spiral galaxies in our sample)
and used all other default settings for the models.  As Starburst99
does not calculate flux densities in the IRAC bands, we derived these
values by following the instructions given in the Spitzer Data
Analysis Cookbook
\citep{ssc12}\footnote{http://irsa.ipac.caltech.edu/data/SPITZER/docs/dataanalysistools\\ /cookbook/}.

\begin{figure}
\includegraphics[width=0.45\textwidth]{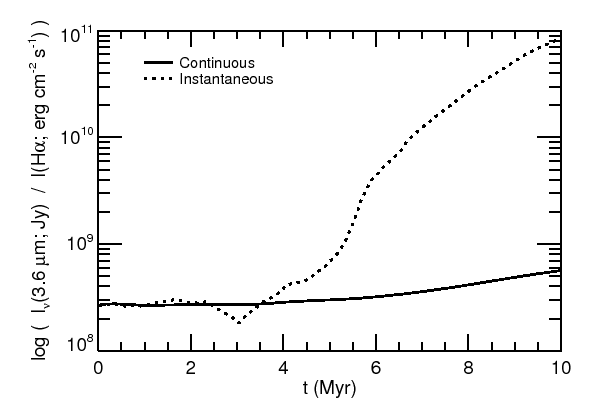}
\caption{Plot of the 3.6~$\mu$m to H$\alpha$ surface brightness ratio
  as a function of time since the onset of star formation as
  determined using Starburst99.  The solid line represents the continuous
  star formation scenario, while the dotted line represents the instantaneous 
  burst scenario.}
\label{f_iracharatio}
\end{figure}

The 3.6~$\mu$m/H$\alpha$ ratios from the Starburst99 simulation are
plotted in Figure~\ref{f_iracharatio}.  In the first 3 Myr after an
instantaneous burst of star formation or after the onset of continuous
star formation, the ratio of the 3.6~$\mu$m flux density to H$\alpha$
flux stays at $(2.7 \pm 0.2)\times 10^{8}$ Jy (erg s$^{-1}$
cm$^{-2}$)$^{-1}$ regardless of which star formation scenario is used.
After 3 Myr, photoionising stars evolve off of the main sequence, and
the 3.6~$\mu$m/H$\alpha$ ratio changes notably, although how it
changes depends on which star formation scenario we use.  In an
instantaneous burst, the H$\alpha$ emission drops off significantly
after 3 Myr, and the 3.6~$\mu$m/H$\alpha$ ratio increases
exponentially to $\sim 8.5 \times 10^{10}$ Jy (erg s$^{-1}$
cm$^{-2}$)$^{-1}$.  However, it is unclear that we would detect
H$\alpha$ emission from stellar populations much older than 3~Myr, and
if we did, then it would seem unlikely that such stellar populations
would produce a significant fraction of the total H$\alpha$ emission.
In the continuous star formation scenario, the ratio rises upwards
more gradually, increasing to $\sim 5.7\times 10^{8}$ Jy (erg s$^{-1}$
cm$^{-2}$)$^{-1}$ at 10 Myr.  While the entire stellar population of a
spiral galaxy may be treated as undergoing continuous star formation,
it may be more appropriate to treat individual regions as
instantaneous bursts of star formation.  We will therefore assume that
the H$\alpha$ emission that we are observing originates mainly from
stars that formed in the past 3~Myr, in which case, we will apply $2.7
\times 10^{8}$ Jy (erg s$^{-1}$ cm$^{-2}$)$^{-1}$ to the H$\alpha$
fluxes to estimate the 3.6~$\mu$m emission from stars that have formed
in the past 3~Myr.

We examined how the corrections affected the 3.6~$\mu$m surface
brightness itself as well as the logarithm of the 3.6~$\mu$m surface
brightnesses in galaxies for which we had H$\alpha$ data.  The
corrections were derived from the extinction-corrected H$\alpha$
intensities calculated using Equation~\ref{e_hacorr}.  We used the
same 24~arcsec binned data that were used for the analysis on the
250/350~$\mu$m surface brightness ratio in
Section~\ref{s_singlesourceanalysis}.  The data are shown in
Table~\ref{t_iracsfrrem}.

\begin{table}
\caption{Statistics from tests of subtracting emission from stars 
    with ages $<$3~Myr from 3.6~$\mu$m emission.}
\label{t_iracsfrrem}
\begin{center}
\begin{tabular}{@{}lcccc@{}}
\hline
Galaxy &   
             \multicolumn{2}{c}{Change in $I_\nu(3.6\mu\mbox{m})$} &
             \multicolumn{2}{c}{Change in $\log(I_\nu(3.6\mu\mbox{m}))$} \\
&
             Median &          Max &
             Median &          Max \\
\hline
NGC 628 &
             0.4 \% &          1.3 \% &
             0.0 \% &          0.1 \% \\
NGC 925 &
             0.9 \% &          4.4 \% &
             0.1 \% &          0.3 \% \\
NGC 2403 &
             0.7 \% &          6.8 \% &
             0.1 \% &          0.6 \% \\
NGC 3031 &
             0.2 \% &          1.6 \% &
             0.0 \% &          0.1 \% \\
NGC 3184 &
             0.5 \% &          1.1 \% &
             0.0 \% &          0.1 \% \\
NGC 3621 &
             0.7 \% &          4.1 \% &
             0.1 \% &          0.3 \% \\
NGC 3631 &
             0.9 \% &          1.5 \% &
             0.1 \% &          0.1 \% \\
NGC 3938 &
             0.5 \% &          2.8 \% &
             0.0 \% &          0.2 \% \\
NGC 3953 &
             0.2 \% &          0.7 \% &
             0.0 \% &          0.1 \% \\
NGC 4254 &
             0.9 \% &          1.7 \% &
             0.1 \% &          0.1 \% \\
NGC 4303 &
             0.6 \% &          1.6 \% &
             0.1 \% &          0.2 \% \\
NGC 4321 &
             0.5 \% &          1.1 \% &
             0.0 \% &          0.1 \% \\
NGC 4501 &
             0.3 \% &          0.5 \% &
             0.0 \% &          0.0 \% \\
NGC 4535 &
             0.6 \% &          2.2 \% &
             0.1 \% &          0.2 \% \\
NGC 4548 &
             0.1 \% &          0.2 \% &
             0.0 \% &          0.0 \% \\
NGC 4579 &
             0.2 \% &          0.4 \% &
             0.0 \% &          0.0 \% \\
NGC 4725 &
             0.2 \% &          0.5 \% &
             0.0 \% &          0.0 \% \\
NGC 5055 &
             0.3 \% &          0.7 \% &
             0.0 \% &          0.1 \% \\
NGC 5236 &
             0.5 \% &          2.0 \% &
             0.0 \% &          0.2 \% \\
NGC 5364 &
             0.4 \% &          1.5 \% &
             0.0 \% &          0.1 \% \\
NGC 5457 &
             0.4 \% &          17.0 \% &
             0.0 \% &          1.4 \% \\
NGC 7793 &
             0.8 \% &          4.0 \% &
             0.1 \% &          0.3 \% \\
\hline
\end{tabular}
\end{center}
\end{table}

As can be seen by these tests, the correction generally has a very
minor impact on the 3.6~$\mu$m emission.  In most cases, the
correction is $<$2\%, indicating that $>$98\% of the total 3.6~$\mu$m
emission from all regions within the optical discs comes from
non-ionising intermediate-aged and evolved stars.  We did see some
stronger changes in a few individual star forming regions.  In
NGC~5471, which is a location containing a bright H{\small II} region
in the outer disc of NGC~5457, we estimate that 17\% of the
3.6~$\mu$m emission may originate from stars with ages $<$3~Myr.  In a
few other star forming regions at the periphery of the disc of
NGC~5457 and a couple of star forming regions in NGC 925, 2403, 3621,
3938, 4535, and 7793, the correction changed the 3.6~$\mu$m emission
by 2-7\%.  However, these few ``extreme'' cases have a $<$1\% effect
on the logarithm of the 3.6~$\mu$m emission, which we use in most of
our analysis.  Moreover, the weighted correlation coefficients between
the logarithms of the uncorrected and corrected 3.6~$\mu$m surface
brightnesses is 1.00 in all cases.

The relatively small effect of this adjustment to the 3.6~$\mu$m data
ultimately has no significant effect on our analysis.  If we attempted
to account for stars older than 3 Myr, then the changes in the
3.6~$\mu$m emission would be higher, but this would depend upon being
able to determine what the star formation history is for these
regions.  For reference, we estimate using the physical dimensions of
the 24~arcsec resolution elements in Table~\ref{t_sample} and assuming
a rotational velocity of 200 km s$^{-1}$ that stars may pass through
the resolution elements on periods ranging from 2 to 12 Myr.  In some
cases, the correction for the 3.6~$\mu$m band that we are already
using (which assumes that the stellar populations are $<$3~Myr in age)
is already applicable.  For older regions, we would need to guess the
star formation history of the regions.  We also have the complicating
issue that the star formation tracers that we are using are imperfect
for estimating this correction.  The H$\alpha$ emission without the
extinction correction would give a correction to the 3.6~$\mu$m band
that is too low, while the H$\alpha$ emission combined with the
24~$\mu$m band may potentially be affected by 24~$\mu$m emission from
diffuse dust heated by the ISRF (see Appendix~\ref{a_ha24comp}).
These issues along with the relatively minor benefit from removing the
emission from young stars from the 3.6~$\mu$m band leads us to the
conclusion that we should simply use the 3.6~$\mu$m surface
brightnesses as observed for our analysis.

\section{Offsets between infrared surface brightness ratios and spiral arms}
\label{a_offsetlinecut}

\begin{figure}
\includegraphics{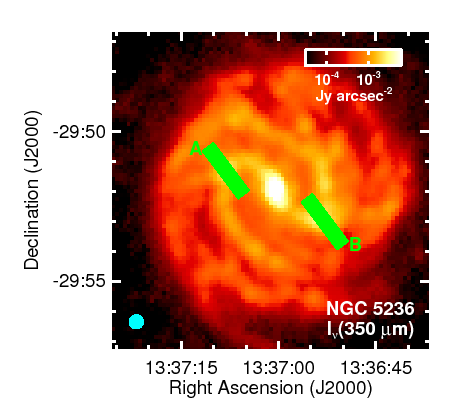}
\includegraphics{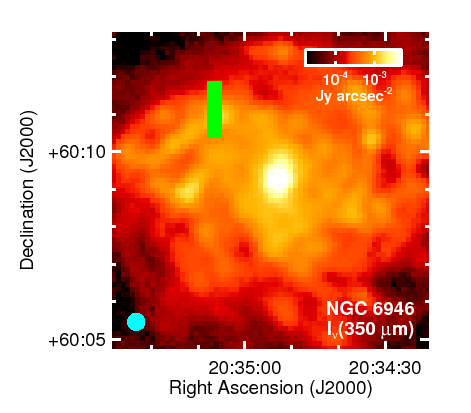}
\caption{The 350~$\mu$m images for NGC~5236 and 6946 with green lines
  indicating locations where surface brightness profiles were measured
  across the spiral arms.  The cyan circles show the FWHM of the PSF.
  The profiles themselves are shown in Figure~\ref{f_linecut}.}
\label{f_linecutmap}
\end{figure}

\begin{figure}
\includegraphics[width=0.45\textwidth]{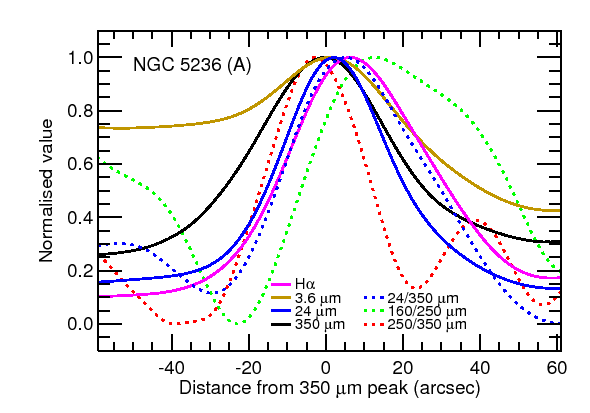}
\includegraphics[width=0.45\textwidth]{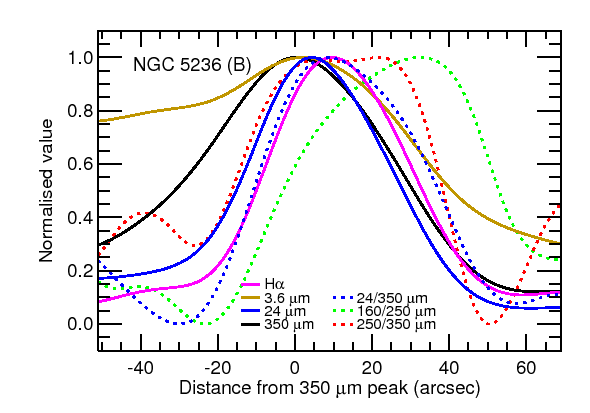}
\includegraphics[width=0.45\textwidth]{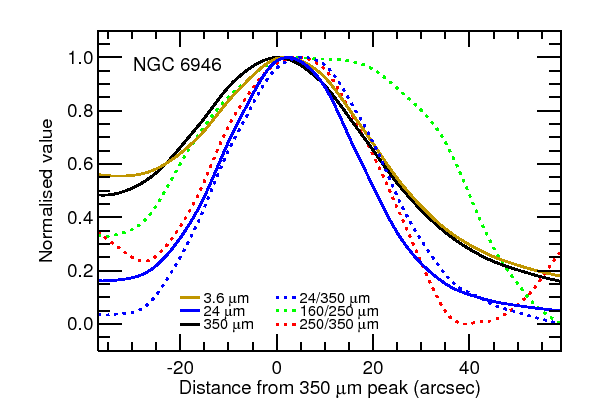}
\caption{Profiles of the surface brightnesses (shown as solid lines
  and normalised so that the peaks equal 1) and surface brightness
  ratios (shown as dotted lines and normalised so that the profiles
  range from 0 to 1) across spiral arms in NGC~5236 and 6946.  The
  locations of these profiles are shown in Figure~\ref{f_linecutmap}.
  Measurements were made at 1~arcsec intervals in 24~arcsec wide
  regions in data where the PSF FWHM is 25~arcsec. The x-axes show the
  distance from the peak in the 350~$\mu$m profile, with positive
  x-values showing distance from the 350~$\mu$m peak on the downstream
  side of the arms.  The uncertainties for the normalised surface
  brightness profiles are $\ltsim$0.01, and the uncertainties for the
  normalised surface brightness ratios are $\sim0.10$.}
\label{f_linecut}
\end{figure}

\addtocounter{section}{1}
\begin{table*}
\begin{minipage}{128mm}
\caption{Weighted correlation coefficients for relations between
  160/250 and 250/350~$\mu$m ratios and tracers of heating sources in data
  with different bin sizes.}
\label{t_reseffect_singlesource}
\begin{tabular}{@{}lccccc@{}}
\hline
Galaxy &
    Bin Size & 
    \multicolumn{4}{c}{Weighted Correlation Coefficients} \\
&
    (arcsec) &
    \multicolumn{2}{c}{$\log(I_\nu(160\mu\mbox{m})/I_\nu(250\mu\mbox{m}))$} &
    \multicolumn{2}{c}{$\log(I_\nu(250\mu\mbox{m})/I_\nu(350\mu\mbox{m}))$} \\
&
    &
    vs. $\log(I(\mbox{H}\alpha))^a$ &
    vs. $\log(I_\nu(3.6\mu\mbox{m}))$ &
    vs. $\log(I(\mbox{H}\alpha))^a$ &
    vs. $\log(I_\nu(3.6\mu\mbox{m}))$ \\
\hline
NGC 3031 &
            24 &
            0.17 &     0.84 &     0.16 &     0.87 \\
&
            120 &
            -0.07 &     0.91 &    0.17 &     0.90 \\
NGC 5457 &
            24 &
            0.89 &     0.57 &     0.40 &     0.89 \\
&
            120 &
            0.92 &     0.57 &     0.53 &     0.94 \\
\hline
\end{tabular}
$^a$ The H$\alpha$ data used here were corrected for extinction using 24~$\mu$m
    data.\\
\end{minipage}
\end{table*}
\addtocounter{section}{-1}

As stated in Section~\ref{s_qualanalysis} we see small offsets between
the star forming regions and the colour temperatures in some of the
images in Figure~\ref{f_maptemp}.  This phenomenon had been seen
previously.  \citet{bendo12a} and \citet{foyle12} illustrated that
regions with enhanced colour temperatures in the spiral arms of M83
appeared offset from the star forming regions, and
\citet{mentuchcooper12} had indicated that such an offset also
appeared to be present in M51 but did not illustrate this offset.  To
examine this further, we measured the profiles of surface brightnesses
across spiral arm segments in NGC~5236 and 6946, the two closest
galaxies where we see this phenomenon.  Figure~\ref{f_linecutmap}
shows the locations where we measured the profiles in these galaxies,
and Figure~\ref{f_linecut} shows the profiles themselves.

These images illustrate similar structures in all three spiral arm
segments.  The 350~$\mu$m data serve as a proxy for the dust mass, as
the band lies in a regime of the SED where it is relatively
insensitive to the temperature variations seen in these data.  We will
refer to the peak of the 350~$\mu$m emission as the dust lane within
these galaxies.  The H$\alpha$ emission, which traces gas photoionised
by star forming regions, and the 24~$\mu$m emission, which traces hot
dust heated by the star forming regions, peak to one side of the
350~$\mu$m emission, although the difference between the peaks is
small compared to the 25~arcsec resolution of the data.  None the
less, this would be consistent with the classical depiction of spiral
density waves in which gas falls into the wave, gets compressed, and
forms stars on the other side of the wave \citep[e.g.][]{robert69,
  elmegreen79}.  In the line segments we have selected, material is
flowing into the concave sides of the spiral density waves, and new
stars are found on the convex side.  The asymmetric distribution of
dust and young stars around the star forming regions has a few
potential effects on dust heating in these locations.

First, we can see in the NGC~5236 data that the H$\alpha$ emission is
offset relative to the 24~$\mu$m emission, although this offset is
small compared to the 25~arcsec resolution of the data.  This will
occur in part because the star forming regions closer to the dust lane
are more obscured, leading to a change in the H$\alpha$/24~$\mu$m
ratio.  The normalised H$\alpha$ emission is also higher than the
normalised 24~$\mu$m emission at locations downstream from the spiral
arms.  Some fraction of this H$\alpha$ emission may originate from gas
in the diffuse ISM photoionised by photons escaping from star forming
regions.  This enhancement in H$\alpha$ emission would be seen on only
one side of the star forming regions because the optical depth is much
higher on the other side of the regions, which means that
photoionising photons can travel much further away from the dust lane
than towards it.

From the perspective of our analysis, the most interesting feature
here is the presence of the strongly enhanced 160/250~$\mu$m ratio on
the opposite side of the star forming regions from the dust lane,
which can be offset by a distance comparable to the 25~arcsec
resolution of the data.  The ratio is clearly not tracing the
temperature of dust heated locally by the star forming regions;
otherwise, the 160/250~$\mu$m ratio would peak at the same location as
the 24/350~$\mu$m ratio.  A similar offset in enhanced polycyclic
aromatic hydrocarbon (PAH) emission in the IRAC 8~$\mu$m band was
found in M83 by Jones et al. (2014, submitted).  They even found that
the 8/250~$\mu$m surface brightness ratio (a measure of PAH excitation
normalised by dust mass) was correlated with the 160/250~$\mu$m ratio,
suggesting that both the dust emission at 160~$\mu$m and the PAH
excitation were linked.

The peak in the 160/250~$\mu$m ratio may be offset for one of two
reasons.  First, the 160~$\mu$m emission from these galaxies may
primarily originate from dust heated by photoionising light escaping
the star forming regions into the diffuse ISM.  Because the light
primarily propagates away from the dust lanes, the peak in the
160/250~$\mu$m ratio appears offset.  The physical distance between
the peak in the 160/250~$\mu$m ratios and the peak in the 350~$\mu$m
emission is $\sim0.5$~kpc.  This distance is in the middle of the
0.1-1~kpc range of the mean free path of V-band photons in galaxies
given by the analysis from \citet{xilouris99} and \citet{bianchi07},
which implies that it is from dust heated by diffuse light from star
forming regions. The other possibility is that emission in the
160~$\mu$m band is from dust in an optically-thin environment heated
locally by B and A stars that have emerged from the dusty star forming
regions 10-100~Myr ago.  These stars would not produce the photons
needed to photoionise hydrogen but would produce substantially high
radiation fields that include large numbers of soft ultraviolet
photons that could warm up the large dust grains.

The 250/350~$\mu$m ratio presents relatively different results in the
three profiles.  In profile A in NGC 5236, the ratio peaks on the side
of the dust lane opposite the side of the star forming region, and in
NGC~6946, the peak corresponds to the location of the peak in the
3.6~$\mu$m emission.  This would be consistent with the 250/350~$\mu$m
ratio tracing the temperature of the dust in the spiral arm that is
shielded from the star forming regions and heated by the radiation
field from the evolved stellar population.  In profile B in NGC~5236,
the 250/350~$\mu$m ratio peaks in two locations: one peak corresponds
to the location of the 24~$\mu$m peak, while the other appears
$\sim15$~arcsec further downstream in a location where the
160/250~$\mu$m ratio is also enhanced.  This could indicate that the
ratio in this location is still sensitive to heating by either light
diffusing from the star forming regions or unobscured B and A stars.
However, the structures in the 250/350~$\mu$m ratio map for NGC~5236
are relatively poorly defined overall and might be affected by
residual striping in the 350~$\mu$m data, so it is difficult to
interpret the significance of the 250/350~$\mu$m profiles for the
galaxy.

As we stated earlier, NGC~5236 and 6946 are the two galaxies in our
sample where we see these structure most prominently.  We do see some
hints of similar structures in the northern part of NGC~628, but the
effect is not particularly strong.  Aside from these galaxies, though,
we do not see any cases where we can see significant offsets between
the spiral arm structures traced by the 24~$\mu$m emission and the
structures traced by the 160/250 and 250/350~$\mu$m ratios.  The
structures appear on scales of $\ltsim$0.5~kpc, so for galaxies at
$\gtrsim$5~Mpc, offsets between the surface brightness ratios and the
star forming regions may not be readily apparent within data at
25~arcsec resolutions.  The remaining galaxies within 5~Mpc may not
exhibit these structures for a large number of reasons.  In NGC~3031,
the 160/250 and 250/350~$\mu$m ratios appear more strongly influenced
by the evolved stellar population, so we would not expect to see this
phenomenon at all.  Meanwhile, NGC~2403 and 7793 are both flocculent
galaxies where, because these two galaxies lack the large shock fronts
created by the spiral density waves, star formation is expected to
occur in clouds that collapse because of local graviational
instabilities.  Dust may be distributed symmetrically around star
forming regions in these galaxies, and simulations by \citet{dobbs10}
suggest that no stellar age gradients will be seen across the spiral
segments.  In the scenario where the 160/250 or 250/350~$\mu$m ratios
are enhanced by light diffusing from star forming regions, the light
would diffuse symmetrically around star forming regions.  In the
scenario where the ratios are enhanced by B and A stars, it may be
possible that the B and A stars are not located preferentially on one
side of current star formation sites.  In either case, the 160/250 and
250/350~$\mu$m ratios will peak in the same locations as the emission
from the star forming regions themselves (if the star forming regions
influence the ratios).

We recommend treating these results cautiously.  Although the offsets
in the peaks of the surface brightness ratio profiles are quite
apparent, the offsets among individual wave bands is relatively small
compared to the 25~arcsec FWHM of the PSF.  Unidentified astrometry
problems could produce some of the patterns seen in these profiles,
although we have checked the astrometry of these data using foreground
stars and background galaxies outside the optical discs of the
galaxies, and the structures in the profiles continue to appear even
when small adjustments are made to the astrometry of the data (or
adjustments are not applied).  It is unlikely that the PSF-matching
steps could have introduced some structures.  This step should match
not only the FWHMs of the PSFs but also the profiles of the PSFs, and
the kernels from \citet{aniano11} that we used are based on
observational data and should contain features related to the
instrument and data processing.  Even though these kernels do not
always work perfectly, the artefacts from this process that appear
around bright sources in the colour temperature maps usually look
symmetric, whereas the structures that we see in associated with the
spiral arms in the 160/250~$\mu$m maps are asymmetric.  It is more
likely that the offset 160/250~$\mu$m features are intrinsic
structures within the galaxies rather than artefacts related to
astrometry or PSF-matching issues.

\section{Tests of resolution effects on the data}
\label{a_reseffect}

Because the galaxies are located at distances that vary by a factor of
7 and also because the same analytical methods have been applied to
the same galaxies using data at different resolutions, we tested the
effects of performing our analyses using data measured in different
sized bins.  For this analysis, we selected NGC~3031 and 5457, the two
galaxies with the largest angular sizes in the sample.  We rebinned
the data for these galaxies into 120~arcsec bins.  This is equivalent
to shifting NGC~3031 from 3.6 to 18.0~Mpc and to shifting NGC~5457
from 6.7 to 33.5~Mpc, which are equivalent to or larger than the
distances for the furthest galaxies in our sample.  We then repeated
the analysis in Sections~\ref{s_singlesourceanalysis} and
\ref{s_decompositionanalysis}.

Table~\ref{t_reseffect_singlesource} shows the changes in the
correlation coefficients in the relations between the infrared surface
brightness ratios and the tracers of different dust heating sources.
We generally see the correlation coefficients increase up to 0.15 when
we use the larger bin size except in the relation of the
160/250~$\mu$m ratio to H$\alpha$ emission, where the relation
exhibited uncorrelated scatter in both the 24 and 120~arcsec binned
data.  We also still generally obtain the same results for these two
galaxies that we did in Section~\ref{s_singlesourceanalysis}.
Regardless of the bin size used, the 160/250 and 250/350~$\mu$m ratios
for NGC~3031 and the 250/350~$\mu$m ratios for NGC~5457 appear much
better correlated with the 3.6~$\mu$m surface brightness, while the
160/250~$\mu$m ratios for NGC~5457 still appear better correlated with
the H$\alpha$ emission.

Table~\ref{t_reseffect_singlesource} shows the resulting $\eta$ after
applying the decomposition analysis in
Section~\ref{s_decompositionanalysis} to data with different bin
sizes.  Except for the $\eta$(250/350~$\mu$m) for NGC~5457, where the
$I (\mbox{SFR})$ term was negligible, the resulting $\eta$ all
decrease.  The decrease in $\eta$(160/250~$\mu$m) for NGC~3031 is
relatively small compared to the uncertainties in the $\eta$ values
for the 120~arcsec bins.  The $\eta$(250/350~$\mu$m) is 0.22, although
this is comparable to the uncertainty in the methodology following the
analysis in Appendix~\ref{a_efractest} (0.15) as well as the
uncertainty of 0.17 in $\eta$(250/350~$\mu$m) for the 120~arcsec
binned data that was determined using the Monte Carlo analysis.  The
$\eta$(160/250~$\mu$m) for NGC~5457 decreases by 0.16,
which is also approximately equivalent to the uncertainty of 0.15 in
the methodology.  The change may have resulted from combining brighter
emission from warmer dust heated locally by star forming regions with
fainter emission from cooler dust heated by older stars.  The dust
emission observed in these coarser bins would appear to be dominated
more by dust heated by star forming regions, an issue also highlighted
by \citet{galliano11}.

In summary, these tests represent the most severe tests of resolution
effects that could affect the data.  The analysis of the correlation
between surface brightness ratios and dust heating sources in
Section~\ref{s_singlesourceanalysis} should be robust against
resolution effects.  The decomposition analysis in
Section~\ref{s_decompositionanalysis} is somewhat affected by
resolution effects, although the changes in the derived $\eta$ values
are generally consistent when taking into account both the 0.15
uncertainty in the methodology and the uncertainties from the Monte
Carlo analysis.  These issues are noted in
Section~\ref{s_decompositionanalysis}, but we will treat the $\eta$ as
though resolution has a minor affect on the data.

\addtocounter{table}{1}
\begin{table}
\caption{Results from fitting Equation~\ref{e_multfit} to the 160/250 and 
  250/350~$\mu$m surface brightness ratios.}
\label{t_reseffect_multfit}
\begin{center}
\begin{tabular}{@{}lccc@{}}
\hline
Galaxy &
    Bin Size &
    \multicolumn{2}{c}{$\eta$} \\
&
    (arcsec) &
    160/250~$\mu$m &
    250/350~$\mu$m \\
\hline
NGC 3031 &
    24 &
    $0.79 \pm 0.02$ &
    $0.95 \pm 0.03$ \\
&
    120 &
    $0.72 \pm 0.11$ &
    $0.73 \pm 0.17$ \\
NGC 5457 &
    24 &
    $0.37 \pm 0.01$ &
    $1.00^a$ \\
&
    120 &
    $0.21 \pm 0.05$ &
    $1.00^a$ \\
\hline
\end{tabular}
\end{center}
$^a$ For these fits, $I (\mbox{SFR})$ in Equation~\ref{e_multfit} was
    found to be negligible compared to $A_1 I_\nu (3.6 \mu \mbox{m})$.
    We therefore fit $\ln(I_\nu(250 \mu \mbox{m})/I_\nu(350 \mu
    \mbox{m}))$ to $\ln(I_\nu (3.6 \mu \mbox{m}))$ using a linear
    function.  The $\alpha$ listed here is the slope, and $A_2$ is the
    constant.\\
\end{table}

\section{Tests of measuring $\eta$ using simulated data}
\label{a_efractest}

\begin{figure*}
\includegraphics{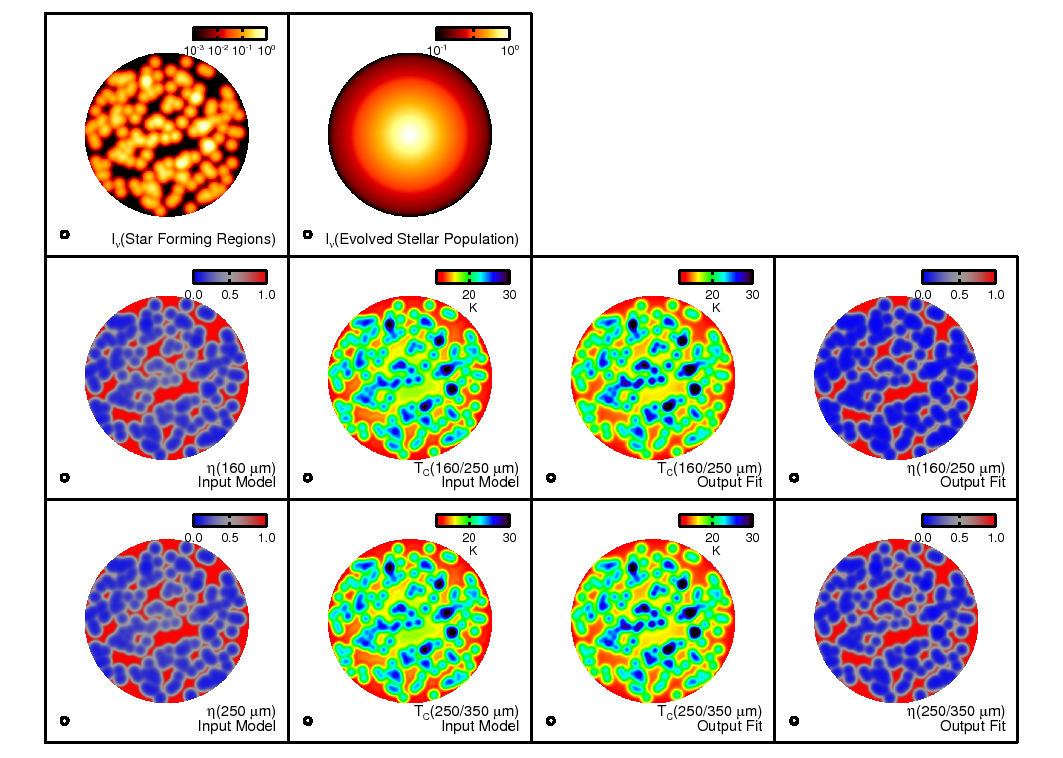}
\caption{Input and output data for tests in Appendix~\ref{a_efractest}
  showing the effectiveness of using Equations~\ref{e_multfit} and
  \ref{e_efrac} and the methodology in
  Section~\ref{s_decompositionanalysis} to measure the fraction of
  dust heated by different populatons of stars by comparing the
  emission from the dust heating sources to the observed infrared
  surface brightness ratios.  All data are for a simulated face-on
  galaxy disc where the emission is truncated at a diameter of
  10~arcmmin and where the resolution of the data is 25~arcsec.  The
  top row shows the model surface brightnesses for star forming and
  evolved stellar populatons (normalised for one).  The bottom two rows
  show, for scenario 5 in Table~\ref{t_efractestresults} (where the
  input $\eta$(250~$\mu$m) is set to 0.50), the input
  $\eta$ and 160 and 250~$\mu$m, the resulting input 160/250
  and 250/350~$\mu$m colour temperature maps, the best fitting output
  colour temperature maps, and the output $\eta$ maps for
  the 160/250 and 250/350~$\mu$m ratios.  The colour temperature maps
  are computed using $\beta$=2.  The output
  $\eta$(160/250~$\mu$m) should be compared to the input
  $\eta$(160~$\mu$m), and the output
  $\eta$(250/350~$\mu$m) should be compared to the input
  $\eta$(250~$\mu$m).}
\label{f_efractestresults}
\end{figure*}

To test the effectiveness of using Equations~\ref{e_multfit} and
\ref{e_efrac} to derive the relative fraction of variation in emission
in a waveband that is related to heating by a specific population of
stars, we created simplified models of dust heating within disc
galaxies.  The dust grains are treated as modified blackbodies where
$E \propto T^6$, which is appropriate for $\beta=2$.  We assume that
the dust is at thermodynamic equilibrium with the local radiation
field and that the temperature depends upon heating from two stellar
populations; this effectively excludes stochastically heated dust as
well as more complex radiative transfer effects.  The distribution of
dust mass is treated as arbitrary.

To represent the dust heated by the two different stellar populations,
we created two circular discs with diameters of 10~arcmin and pixel
sizes of 1~arcsec.  For the energy $E_{ES}$ from evolved stellar
population (including intermediate and older stars), we used a smooth
exponential disc with a scale length of 150~arcsec.  For the energy
$E_{SF}$ from the star forming regions, we created a disc containing
150 randomly-placed unresolved sources scaled by an exponential
function with a scale length of 240~arcsec and five additional
randomly-placed sources that are set to 2 times the peak of the
exponential profile.  This is a simplified representation of the star
forming structures that could be expected in a flocculent spiral
galaxy, and the broader scale length for the star forming regions
compared to the evolved stellar populations reflects how star forming
regions are distributed more broadly in spiral galaxies.  We assume
that the distribution of both stellar populations and the dust
perpendicular to the disc of the galaxy and the scale height of the
dust are similar and therefore do no additional computations for the
vertical propigation of light.

The $E$ functions were then rescaled to compute SEDs based on modified
blackbodies with $E \propto T^6$.  The energy-weighted mean $T$ was
set to 15~K for the evolved stellar population component and 25~K for
the star forming component, which is similar to the temperatures
measured for the separate thermal components in in
Section~\ref{s_discussion_sed}.  The data were then used to create maps
with 1~arcsec pixels of the intensities from these components at 160,
250, and 350~$\mu$m.  The maps were convolved with a 25~arcsec
Gaussian function, which is similar to the PSF of the SPIRE 350~$\mu$m
data.  The maps were then rebinned into 24~arcsec bins for the
quantitative analysis.

We display data created for multiple scenarios where the $E$ values
were rescaled to create input $\eta$(250~$\mu$) ratios
ranging from 0.1 to 0.9.  We then fit the 160/250 and 250/350~$\mu$m
ratios using $E_{ES}$ and $E_{SF}$ in Equation~\ref{e_multfit} and
computed output $\eta$ values for these ratios using
Equation~\ref{e_efrac}, which we then compare to the input
$\eta$ measured at the shorter wavelength.

\begin{table}
\caption{Comparison of input and output $\eta$ results.}
\label{t_efractestresults}
\begin{center}
\begin{tabular}{@{}lcccc@{}}
\hline
Scenario &
    \multicolumn{2}{c}{Input $\eta$} &
    \multicolumn{2}{c}{Output $\eta$} \\
Number &
    160/250~$\mu$m &          250/350~$\mu$m &
    160/250~$\mu$m &          250/350~$\mu$m \\
\hline
1 &
    0.11 &                    0.10 &
    0.13 &                    0.12 \\
2 &
    0.19 &                    0.20 &
    0.18 &                    0.19 \\
3 &
    0.26 &                    0.30 &
    0.22 &                    0.25 \\
4 &
    0.33 &                    0.40 &
    0.26 &                    0.32 \\
5 &
    0.41 &                    0.50 &
    0.31 &                    0.40 \\
6 &
    0.49 &                    0.60 &
    0.37 &                    0.49 \\
7 &
    0.58 &                    0.70 &
    0.46 &                    0.61 \\
8 &
    0.69 &                    0.80 &
    0.58 &                    0.73 \\
9 &
    0.82 &                    0.90 &
    0.75 &                    0.86 \\
\hline
\end{tabular}
\end{center}
\end{table}

Mean input and output $\eta$ values from these tests are
given in Table~\ref{t_efractestresults}.
Figure~\ref{f_efractestresults} shows the model surface brightnesses
for the star forming and evolved stellar populatons, the input
$\eta$ and 160 and 250~$\mu$m, the resulting 160/250 and
250/350~$\mu$m colour temperature maps, and the output
$\eta$ maps for the 160/250 and 250/350~$\mu$m ratios for
scenario 5 (where the input $\eta$(250~$\mu$m) is set to
0.50).  

The input and output maps agree within $\sim1$~K in most locations,
but the brighter star forming regions tend to be $\sim5$~K hotter in
the output images than in the input images.  This illustrates the
limitations of using Equation~\ref{e_multfit} to relate an observed
colour temperature to emission from multiple dust heating sources, at
least over large temperature ranges.  The measured $\eta$ in
individual locations may be within 0.20 of the input value, but we
typically measure a global mean $\eta$ that is within 0.10
of the input value.  We get input and output $\eta$ values
within 0.01 when $\eta\ltsim$0.20.  On the other hand, the
difference between input and output $\eta$ values is
0.10-0.15 in situations where $\eta$ is between 0.50 and
0.70.  The output $\eta$ value is always lower than the
input $\eta$ value, which demonstrates that the methodology
exhibits a systematic bias that could potentially be corrected to
improve the accuracy of the measured $\eta$.

Given these results, we will use 0.15 as the uncertainties in the
$\eta$ values from the analysis in
Section~\ref{s_decompositionanalysis} when rescaling global flux
densities to create SEDs for separate thermal components.  However, we
will point out that the uncertainties could be higher in multiple
circumstances.  If the difference between the dust temperatures is
larger, then the assumptions behind Equation~\ref{e_multfit} no longer
apply and the uncertainty in $\eta$ could increase.  As
mentioned elsewhere in the text, the uncertainty in $\eta$
will be high if the star forming regions and evolved stellar populations
have similar distributions or if the mean free path of light heating
the dust is larger than the resolution elements of our data.  

Although we only show results for 160/250 and 250/350~$\mu$m ratios,
we did find that the results may still be reliable when creating
surface brightness ratios based on any two wavebands where the longer
waveband samples emission on the Rayleigh-Jeans side of the SED.
However, the $\eta$ values could become more uncertain if
both wavebands sample emission from the peak or Wein side of the dust
SED.  This should be taken into account when attempting to apply this
methodology to other datasets.

\section{Colour temperature images}
\label{a_fig}

This section includes the full versions of Figures~\ref{f_maptemp} in
Section~\ref{s_singlesourceanalysis} and \ref{f_multfit} in
Section~\ref{s_decompositionanalysis}.  These figures will appear only
online in the final published version of the paper.

\begin{figure*}
\includegraphics[width=0.95\textwidth]{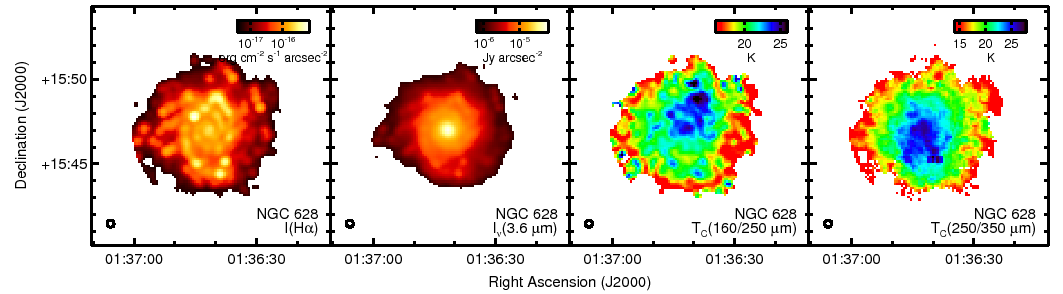}
\includegraphics[width=0.95\textwidth]{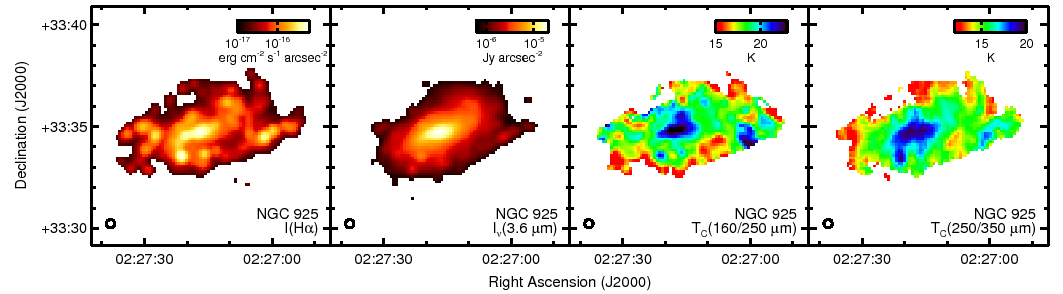}
\includegraphics[width=0.95\textwidth]{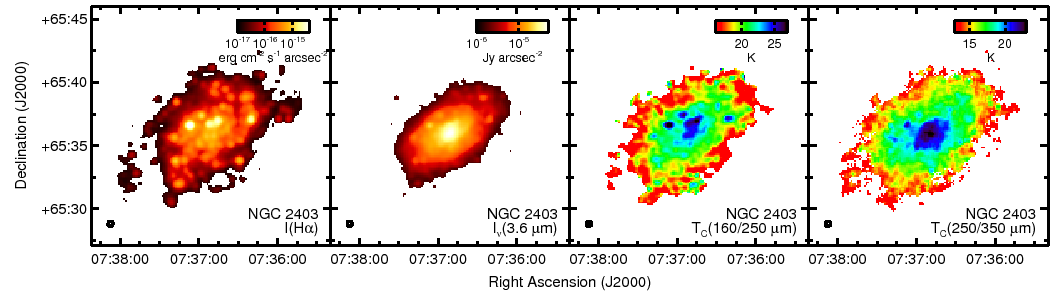}
\includegraphics[width=0.95\textwidth]{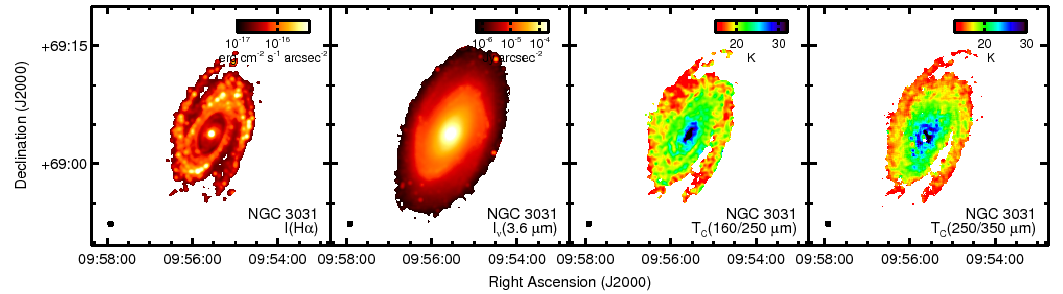}
\caption{H$\alpha$ (or 24~$\mu$m), 3.6~$\mu$m, 160/250~$\mu$m, and
  250/350~$\mu$m images of all galaxies in the sample.  All images are
  based on data where the PSF has been matched to the PSF of the
  350~$\mu$m image, which has a FWHM of 25~arcsec.  The 160/250 and
  250/350~$\mu$m ratios are shown as colour temperatures (based on
  modified blackbodies with $\lambda^{-2}$ emissivities) so that the
  colours can be related to the SED shape, although the actual
  temperatures may vary.  Only data within approximately the optical
  disc of each galaxy detected at the $5\sigma$ level in the
  H$\alpha$, 24~$\mu$m, or 3.6~$\mu$m maps or above the $3\sigma$
  level in the colour temperature maps is shown; other parts of the
  image are left blank.}
\label{f_maptempall}
\end{figure*}

\addtocounter{figure}{-1}
\begin{figure*}
\includegraphics[width=0.95\textwidth]{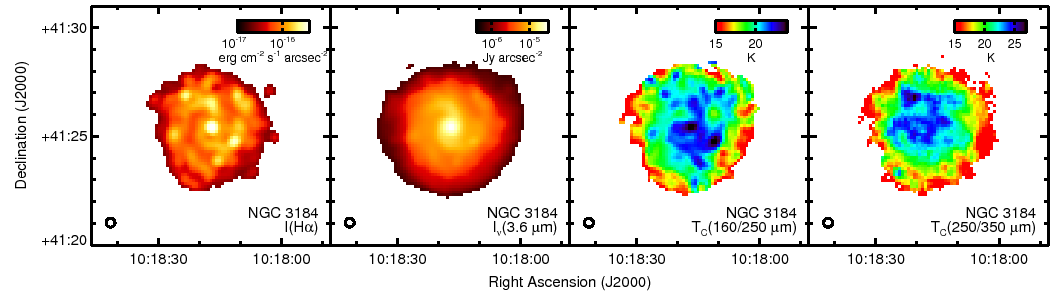}
\includegraphics[width=0.95\textwidth]{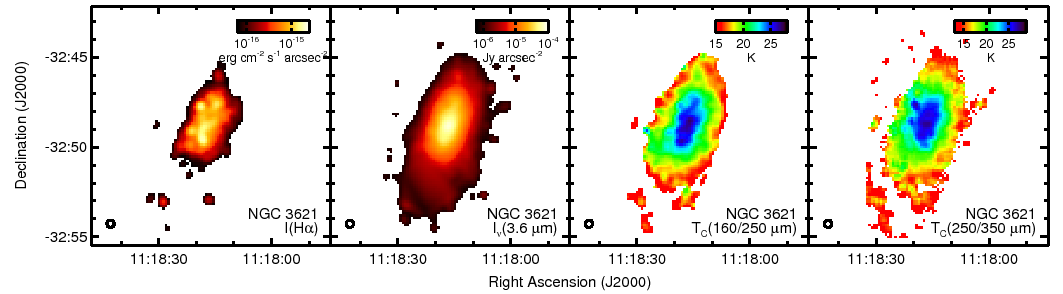}
\includegraphics[width=0.95\textwidth]{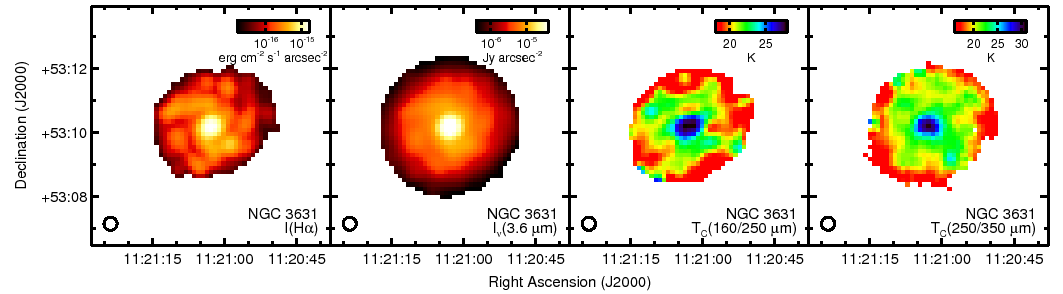}
\includegraphics[width=0.95\textwidth]{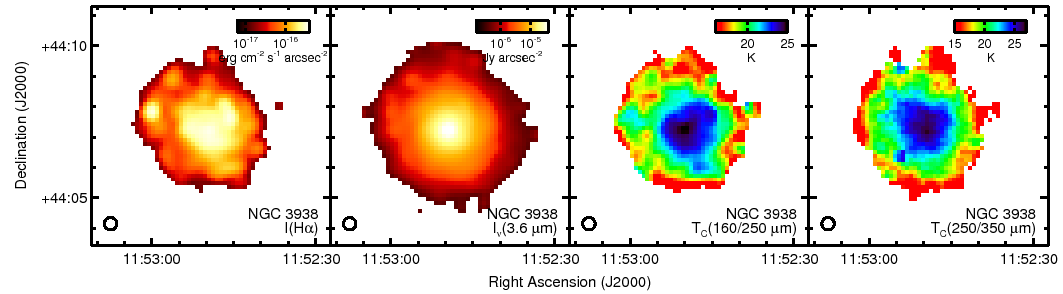}
\caption{Continued.}
\end{figure*}

\addtocounter{figure}{-1}
\begin{figure*}
\includegraphics[width=0.95\textwidth]{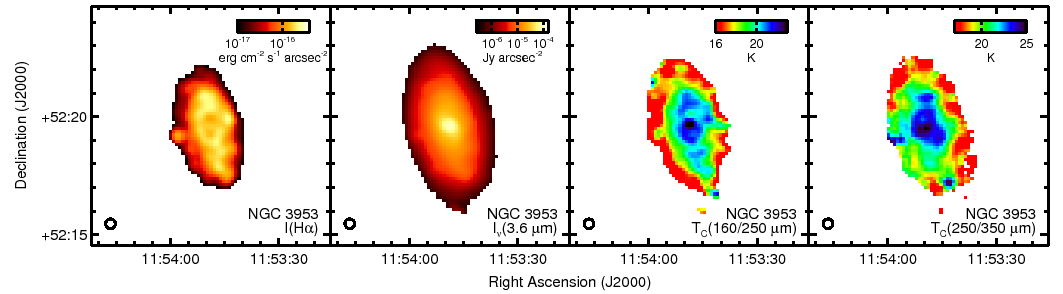}
\includegraphics[width=0.95\textwidth]{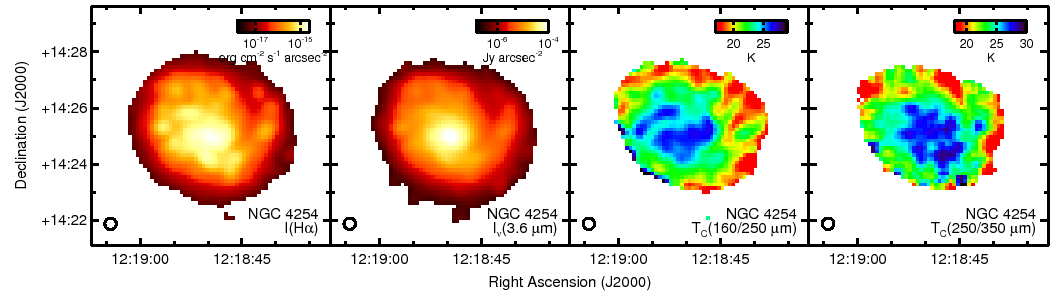}
\includegraphics[width=0.95\textwidth]{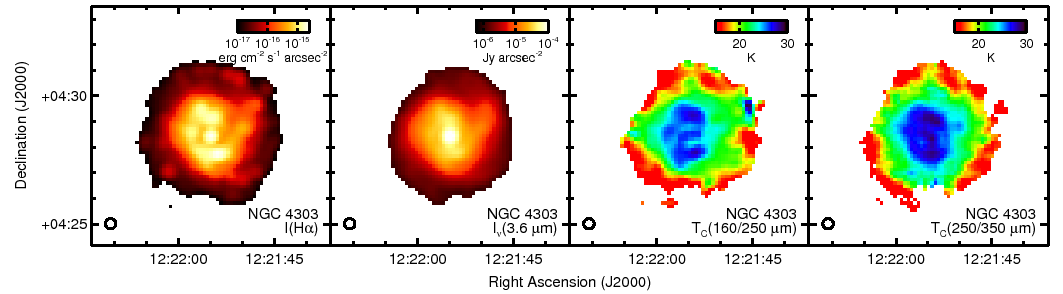}
\includegraphics[width=0.95\textwidth]{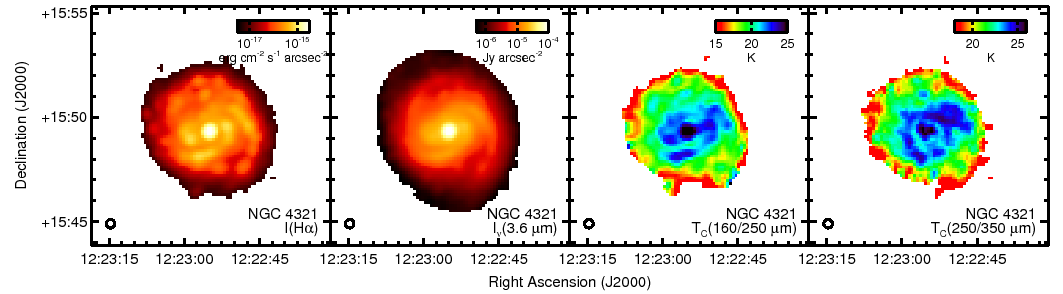}
\caption{Continued.}
\end{figure*}

\addtocounter{figure}{-1}
\begin{figure*}
\includegraphics[width=0.95\textwidth]{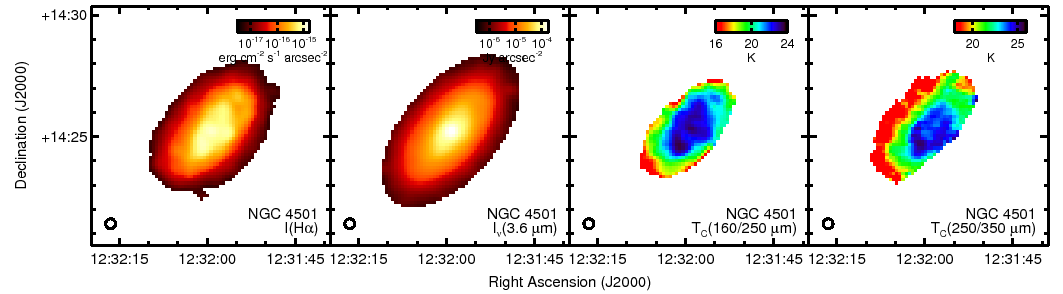}
\includegraphics[width=0.95\textwidth]{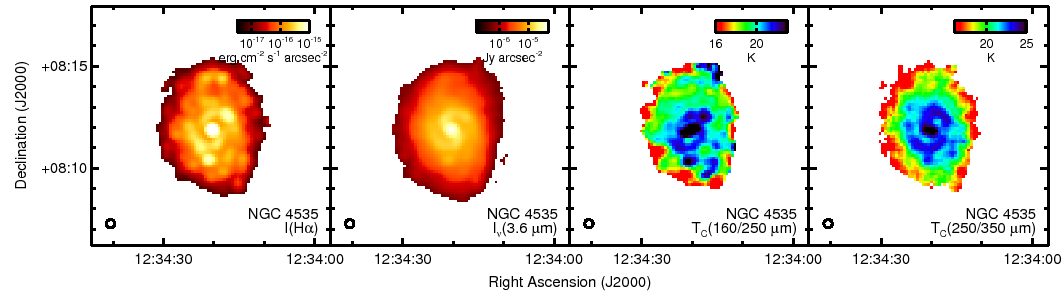}
\includegraphics[width=0.95\textwidth]{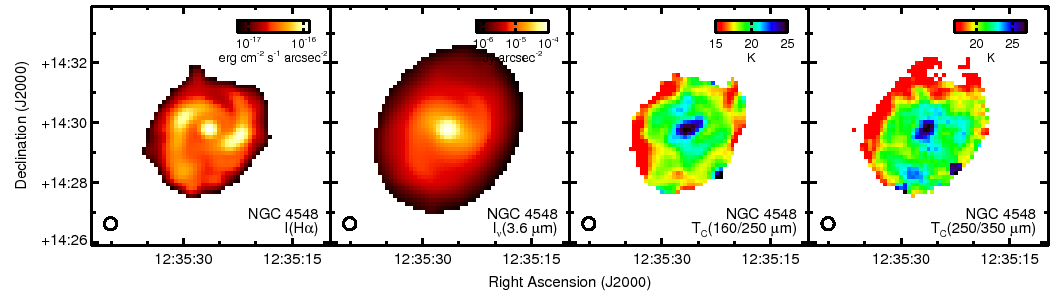}
\includegraphics[width=0.95\textwidth]{bendogj_fig01b.png}
\caption{Continued.}
\end{figure*}

\addtocounter{figure}{-1}
\begin{figure*}
\includegraphics[width=0.95\textwidth]{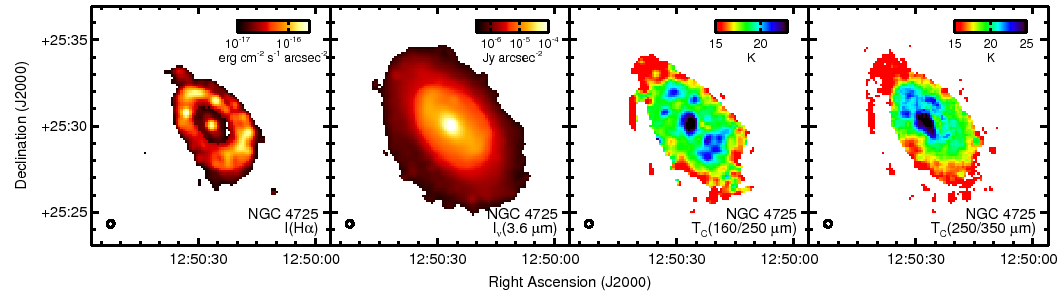}
\includegraphics[width=0.95\textwidth]{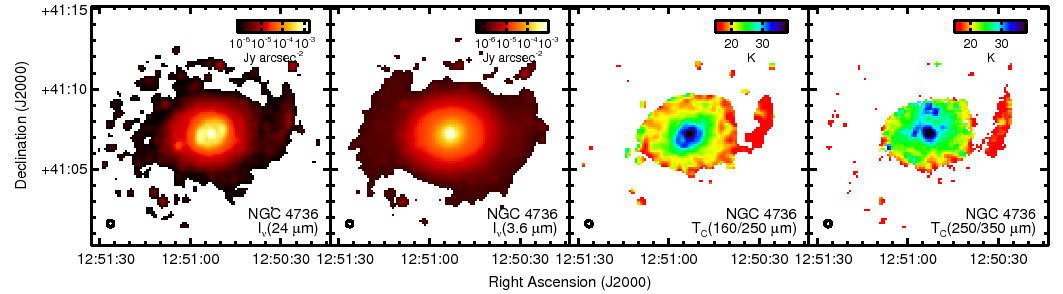}
\includegraphics[width=0.95\textwidth]{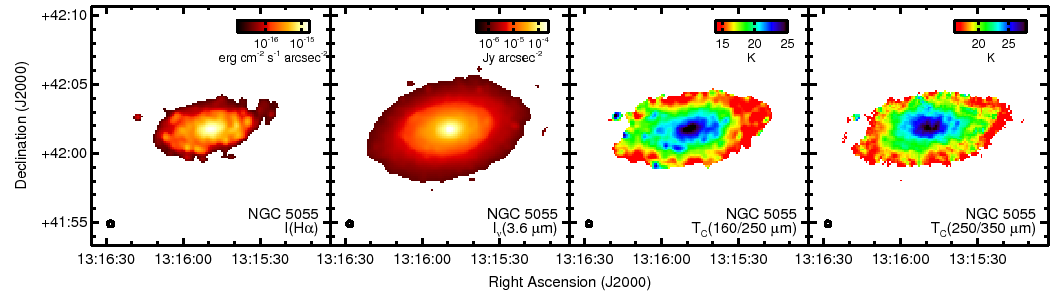}
\includegraphics[width=0.95\textwidth]{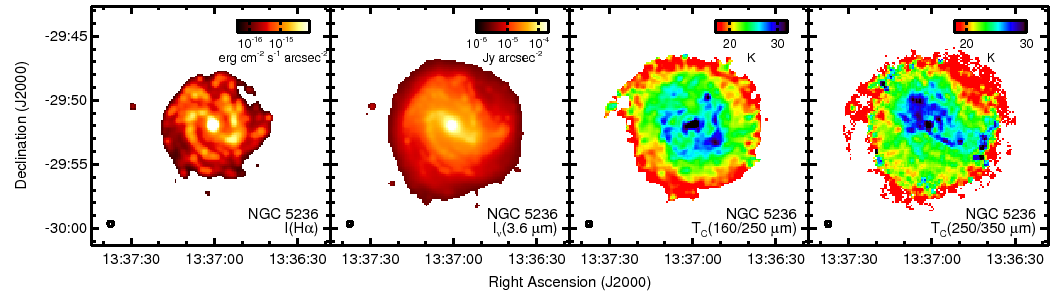}
\caption{Continued.}
\end{figure*}

\addtocounter{figure}{-1}
\begin{figure*}
\includegraphics[width=0.95\textwidth]{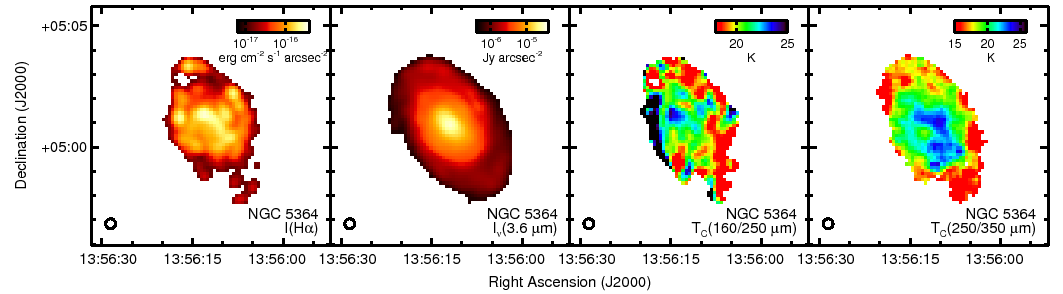}
\includegraphics[width=0.95\textwidth]{bendogj_fig01a.png}
\includegraphics[width=0.95\textwidth]{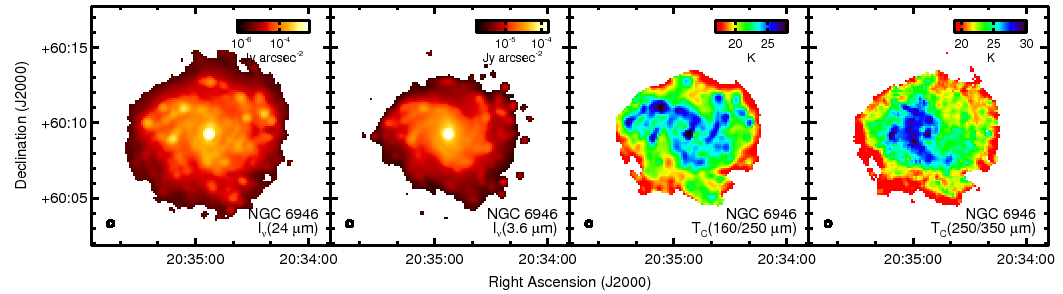}
\includegraphics[width=0.95\textwidth]{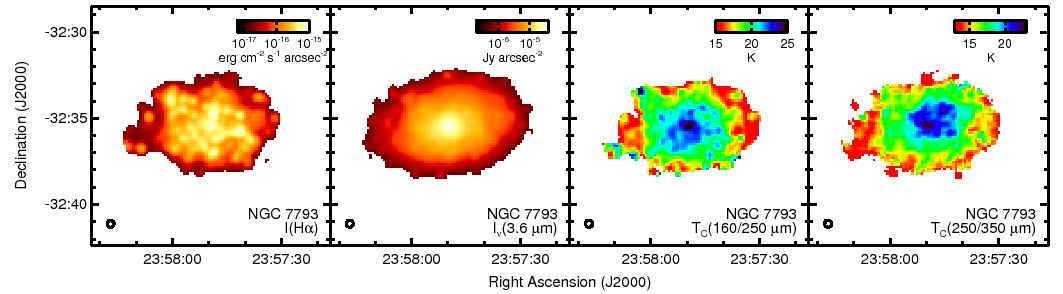}
\caption{Continued.}
\end{figure*}

\begin{figure*}
\includegraphics[width=0.8\textwidth]{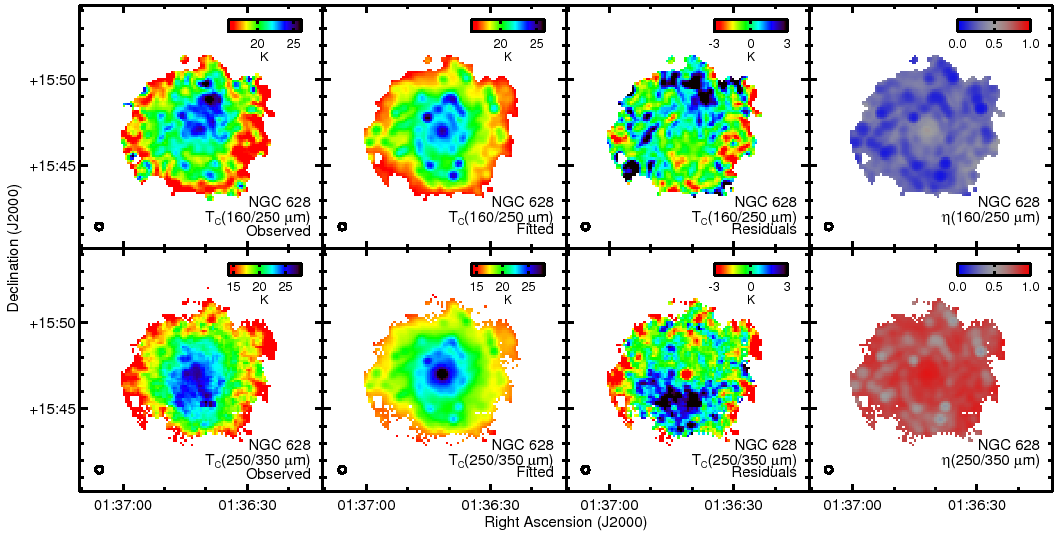}
\includegraphics[width=0.8\textwidth]{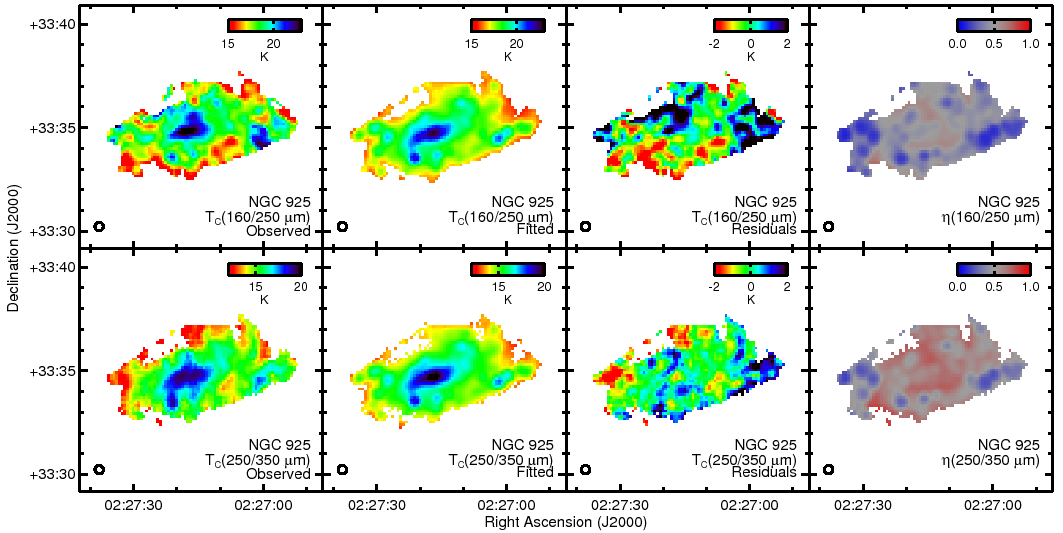}
\includegraphics[width=0.8\textwidth]{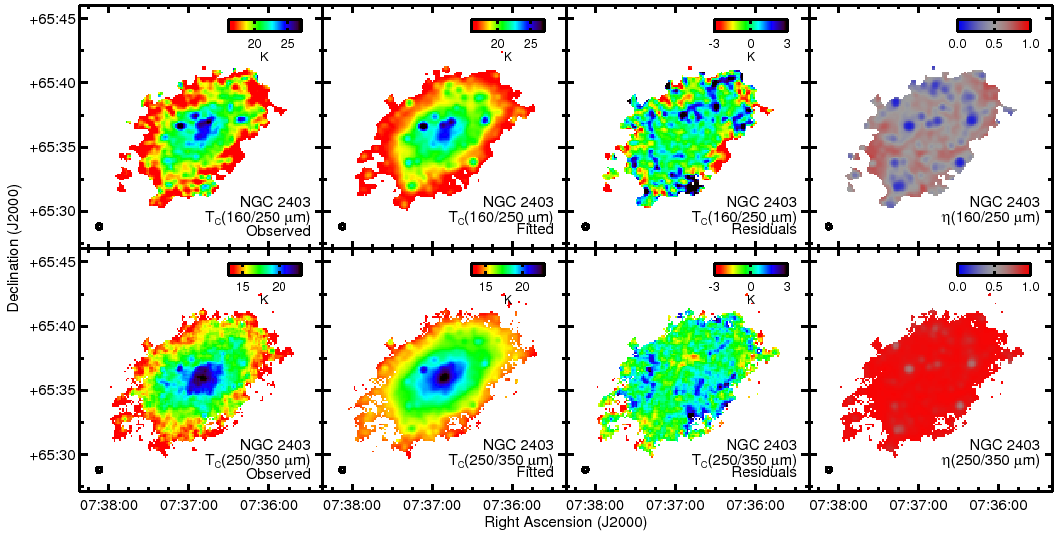}
\caption{Examples of the results from using Equation~2 to fit the
  160/250 and 250/350~$\mu$m surface brightness ratios as a function
  of both the H$\alpha$ and 3.6~$\mu$m data.  The first columns of
  images on the left show the observed colour temperatures.  The
  second column of images show the colour temperatures produced by
  adding together the H$\alpha$ (or 24~$\mu$m) and 3.6~$\mu$m images
  using Equation~2 and the best fitting parameters from Table~5.  The
  third column shows the residuals from subtracting the fitted colour
  temperature map from the observed colour temperature map.  The
  fourth column shows the fraction of dust heating by star forming
  regions predicted by the parameters in Table~5 and Equation~3.  The
  images are formatted in the same way as the colour temperature
  images in Figure~1.}
  \label{f_multfitall}
\end{figure*}

\addtocounter{figure}{-1}
\begin{figure*}
\includegraphics[width=0.8\textwidth]{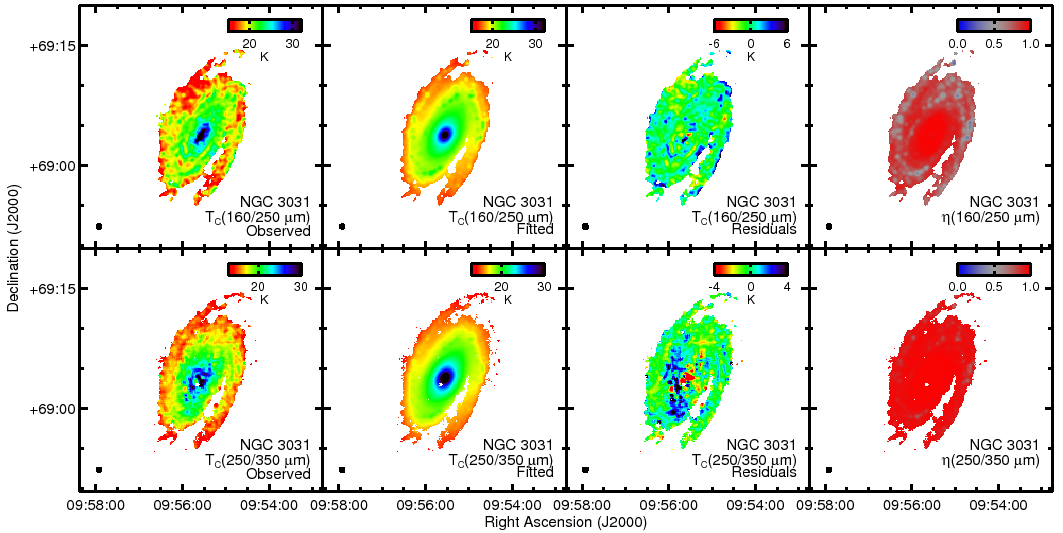}
\includegraphics[width=0.8\textwidth]{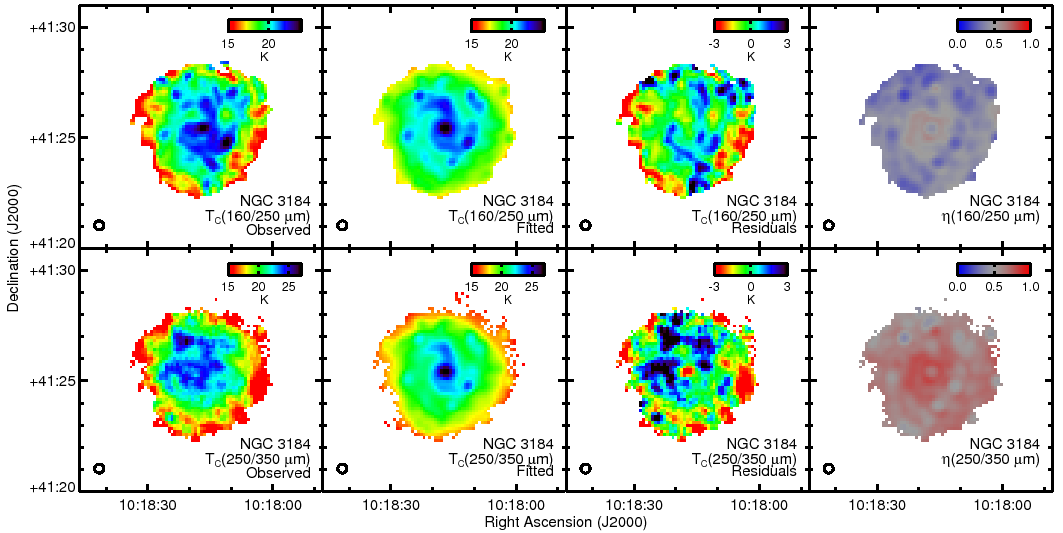}
\includegraphics[width=0.8\textwidth]{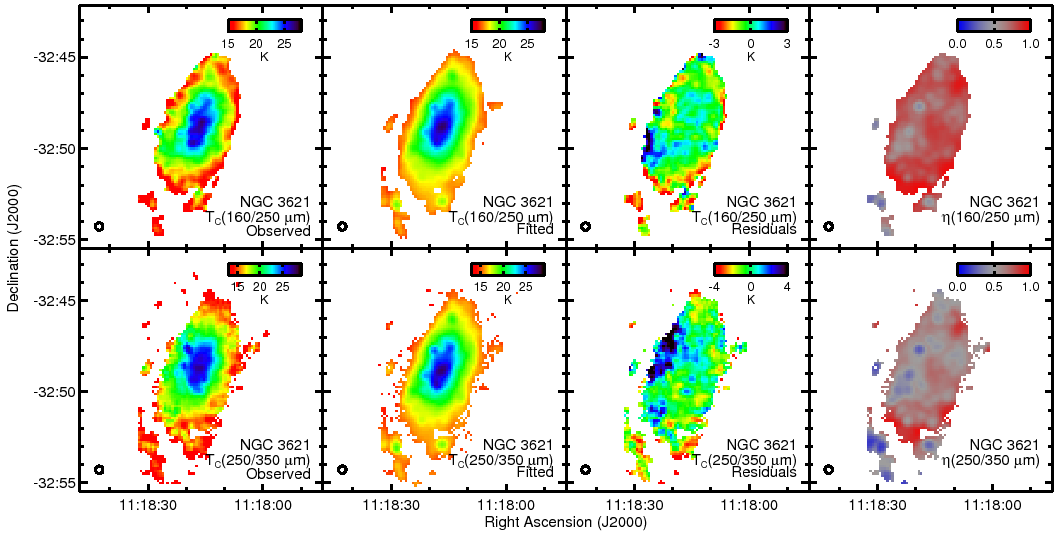}
\caption{Continued.}
\end{figure*}

\addtocounter{figure}{-1}
\begin{figure*}
\includegraphics[width=0.8\textwidth]{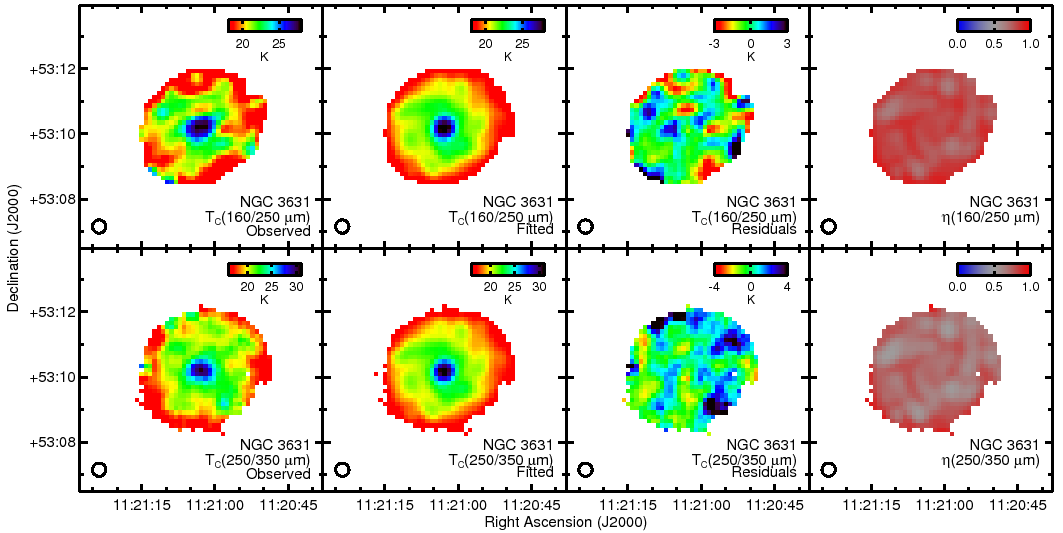}
\includegraphics[width=0.8\textwidth]{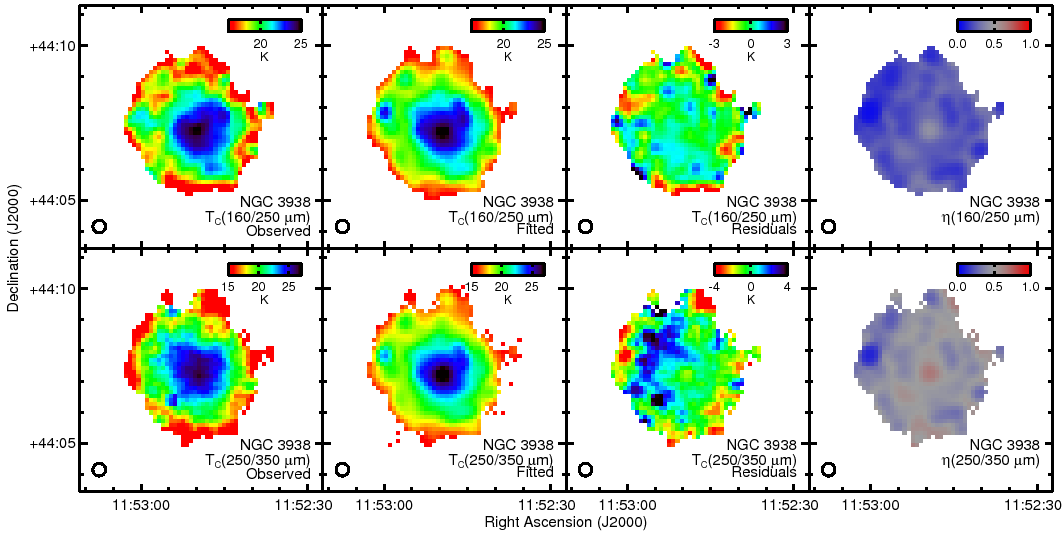}
\includegraphics[width=0.8\textwidth]{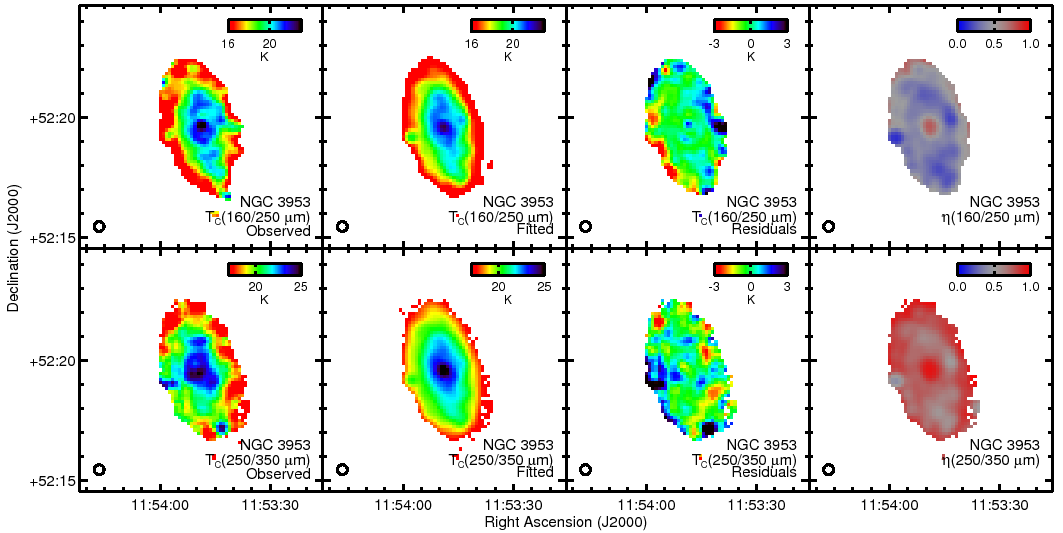}
\caption{Continued.}
\end{figure*}

\addtocounter{figure}{-1}
\begin{figure*}
\includegraphics[width=0.8\textwidth]{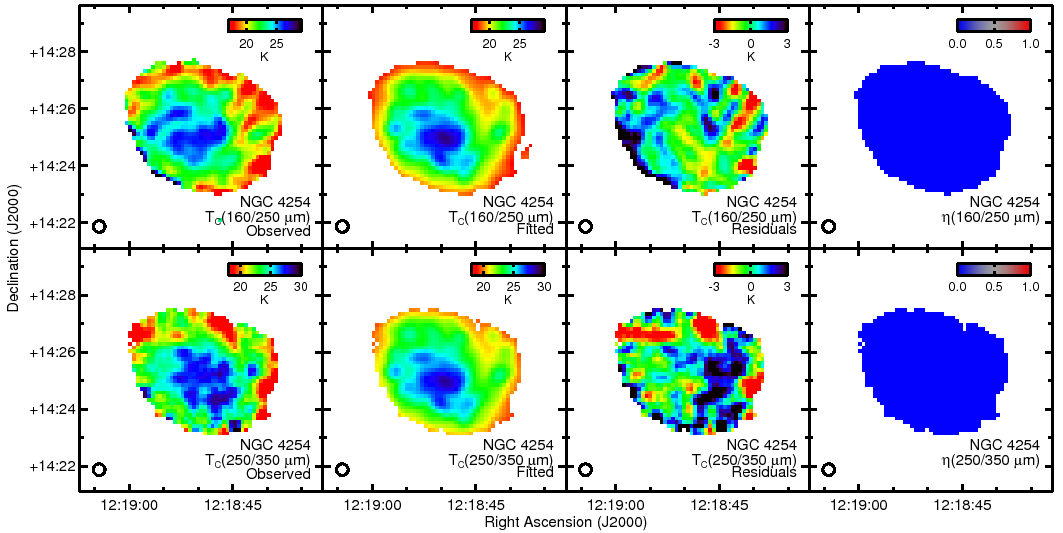}
\includegraphics[width=0.8\textwidth]{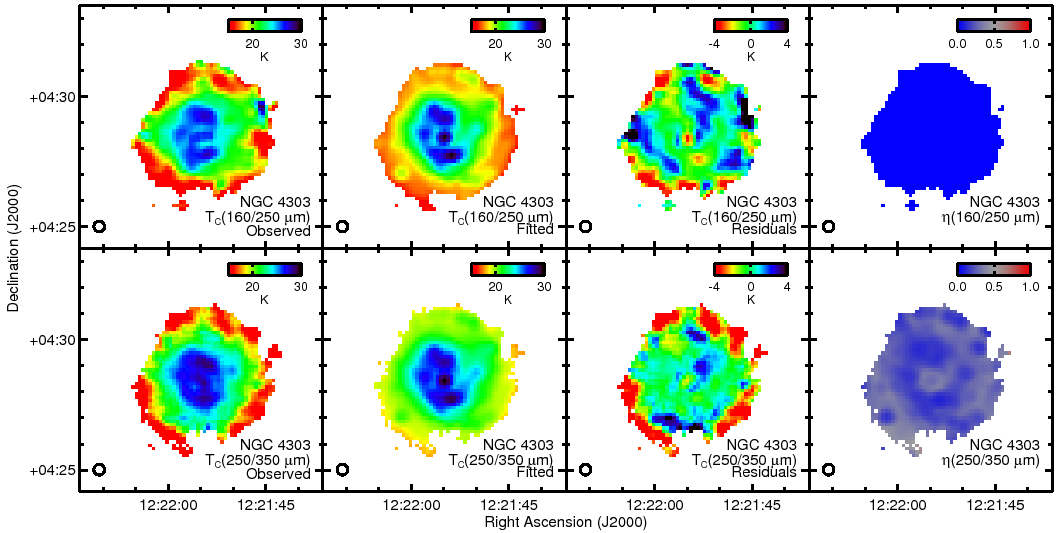}
\includegraphics[width=0.8\textwidth]{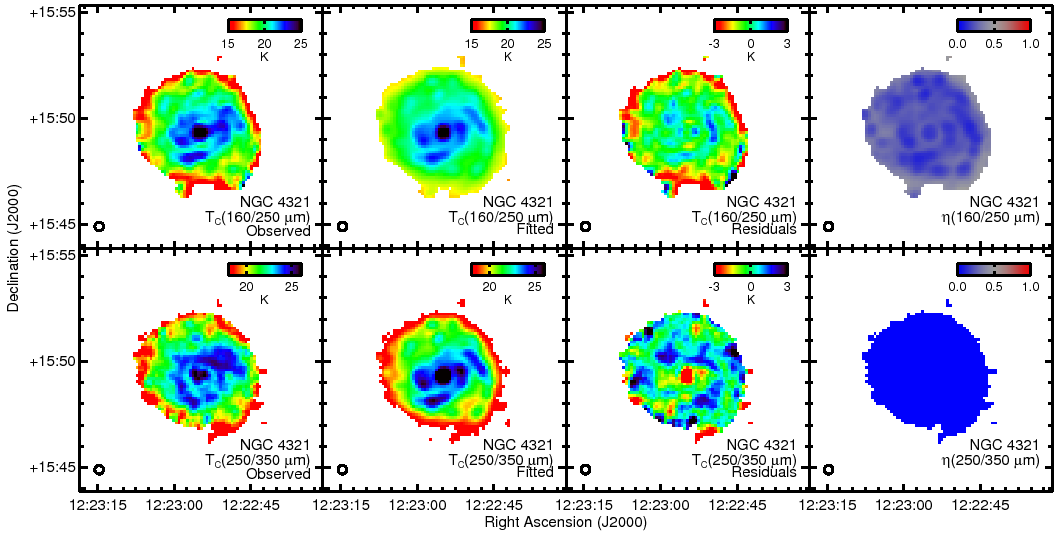}
\caption{Continued.}
\end{figure*}

\addtocounter{figure}{-1}
\begin{figure*}
\includegraphics[width=0.8\textwidth]{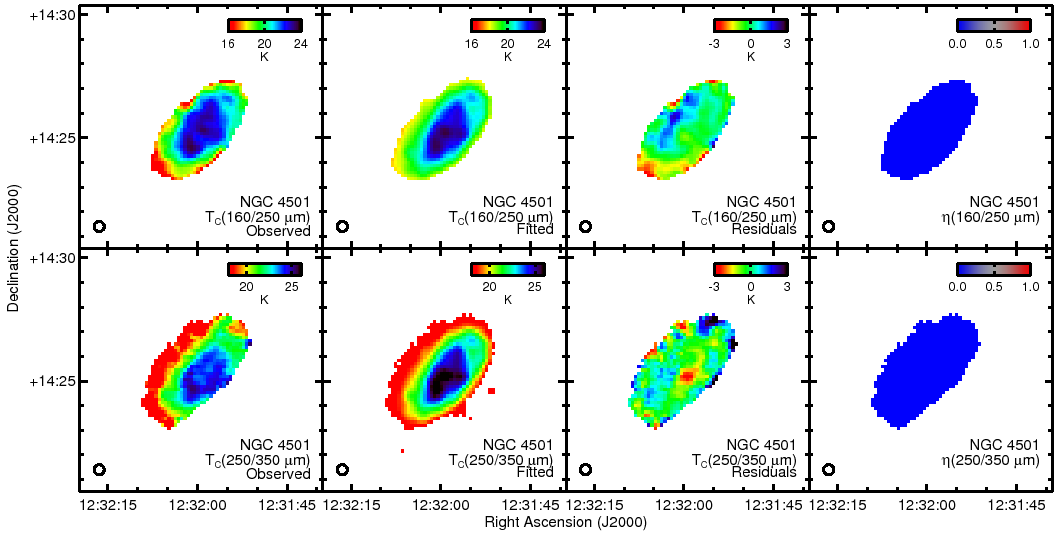}
\includegraphics[width=0.8\textwidth]{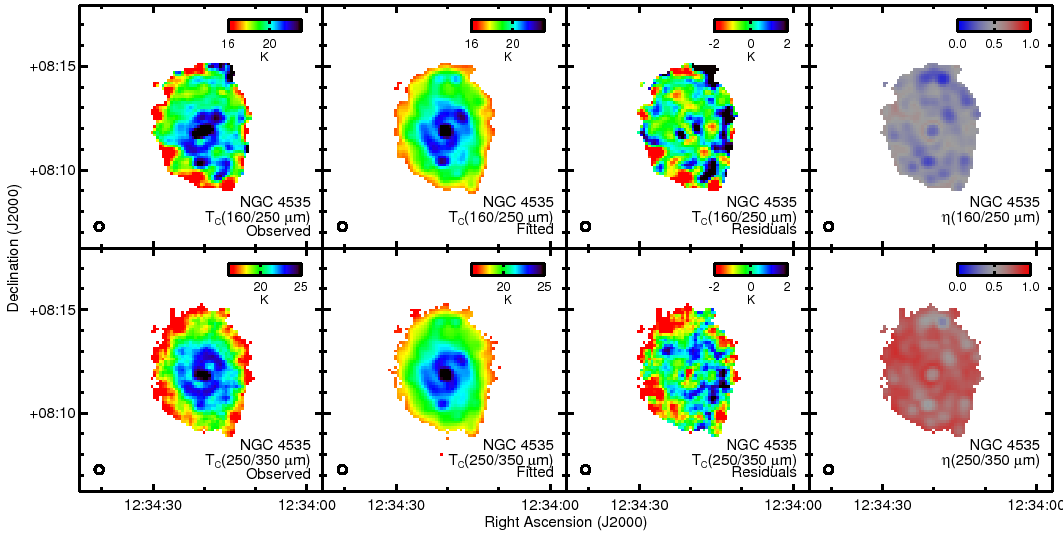}
\includegraphics[width=0.8\textwidth]{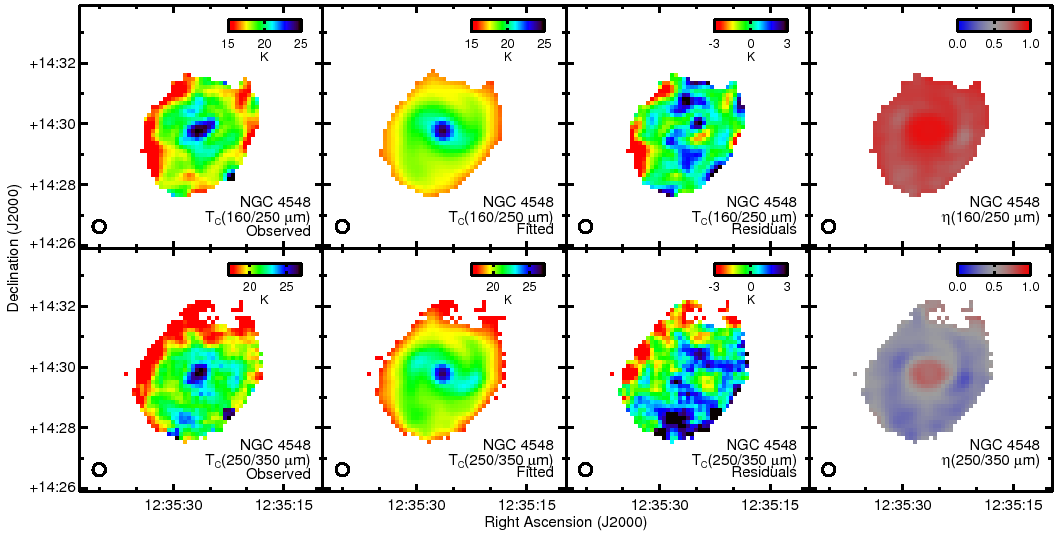}
\caption{Continued.}
\end{figure*}

\addtocounter{figure}{-1}
\begin{figure*}
\includegraphics[width=0.8\textwidth]{bendogj_fig07b.png}
\includegraphics[width=0.8\textwidth]{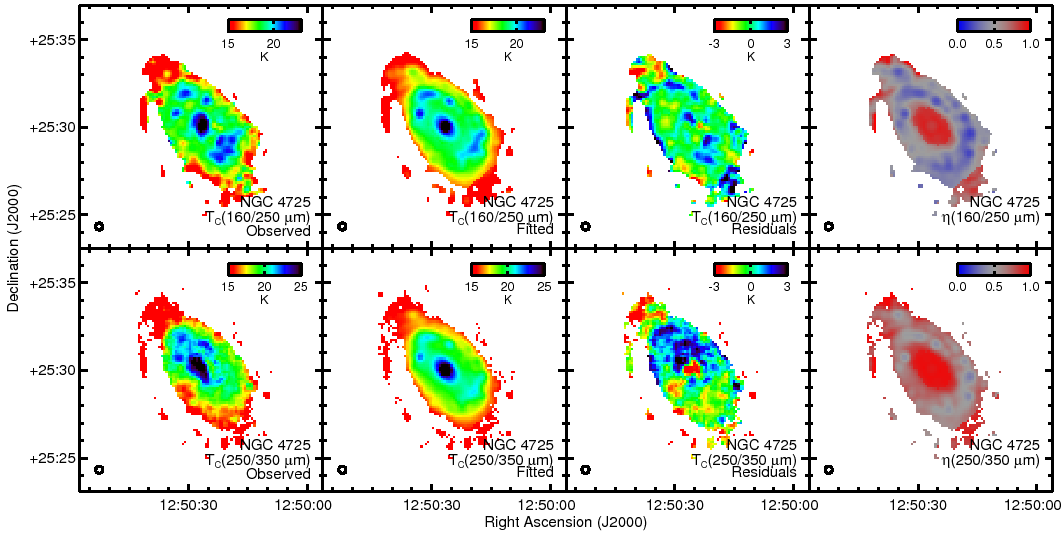}
\includegraphics[width=0.8\textwidth]{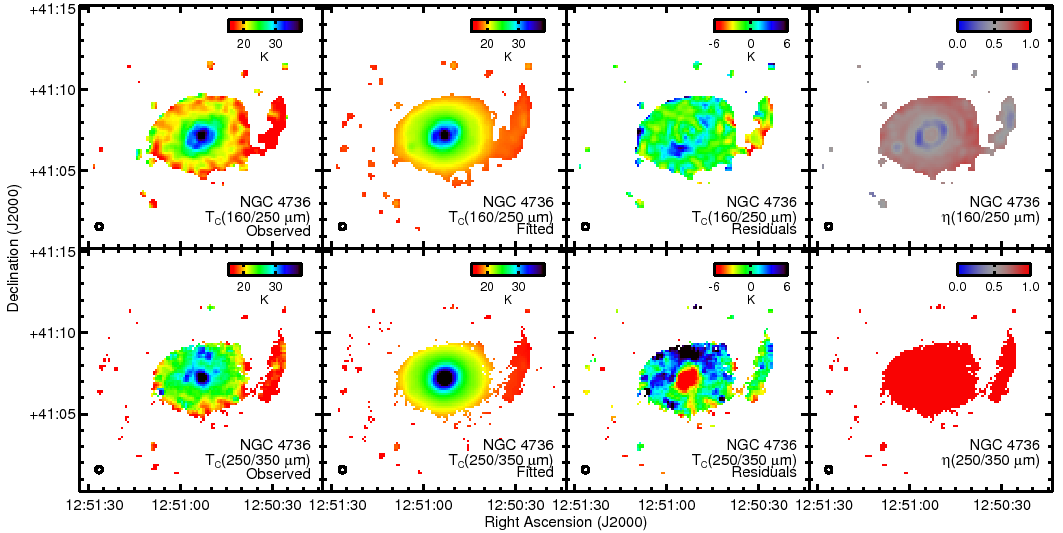}
\caption{Continued.}
\end{figure*}

\addtocounter{figure}{-1}
\begin{figure*}
\includegraphics[width=0.8\textwidth]{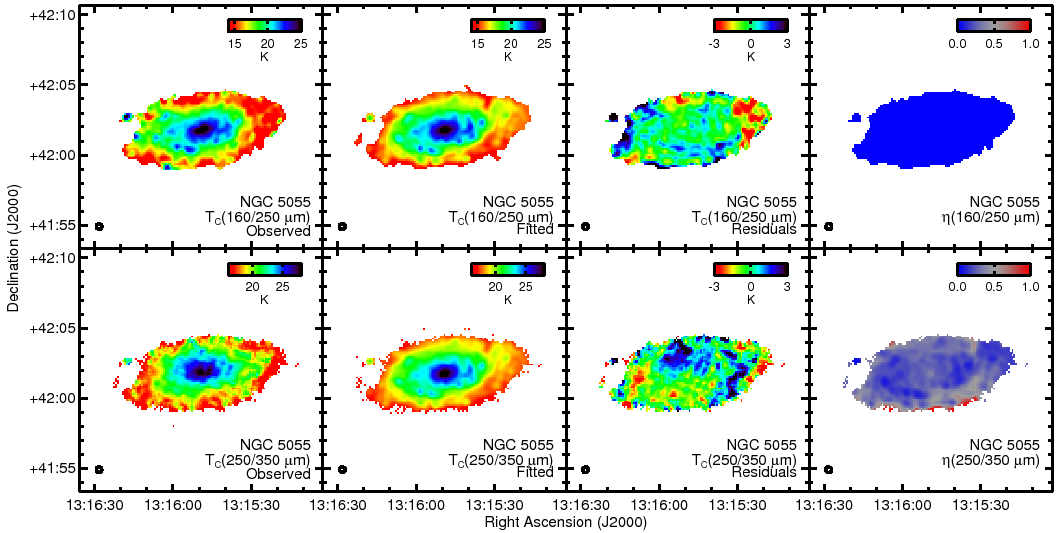}
\includegraphics[width=0.8\textwidth]{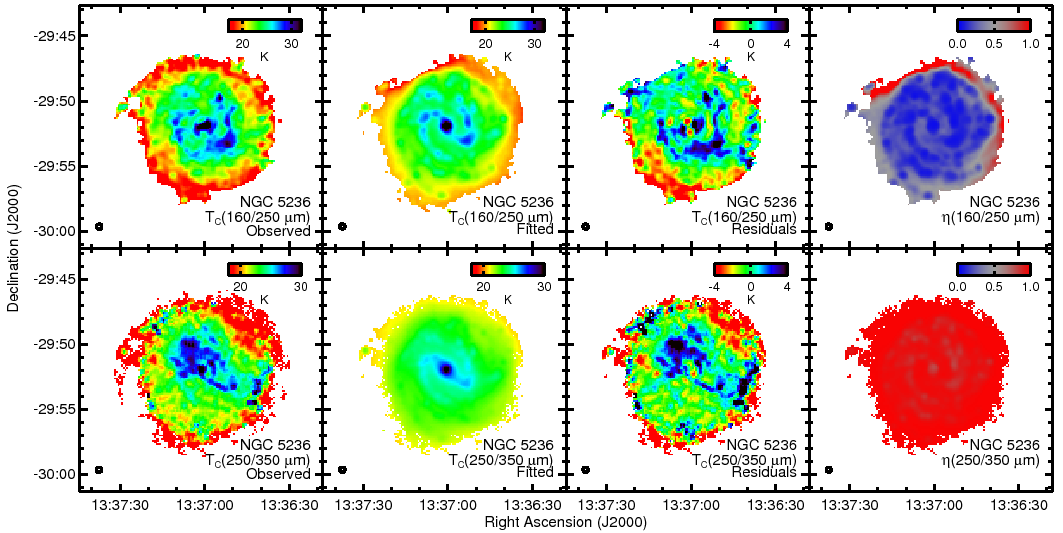}
\includegraphics[width=0.8\textwidth]{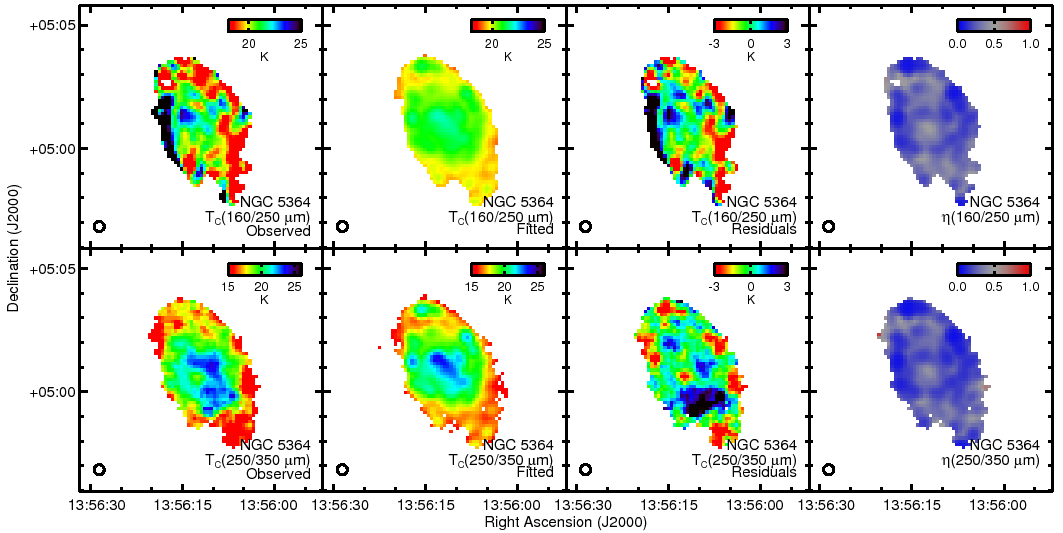}
\caption{Continued.}
\end{figure*}

\addtocounter{figure}{-1}
\begin{figure*}
\includegraphics[width=0.8\textwidth]{bendogj_fig07a.png}
\includegraphics[width=0.8\textwidth]{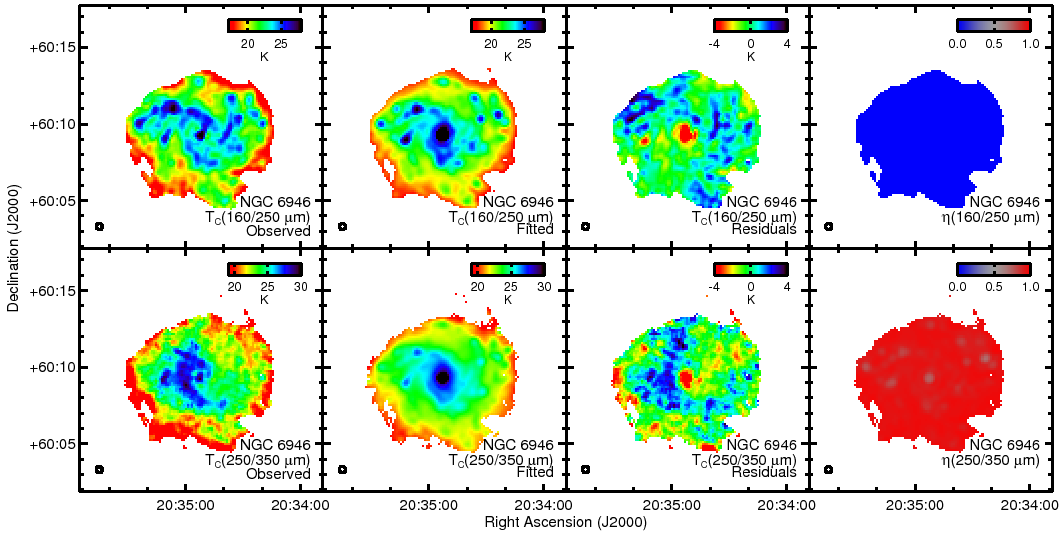}
\includegraphics[width=0.8\textwidth]{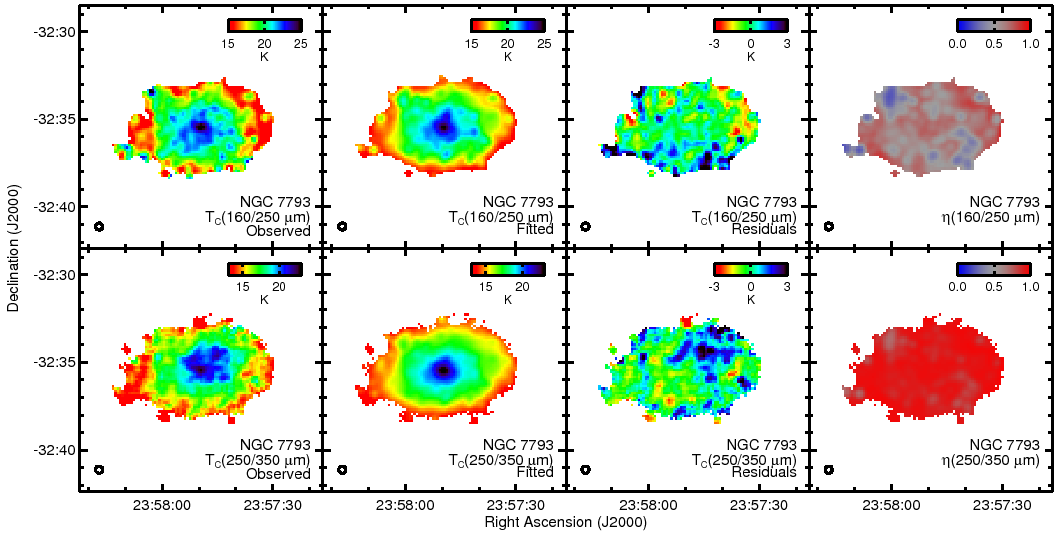}
\caption{Continued.}
\end{figure*}

\label{lastpage}

\end{document}